\documentclass[a4paper,11pt]{article}

\usepackage{jcappub} 

\usepackage[T1]{fontenc} 

\usepackage{amsmath,amssymb}
\usepackage[tight,footnotesize]{subfigure}
\usepackage[usenames]{color}
\usepackage[dvipsnames]{xcolor}
\usepackage{float}
\usepackage{wrapfig}
\usepackage{hyperref}
\usepackage{mathtools}
\usepackage[utf8]{inputenc}
\usepackage{graphicx}
\usepackage{soul}
\graphicspath{ {./images/} }

\usepackage{xcolor}

\begin{document}

\newcommand{\com}{{\cal C}}
\newcommand{\fdl}[1]{{\textcolor{violet}{#1}}}
\newcommand{\SC}[1]{{\textcolor{blue}{#1}}}

\newcommand{\tcb}{\textcolor{blue}}
\newcommand{\dd}{\mathrm{d}} 

\def\ga{\mathrel{\raise.3ex\hbox{$>$\kern-.75em\lower1ex\hbox{$\sim$}}}}
\def\la{\mathrel{\raise.3ex\hbox{$<$\kern-.75em\lower1ex\hbox{$\sim$}}}}

\def\be{\begin{equation}}
\def\ee{\end{equation}}
\def\bea{\begin{eqnarray}}
\def\eea{\end{eqnarray}}

\def\betap{\tilde\beta}
\def\del{\delta_{\rm PBH}^{\rm local}}
\def\Msun{M_\odot}
\def\Rcl{R_{\rm clust}}
\def\fPBH{f_{\rm PBH}}

\newcommand{\Mpl}{M_P} 
\newcommand{\mpl}{m_\mathrm{pl}} 

\definecolor{MONZA}{HTML}{CF000F}
\definecolor{DARKBLUE}{HTML}{00008b}
\definecolor{DARKMAGENTA}{HTML}{8b008b}
\definecolor{DARKCYAN}{HTML}{008B8B}
\definecolor{DARKORANGE}{HTML}{FF8C00}

\newcommand{\Red}[1]{\textcolor{MONZA}{\sffamily #1}}
\newcommand{\Blue}[1]{\textcolor{DARKBLUE}{\sffamily #1}}
\newcommand{\Cyan}[1]{\textcolor{DARKCYAN}{\sffamily #1}}
\newcommand{\Mag}[1]{\textcolor{DARKMAGENTA}{\sffamily #1}}
\newcommand{\mathblue}[1]{\textcolor{DARKBLUE}{#1}}
\newcommand{\mathmag}[1]{\textcolor{DARKMAGENTA}{#1}}

\newcommand{\AEM}[1]{\textcolor{DARKMAGENTA}{\sffamily [AEM: #1]}}
\newcommand{\EB}[1]{\textcolor{green}{\sffamily [EB: #1]}}

\title{Simulations of PBH formation at the QCD epoch and comparison with the GWTC-3 catalog}

\author[a]{Albert Escriv\`a}

\author[a]{Eleni Bagui}

\author[a]{Sebastien Clesse}

\affiliation[a]{Service de Physique Th\'eorique, Universit\'e Libre de Bruxelles (ULB), Boulevard du Triomphe, CP225, 1050 Brussels, Belgium}

\emailAdd{albert.escriva@ulb.be}
\emailAdd{eleni.bagui@ulb.be}
\emailAdd{sebastien.clesse@ulb.be}

\date{\today}

\abstract{The probability of primordial black hole (PBH) formation is known to be boosted during the Quantum Chromodynamics (QCD) crossover due to a slight reduction of the equation of state.  This induces a high peak and other features in the PBH mass distribution.  But the impact of this variation during the PBH formation has been so far neglected. In this work we simulate for the first time the formation of PBHs by taking into account the varying equation of state at the QCD epoch, compute the over-density threshold using different curvature profiles
and find that the resulting PBH mass distributions are significantly impacted. The expected merger rate distributions of early and late PBH binaries is comparable to the ones inferred from the GWTC-3 catalog for dark matter fractions in PBHs within $0.1 < f_{\rm PBH} <1 $.  The distribution of gravitational-wave events estimated from the volume sensitivity could explain mergers around $30-50 M_\odot$, with asymmetric masses like GW190814, or in the pair-instability mass gap like GW190521.  However, none of the considered cases leads to a multi-modal distribution with a secondary peak around $8-15 M_\odot$, as suggested by the GWTC-3 catalog, possibly pointing to a mixed population of astrophysical and primordial black holes. }

\maketitle
\flushbottom

\section{Introduction}
The field of primordial black holes (PBHs) is nowadays one of the most fascinating areas of research in cosmology.  The first detection of gravitational waves from a binary black hole merger by LIGO/Virgo~\cite{LIGO} has triggered a renewed interest for PBHs after it was suggested that they could be primordial~\cite{Bird,Clesse:2016vqa,Sasaki:2016jop}. PBHs, which were first proposed in~\cite{acreation1,hawking1,hawking2,Chapline:1975ojl}, are black holes that could have formed in the very early Universe due to the collapse of large non-linear peaks in the primordial density fluctuations. PBHs could have a lot of implications in our Universe, like being the seeds of supermassive black holes~\cite{PhysRevD.66.063505,Clesse:2015wea,Bernal:2017nec}, generating large-scale structures~\cite{Meszaros:1975ef,darkmatter6}, ultra-faint dwarf galaxies~\cite{Clesse:2017bsw} or changing the thermal history of the Universe~\cite{10.1093/mnras/194.3.639}. But probably the most remarkable implication is that they can constitute a significant fraction  or even the totality of the dark matter~\cite{Carr1,darkmatter1,Bird,Clesse:2016vqa,Sasaki:2016jop,Sasaki:2018dmp,Clesse:2018ogk,Tada:2019amh,Carr:2020gox,Carr:2019kxo,Clesse:2020ghq,Hall:2020daa,Jedamzik:2020ypm,Jedamzik:2020omx,Franciolini:2021tla}. 

If PBHs formed from the collapse of inflationary perturbations~\cite{darkmatter1} (see~\cite{Carr1,Carr:2020xqk,Escriva:2021aeh} for reviews of other mechanisms), their abundance is exponentially sensitive to the over-density threshold leading to the collapse, $\delta_{\rm c}$~\cite{carr75}.
In order to obtain the necessary precision on $\delta_{\rm c}$, simulations of PBH formation based on numerical relativity are needed, although some useful analytical estimations have been pointed out in the literature~\cite{carr75,harada,Escriva:2020tak,universal1}, where the last two take into account the specific shape of the fluctuation.  So far, all these works assume a constant equation of state \emph{during} PBH formation. 
During the radiation epoch, large cosmological fluctuations collapse and form PBHs when they re-enter inside the cosmological horizon~\cite{carr75,Carr1} (see also~\cite{Harada:2017fjm,Harada:2016mhb,Harada:2017fjm,KHLOPOV1980383} for a matter-dominated phase).  However, in the very early Universe, there are different epochs when the equation of state $w$ can change with time. This can be due to a still unknown phase transition or to the standard thermal history of the Universe~\cite{Allahverdi:2020bys} when particles become non-relativistic, and especially at the QCD crossover when protons, neutrons and pions are formed. Because the PBH abundance is exponentially sensitive to the threshold $\delta_{\rm c}$, which itself depends on $w$, even a slight change in $w$ has important effects on the resulting PBH mass distribution~\cite{Escriva:2020tak,musco2013}.  
Remarkably interesting is the effect on the formation of PBHs during the QCD crossover phase, about $ \sim 10^{-5} {\rm s}$ after the Big Bang~\cite{Schmid:1998mx,Laine:2006cp,Widerin:1998my,Boeckel:2010bey,Jedamzik:1996mr,Bhattacharya:2014ara,Carr:2019hud,Laine:2015kra,Borsanyi:2016ksw,Byrnes:2018clq,Saikawa:2018rcs,Juan:2022mir}, from horizon-size fluctuations of order of stellar masses.  At this epoch, the equation of state exhibits a transient reduction from $w =1/3$ down to $w \approx 0.23$, inducing a net reduction of $\delta_{\rm c}$ due to smaller pressure gradients~\cite{musco2013,Escriva:2020tak}.  The probability of PBH formation is therefore inevitably boosted at the QCD epoch and can lead to a high peak in the PBH distribution at the solar-mass scale, likely to be about a hundred times larger in comparison with the standard scenario of a radiation epoch~\cite{Sobrinho:2016fay,Byrnes:2018clq,Carr:2019kxo,Gao:2021nwz,Abe:2020sqb,Clesse:2020ghq}, as well as a  bump covering the mass range between $20 M_\odot$ and $100 M_\odot$, linked to the formation of pions.  Such features induced by the QCD transition could explain GW observations of compact binary coalescences~\cite{Carr:2019kxo,Clesse:2020ghq,Jedamzik:2020omx,Jedamzik:2020ypm,DeLuca:2020sae}, in particular intriguing exceptional events like GW190521, which has at least one black hole progenitor in the pair-instability mass gap~\cite{LIGOScientific:2020iuh}, GW190814 that has a very low mass ratio and a non spinning primary component~\cite{LIGOScientific:2020zkf}, or GW190425~\cite{LIGOScientific:2020aai} with at least one component mass too low for being an astrophysical black hole but larger than all the precisely known neutron star masses.

All the previous works on the effect of the QCD crossover~\cite{Byrnes:2018clq,Carr:2019kxo,Clesse:2020ghq,Jedamzik:2020omx,Jedamzik:2020ypm,Bagui:2021dqi,Juan:2022mir} are based on analytical or numerical results assuming a constant $w$ during PBH formation, which highlights the need of more accurate simulations in numerical relativity~\cite{escriva_solo,refrencia-extra-jaume,new2,Yoo:2021fxs,musco2013,Escriva:2020tak,  musco2009, musco2005, hawke2002,Niemeyer1,Niemeyer2} to capture the highly non-linear dynamics of the process, with the expected time-dependent equation of state. 
This is the main goal of the present work, leading us to
i) analyse how the PBH formation dynamics are affected in the presence of the QCD crossover for different profiles; ii) compute more accurately the threshold for PBH formation; iii) study how the PBH mass can change; iv) better estimate the resulting PBH mass function.  Bi-products of this analysis are better determinations of the expected PBH merger rates and of possible distributions of GW detections.  These are compared to the latest data from the third observing run by LIGO/Virgo, in particular to the latest GWTC-3 catalog~\cite{LIGOScientific:2021djp,LIGOScientific:2021psn}.

The paper is organized as follows:  In Sec.~\ref{sec:basics} we introduce the mathematical setup.  The initial conditions are specified in Sec.~\ref{sec:initial_conditios} and the numerical technique is described in Sec.~\ref{sec:numerics}. The simulated dynamics before and after the formation of the PBH horizon are studied in Secs.~\ref{sec:dynamics} to \ref{sec:pbhmass}. The impact on the PBH mass distributions is analysed in Sec.~\ref{sec:pbh_distribution}.  We compute the merger rates of early and late PBH binaries in Sec.~\ref{sec:merger_rates} and the expected distributions of GW detections for the sensitivity of the third observing run of LIGO/Virgo in Sec.~\ref{sec:catalog}.  Our conclusions are presented and discussed in Sec.~\ref{sec:conclusions}.  We include at the end a final note about a paper by Franciolini et al~\cite{Franciolini:2022tfm} on the same topic that was released simultaneously with our work, but written in total independence. We shortly comment on our respective results.

\section{Theoretical and numerical set-up}\label{sec:basics}

\subsection{Misner-Sharp equations}
\begin{figure}[t]
\centering
\includegraphics[width=3.5 in]{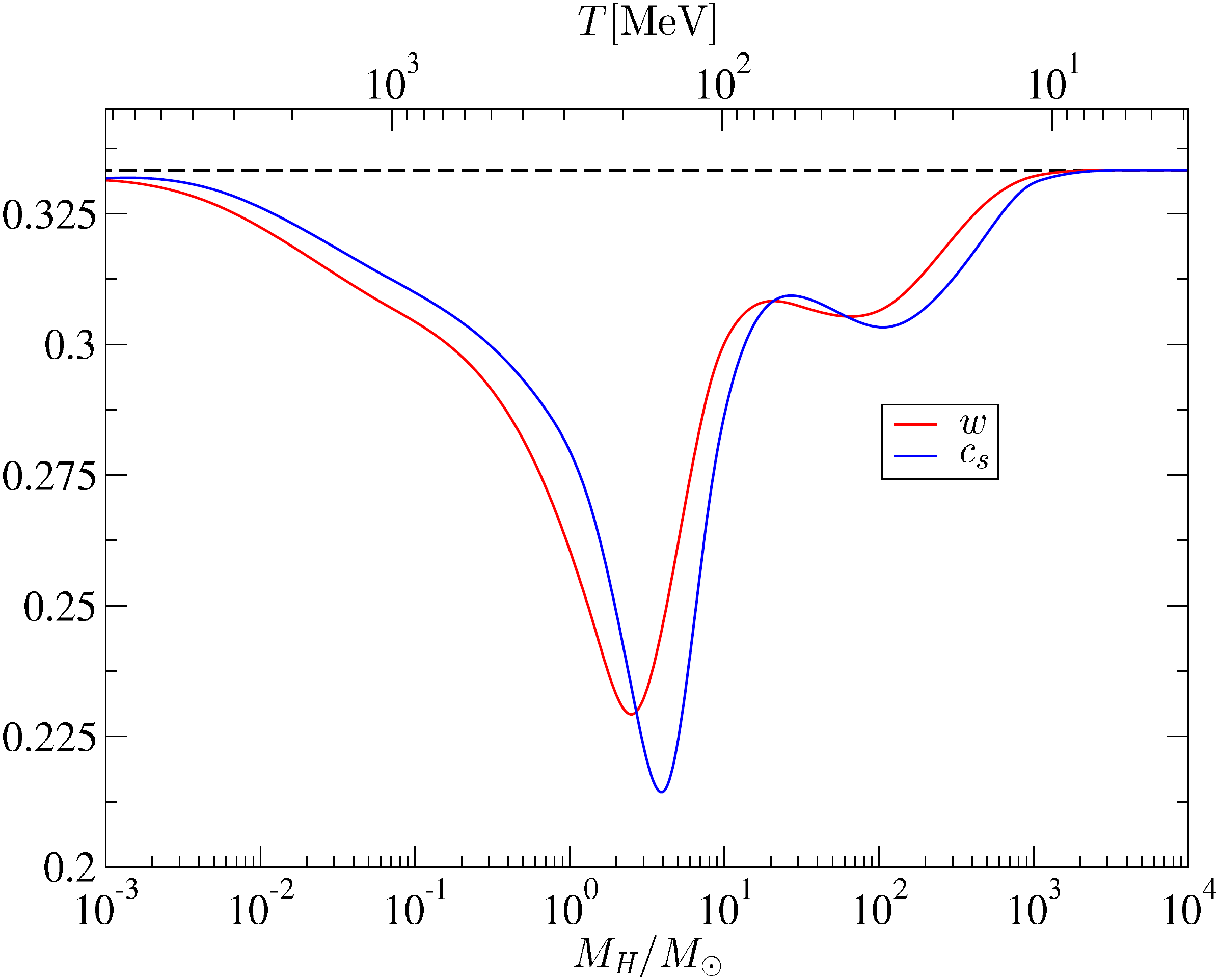}
\caption{Evolution of $w$, $c^{2}_{\rm s}$ (red and blue lines respectively) at the QCD epoch, from the tabulated values of~\cite{qcd_table}, as a function of the horizon mass $M_{\rm H}$ in solar masses.  On the top axis we show the corresponding temperature $T$ in MeV. The dashed black line represents the value for a radiation dominated Universe $w=c^{2}_{\rm s}=1/3$.}
\label{fig:tabulated_data}
\end{figure}

We consider the formation of PBHs assuming spherical symmetry, in a Universe filled with a perfect fluid, whose energy momentum tensor is given by:
\begin{equation}
T^{\mu \nu} = (p+\rho)u^{\mu}u^{\nu}+pg^{\mu\nu},
 \label{eq:tensor_energy}
\end{equation}
and with the following spacetime metric,
\begin{equation}
\label{2_metricsharp}
{\rm d} s^2 = -A(r,t)^2 {\rm d} t^2+B(r,t)^2 {\rm d} r^2 + R(r,t)^2 {\rm d}\Omega^2,
\end{equation}
where $t$ is the cosmic time, $R(r,t)$ is the areal radius, $A(r,t)$ is the lapse function, and ${\rm d} \Omega^{2} = {\rm d}\theta^2+\sin^2(\theta) {\rm d}\phi^2$.
The components of the four-velocity $u^{\mu}$ are given by $u^{t}=1/A$ and $u^{i}=0$ for $i=r,\theta,\phi$, since we are considering comoving coordinates (comoving gauge). We use geometrized units, $G=c=1$. We consider tabluated values of the equation of state $w = p/\rho $, where $p$ is the pressure and $\rho$ is the energy density, during the QCD crossover transition from~\cite{qcd_table} and coming from lattice QCD calculations.  We use a cubic spline to obtain values between the reference data points. The energy density $\rho$, the entropy density $s$ and the equation of state $w$ are given in terms of the relativistic particle degrees of freedom $g_{\rho}$ and $g_{\rm s}$, which depend on the temperature $T$ through
\begin{align}
\label{eq:freedom}
& g_{\rho}(T) = \frac{30 \rho}{\pi^{2}T^{4}}, \hspace{1.5cm}  g_{\rm s}(T) = \frac{45 s}{2 \pi^{2} T^{3}} \nonumber\\
& w(T) = \frac{4 g_{\rm s}(T)}{3 g_{\rho}(T)}-1, \hspace{0.6cm}  c^{2}_{\rm s} = \frac{\partial p}{\partial \rho} =\frac{4(4 g_{\rm s}+ T g'_{\rm s} )}{3(4 g_{\rho} + T g'_{\rho} )}-1,
\end{align}
where $c^{2}_{\rm s}$ is the sound speed of the fluid. Using the previous equations, we can relate $\rho$ to the horizon mass of the Friedmann-Lema\^itre-Robertson-Walker (FLRW) background dynamics, defined as $M_{\rm H} = 4 \pi \rho R^{3}_{\rm H}/3$ where $R_{\rm H}=1/H$ is the Hubble radius.  The minimum value $w \approx 0.23$ is reached when $M_{\rm H} \approx 2.5 M_{\odot}$.
The evolution of $w(T)$ and $c^{2}_{\rm s}(T)$ from~\cite{qcd_table} as a function of the horizon mass is shown in Fig.~\ref{fig:tabulated_data}.

To study the formation process of PBHs in spherical symmetry and using the comoving gauge, we solve numerically the Misner-Sharp (MS) equations~\cite{misnersharp}, which are the Einstein field equations in spherical symmetry for a relativistic fluid in the comoving gauge:
\begin{align}
\label{eq:u_simply}
\dot{U} &= -A\left[\frac{c^{2}_{\rm s}(\rho)}{1+w(\rho)}\frac{\Gamma^2}{\rho}\frac{\rho'}{R'} + \frac{M}{R^{2}}+4\pi R w(\rho) \rho \right], \\
\label{eq:r_simply}
\dot{R} &= A U, \\
\label{eq:rho_simply}
\dot{\rho} &= -A \rho \left[1+w(\rho)\right] \left(2\frac{U}{R}+\frac{U'}{R'}\right), \\
\label{eq:m_simply}
\dot{M} &= -4\pi A w(\rho) \rho U R^{2}, \\
\label{eq:lapse}
A' &=  -A \frac{\rho'}{\rho} \frac{c^{2}_{\rm s}(\rho)}{1+w(\rho)}, \\
\label{eq:radial_M}
M' &= 4 \pi \rho R^{2} R'.
\end{align}
where a dot denotes time derivatives with respect to the cosmic time $t$ and a prime the derivatives with respect to the radius $r$. The last equation is the Hamiltonian constraint, used to check the validity and accuracy of the simulations. $U$ is the radial component of the four-velocity associated with an Eulerian frame (not comoving), which measures the radial velocity of the fluid with respect to the origin of the coordinates. The Misner-Sharp mass $M(r,t)$ is defined as
\begin{equation}
M(R) \equiv \int_{0}^{R} 4\pi \tilde{R}^{2} \rho \, {\rm d}\tilde{R}\, ,
\end{equation}
which is related to $\Gamma$, $U$ and $R$ though the constraint:
\begin{equation}
\Gamma = \sqrt{1+U^2-\frac{2 M}{R}},
\end{equation}
where $\Gamma$ is the so-called generalised Lorentz factor that comprises the gravitational potential energy and the kinetic energy per unit mass.  In the Newtonian limit, one has
\begin{equation}
\Gamma \simeq 1+\frac{1}{2}\frac{v^{2}}{c^{2}}-\frac{GM}{c^{2}R}.
\end{equation} 
For the boundary conditions, one uses $M(r=0,t) = R(r=0,t) = U(r=0,t) = 0 $ and $\rho'(r=0,t)=0$. 

If we compare our approach to the case of a constant $w$ in time, there is a new contribution from the pressure gradients associated to the variations of $w$. This is included in Eqs.~\ref{eq:u_simply} and \ref{eq:lapse} with the following terms:
\begin{equation}
\label{eq:presure_gradient}
p' = \rho' \left[w(\rho)+\rho \frac{\partial w(\rho)}{\partial \rho} \right] = \rho' c^{2}_{\rm s}.
\end{equation}
When $w $ is constant, the sound speed is simply given by $c^{2}_{\rm s} = w $. One therefore expects a net difference with the case of a radiation-dominated Universe, as well as with the cases where $w$ takes the correct value for a QCD crossover transition when the fluctuation re-enters the Hubble horizon but remains constant in the simulations of PBH formation that are used to compute the threshold value $\delta_{\rm c}$. This difference does not only come from the time-dependent equation of state, but also from the radial dependence implying that $c^{2}_{\rm s} \neq w $. Solving the MS equations when considering no curvature fluctuations allows us to recover the FLRW solution, which is obtained numerically by solving:
\begin{align}
\label{eq:background_solution}
\dot{\rho_{\rm b}}&+\sqrt{24 \pi}\rho^{3/2}_{\rm b}\left[1+w(\rho_{\rm b})\right] =0, \\
\frac{\dot{a}}{a} &- \sqrt{\frac{8 \pi \rho_{b}}{3}}=0, \nonumber
\end{align}
where we used $H^{2} = 8\pi \rho_{\rm b}/3$. 
A useful and commonly used estimator for the "strength" of a spherically symmetric perturbation is the compaction function~\cite{Shibata:1999zs}, defined as the mass excess $\delta M (R) = M-M_{\rm b}$ enclosed in the areal radius $R(r, t)$ relative to the case of a FLRW homogeneous background $M_{\rm b}(r,t)=4 \pi \rho_{\rm b} R^{3}/3$, divided by the areal radius,
\begin{equation}
\com(r,t) = \frac{2 \left[M(r,t)-M_{\rm b}(r,t)\right]}{R(r,t)}~.
\label{2_compactionfunction}
\end{equation}
The compaction function is essential to define the threshold for PBH formation~\cite{Shibata:1999zs}, as we will see in the next Section. The compaction function can also be written in terms of the averaged density contrast $\bar \delta$ over a given volume~\cite{refrencia-extra-jaume},
\begin{equation}
    \mathcal{C}(r,t) =  \bar{\delta} (H  R)^{2} ,
\end{equation}
where 
\begin{equation}
    \bar{\delta} = \frac{3}{R^{3}}\int_{0}^{R}\frac{\delta \rho}{\rho_{\rm b}} \tilde{R}^{2} d\tilde{R}.
\end{equation}

\subsection{Gradient expansion approximation and problem of initial conditions }\label{sec:initial_conditios}

\begin{figure}[t]
\centering
\includegraphics[width=1.9 in]{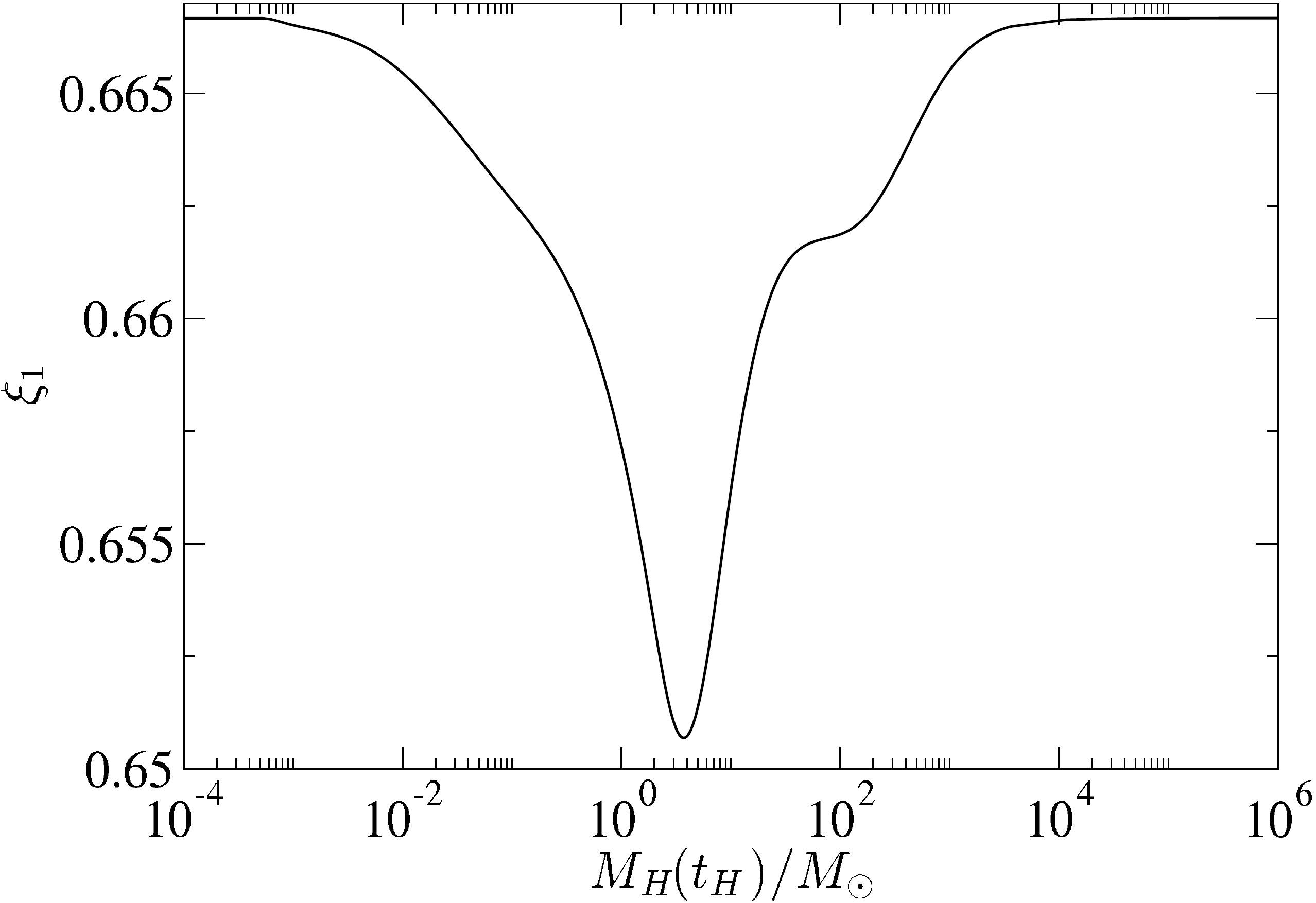}
\includegraphics[width=1.9 in]{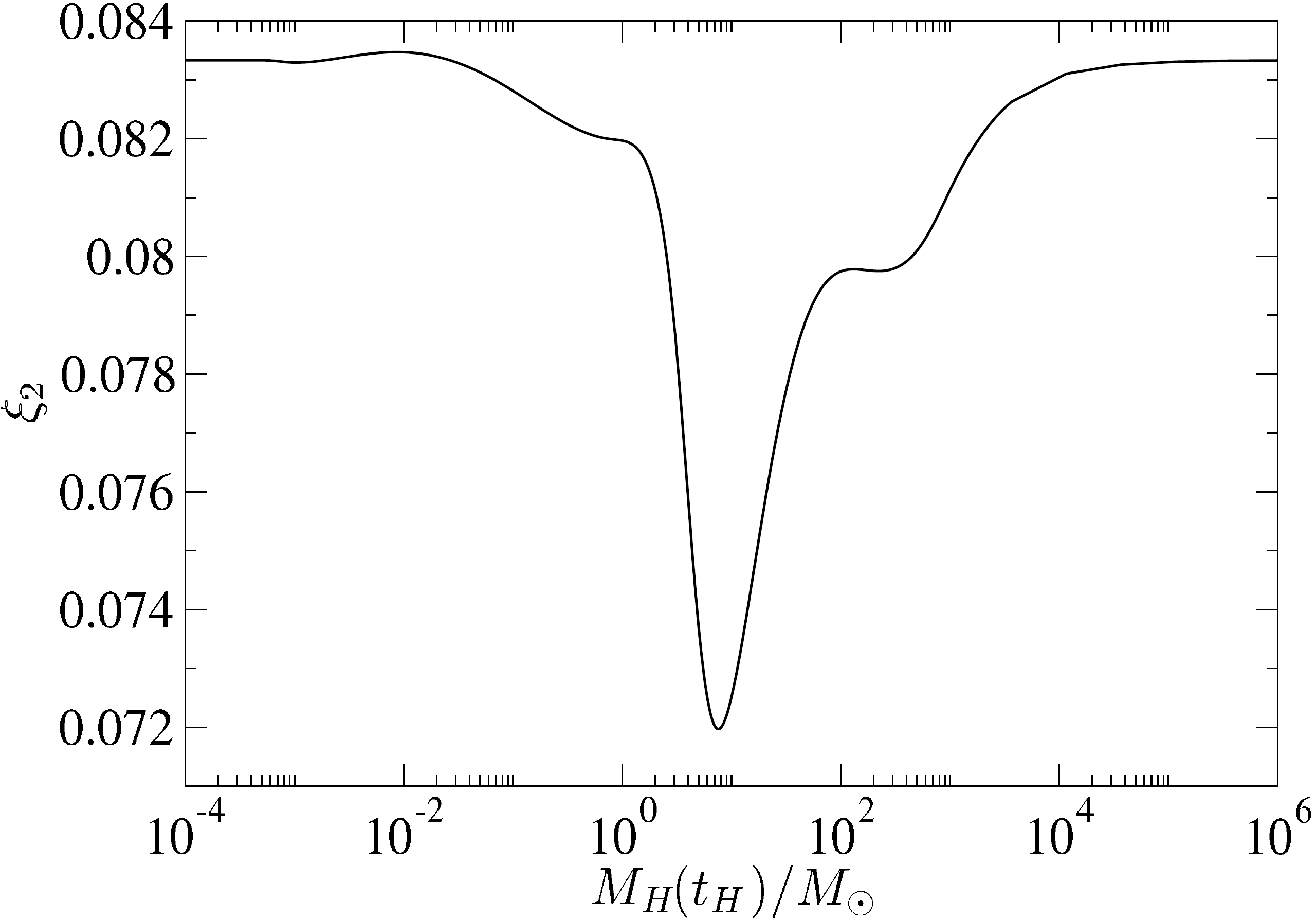}
\includegraphics[width=1.9 in]{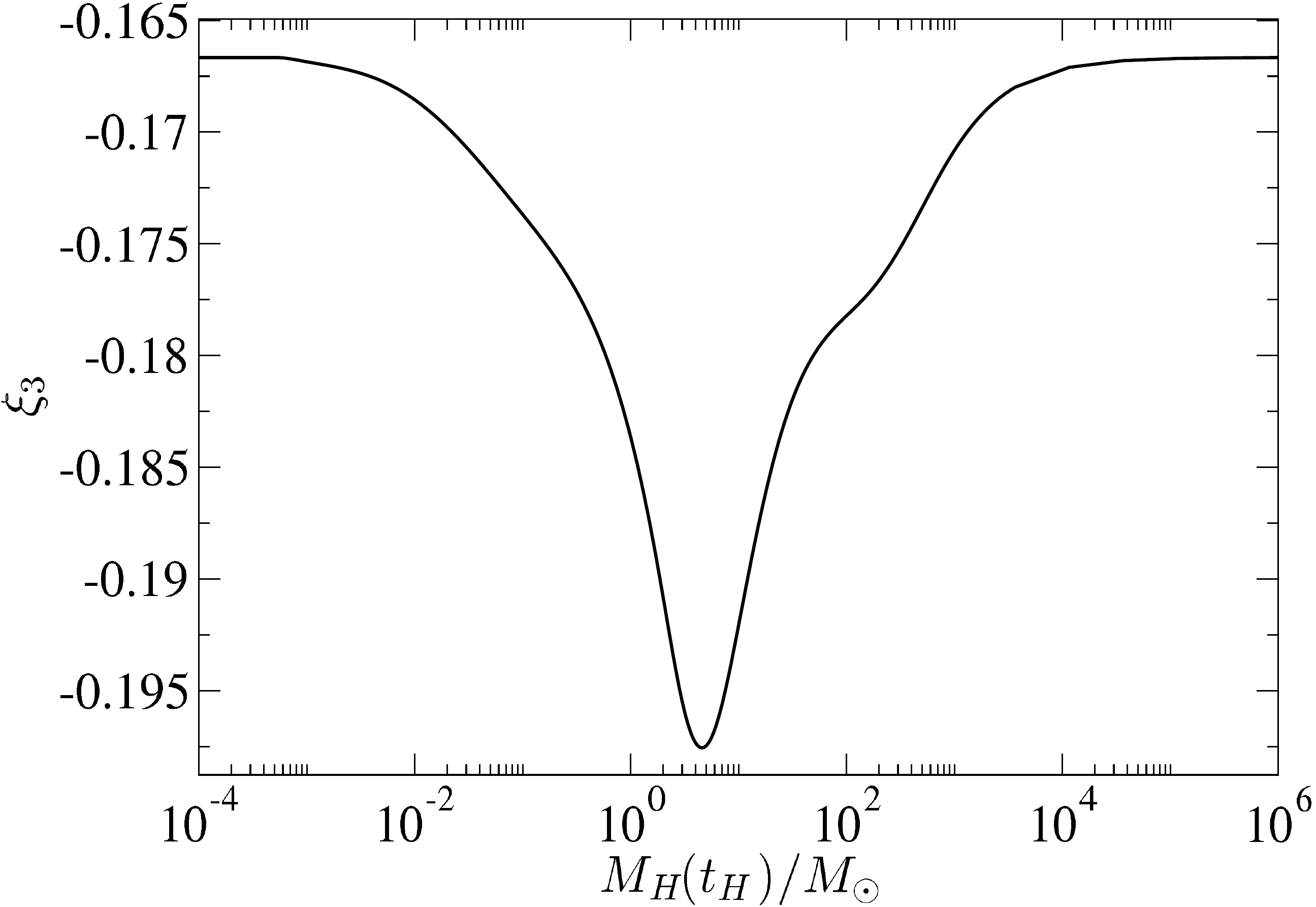}
\caption{Evolution of $\xi_1$, $\xi_2$ and $\xi_3$ from Eqs.~\ref{eq:initial_functions} as a function of $M_{\rm H}/M_{\odot}$ during the QCD crossover transition.}
\label{fig:initial_functions}
\end{figure}

In order to simulate numerically the formation of PBHs from a cosmological fluctuation, it is necessary to specify in a consistent way the initial conditions of such a fluctuation, when it is still super-horizon~\cite{Shibata:1999zs}. We follow the gradient expansion method~\cite{PhysRevD.42.3936}, also called the long-wavelength approximation, that has been already used in previous works, e.g. ~\cite{Shibata:1999zs,musco2007,refrencia-extra-jaume}.   This technique consists in expanding the spatial density gradients in terms of a parameter $\epsilon$ that relates the Hubble horizon $R_{\rm H}$ and the comoving length-scale of the perturbation $R_{\rm m} = a(t) r_{\rm m}$,
\begin{equation}
\epsilon = \frac{R_{\rm H}(t)}{R_{\rm m}(t)}.
\label{eq:epsilon}
\end{equation}
At super-horizon scales one has $\epsilon \ll 1$, and in the limit $\epsilon \rightarrow 0$ the space-time is described locally by a FLRW metric when the fluctuation is smoothed out over a sufficiently large scales $R_{\rm m}$. At leading order in the gradient expansion, the space-time metric at super-horizon scales can be written as a FLRW metric with a non-constant curvature $K(r)$~\cite{Shibata:1999zs},
\begin{equation}
\label{2_FLRWmetric5}
{\rm d}s^2 = -{\rm d}t^2 + a^2(t) \left[\frac{{\rm d}r^2}{1-K(r) r^2}+r^2 {\rm d}\Omega^2 \right],
\end{equation}
and one can use the curvature profile $K(r)$ to characterize a cosmological fluctuation~\footnote{But the metric of Eq.~\ref{2_FLRWmetric5} can also be written in terms of $\zeta(\tilde{r})$ with an appropriate change of coordinates~\cite{musco2018,enea-sasaki-alexei,Hidalgo:2008mv,Garriga,Musco:2020jjb}}. For adiabatic fluctuations (in the case of isocurvature fluctuations leading to PBH formation see~\cite{PhysRevD.105.103530,Yoo:2021fxs}), the curvature $K(r)$ is frozen at super-horizon scales, even when one considers a perfect fluid with a time-dependent equation of state~\cite{PhysRevD.28.679,Ijjas:2020cyh} \footnote{A.E thanks Jaume Garriga for a clarification about this point.}. The gradient expansion method then consist in expanding the MS equations in terms of the $\epsilon^2(t)$ parameter, as in~\cite{musco2007}:
\begin{align}
\label{2_expansion}
A(r,t) &= 1+\epsilon^2(t) \tilde{A},\nonumber\\
R(r,t) &= a(t)r(1+\epsilon^2(t) \tilde{R}),\nonumber\\ 
U(r,t) &= H(t) R(r,t) (1+\epsilon^2(t) \tilde{U} ),\\ 
\rho(r,t) &= \rho_{\rm b}(t)(1+\epsilon^2(t)\tilde{\rho}),\nonumber\\ 
M(r,t) &= \frac{4\pi}{3}\rho_{\rm b}(t) R(r,t)^3 (1+\epsilon^2(t) \tilde{M} ).\nonumber 
\end{align}
One recovers the FLRW solution for $\epsilon \rightarrow 0$. The solution of Eqs.~\ref{2_expansion} represents the linear evolution of the hydrodynamic variables at first order in the gradient expansion.  The perturbations of the tilde variables are obtained by injecting Eqs.~\ref{2_expansion} into the MS equations and taking the first order term in $\mathcal{O}$($\epsilon^{2}$).  The solution for a constant $w$ or even a time-dependent $w(t)$ was derived in~\cite{musco2007} but without explicitly using the time-dependence of the QCD epoch and the corresponding radial dependence.  In our case, $w(\rho)$ has a temporal and radial dependence, leading to $w(\rho) \neq c^{2}_{\rm s}(\rho)$. Indeed, at first order in $\epsilon^2$ we have
\begin{align}
    w(\rho) & \approx w(\rho_{\rm b})+\rho_{\rm b} \frac{\partial w}{\partial \rho} (\epsilon^{2} \tilde{\rho}) +O(\epsilon^4),\nonumber \\
    c^{2}_{\rm s}(\rho) & \approx c^{2}_{\rm s}(\rho_{\rm b})+\rho_{\rm b} \frac{\partial c^{2}_{\rm s}}{\partial \rho} (\epsilon^{2} \tilde{\rho}) +O(\epsilon^4),\nonumber 
\end{align}
which only depend on the background density $\rho_{\rm b}$. Introducing Eqs.~\ref{2_expansion} into Eqs.~\ref{eq:u_simply} - \ref{eq:radial_M} and using the following useful relations
\begin{align}
    \frac{\dot{\epsilon}}{\epsilon} &= \frac{1}{2}H \left[1+3 w(\rho)\right],\nonumber \\
    \frac{\dot{\rho_{\rm b}}}{\rho_{\rm b}} & = -3 H \left[1+w(\rho)\right] ,\nonumber 
\end{align}
one obtains the perturbation variables
\begin{align}
\label{eq:2_perturbations}
\tilde{\rho}&= \xi_{1}(\rho_{\rm b})\left[K(r)+\frac{r}{3}K'(r)\right] r^2_{\rm m},\nonumber \\
\tilde{U} &= \frac{1}{2} \left[\xi_{1}(\rho_{\rm b})-1\right]K(r)r^{2}_{\rm m},\nonumber\\
\tilde{A} &= -\xi_{1}(\rho_{\rm b}) \frac{c^{2}_{\rm s}(\rho_{\rm b})}{1+w(\rho_b)}\left[K(r)+\frac{r}{3}K'(r)\right] r^2_{\rm m},\\
\tilde{M} &= \xi_{1}(\rho_{\rm b}) K(r) r^{2}_{\rm m},\nonumber\\
\tilde{R} &= -\xi_{2}(\rho_{\rm b})\left[K(r)+\frac{r}{3}K'(r)\right]r^2_{\rm m} + \xi_{3}(\rho_{\rm b})\frac{K(r)}{2}r^{2}_{\rm m},\nonumber 
\end{align}
where $\xi_{1}(\rho_{\rm b})$, $\xi_{2}(\rho_{\rm b})$ and $\xi_{3}(\rho_{\rm b})$ are functions of the energy density of the FLRW background, obeying the following differential equations:
\begin{align}
\label{eq:initial_functions}
\frac{{\rm d} \xi_{1}(\rho_{\rm b})}{{\rm d} \rho_{\rm b}} &= -\frac{1}{2 \rho_{\rm b}} + \frac{5+3w(\rho_{\rm b})}{2\left[1+w(\rho_{\rm b})\right]} \frac{\xi_{1}(\rho_{\rm b})}{3 \rho_{\rm b}},\nonumber \\
\frac{{\rm d} \xi_{2}(\rho_{\rm b})}{{\rm d} \rho_{\rm b}} &= -\frac{c^{2}_{\rm s}(\rho_{\rm b})}{3\left[1+w(\rho_{\rm b})\right]^{2}}  \frac{\xi_{1}(\rho_{\rm b})}{\rho_{\rm b}} + \frac{\left[1+3w(\rho_{\rm b})\right]}{3\left[1+w(\rho_{\rm b})\right]}\frac{\xi_{2}(\rho_b)}{\rho_b},\\
\frac{{\rm d} \xi_{3}(\rho_{\rm b})}{{\rm d} \rho_{\rm b}} &= \frac{-1}{3\left[1+w(\rho_{\rm b})\right]} \frac{\left[\xi_{1}(\rho_{\rm b})-1 \right]}{\rho_{\rm b}} + \frac{\left[1+3w(\rho_{\rm b})\right]}{3\left[1+w(\rho_{\rm b})\right]}\frac{\xi_{3}(\rho_{\rm b})}{\rho_{\rm b}}.\nonumber
\end{align}
Eqs.~\ref{eq:initial_functions} are solved numerically using the tabulated values of $w(\rho_{\rm b})$, $c^{2}_{\rm s}(\rho_{\rm b})$, and with the initial conditions such that 
\be
\frac{{\rm d} \xi_1(\rho_{{\rm b},0})}{{\rm d} \rho_{\rm b}} = \frac{{\rm d} \xi_{2}(\rho_{{\rm b},0})}{{\rm d} \rho_{\rm b}} = \frac{{\rm d} \xi_{3}(\rho_{{\rm b},0})}{{\rm d} \rho_{\rm b}}=0~,
\ee 
where $\rho_{\rm b,0}= \rho_{\rm b}(t_0)$ and $t_0$ is the initial time.  On Fig.~\ref{fig:initial_functions}, we have represented the time evolution of $\xi_1, \xi_2, \xi_3$. 
When the Universe is radiation dominated with $w=1/3$, we recover $\xi_1 =2/3 $, $\xi_2 = 1/12$ and $\xi_3 = -1/6$.
As shown in~\cite{Polnarev:2012bi}, where the gradient expansion formalism was applied beyond first order using an iterative scheme, the first order approximation is accurate as long as a sufficiently small $\epsilon$ parameter (specifically $\epsilon<0.1$) is chosen.  See also~\cite{Polnarev:2012bi} for the effects of adding higher-order terms in the $\epsilon$ expansion. Therefore, we will use Eqs.~\ref{2_expansion},~\ref{eq:2_perturbations} to set up consistently the initial conditions at super-horizon scales and Eqs.~\ref{eq:u_simply}-\ref{eq:radial_M} to proceed with the full non-linear evolution of the gravitational collapse.

The formation of a primordial black hole for a given initial condition can be inferred from the dynamics of perturbations
that continue growing (i.e, which do not dissipate) after entering the horizon until the formation of an apparent horizon \cite{penrose}. In spherical symmetry, this condition is satisfied when $2M=R$, which implies that for $\mathcal{C}=1$ an apparent horizon has been already formed.

\subsection{Threshold for PBH formation}

Let us now focus on the definition of the over-density threshold leading to PBH formation. As already mentioned, the compaction function initially defined on super-horizon scales plays a crucial role.
At leading order in the gradient expansion and taking into account the QCD crossover, $\mathcal{C}$ is given by
\begin{equation}
    \mathcal{C}(\rho_{\rm b},r) \approx \tilde{M} \left (\frac{r}{r_{\rm m}} \right)^{2} +O(\epsilon^4) = \xi_{1}(\rho_{\rm b}) K(r) r^{2} +O(\epsilon^4),
\end{equation}
which has been obtained by introducing Eqs.~\ref{2_expansion},~\ref{eq:2_perturbations} into Eq.~\ref{2_compactionfunction} and expanding in $\epsilon$, and where $r_{\rm m}$ corresponds to the location of the peak value of $\mathcal{C}(\rho_{\rm b},r)$, i.e $r_{\rm m}$ must satisfy
\begin{equation}
\label{2_cmax}
K(r_{\rm m})+\frac{r_{\rm m}}{2}K'(r_{\rm m}) = 0.
\end{equation}
When the fluctuation is super-horizon and assuming $w=\text{const.}$, the compaction function is time-independent.  This allows us to define consistently the amplitude of a cosmological density fluctuation as $\delta_{\rm m} \equiv \delta(r_{\rm m})\equiv  \com(r_{\rm m})$ \cite{musco2018,refrencia-extra-jaume,Shibata:1999zs,universal1}.   When $w$ is constant, one therefore has $\delta_{\rm m}= \xi_{1, w={\rm cst}} K(r_{\rm m})r^{2}_{\rm m}$ with $\xi_{1, w={\rm cst}} = 3(1+w)/(5+3w)$. The over-density threshold leading to PBH formation corresponds to the peak value of the critical compaction function $\delta_{\rm c} \equiv \com_{\rm c}(r_{\rm m}) $. 
Perturbations with $\delta_{\rm m}>\delta_{\rm c}$  collapse and form a PBH, and perturbations with $\delta_{\rm m} < \delta_{\rm c}$ are diluted and hence do not lead to black hole formation.

The situation is slightly different at the QCD epoch, because the compaction function on super-horizon scales is now time-dependent and proportional to $\xi_{1}(\rho_{\rm b})$. Nevertheless, the curvature $K(r)$ remains frozen on super-horizon scales during the QCD crossover transition, which allows us to define the amplitude $\delta_{\rm m}$ of a fluctuation when $w(\rho_{\rm b})=c^{2}_{\rm s}(\rho_{\rm b})=1/3$ through $\delta_{\rm m} \equiv (2/3) K(r_{\rm m})r^{2}_{\rm m}$ before the QCD crossover starts, (i.e, during the radiation-dominated epoch), even if we start our numerical simulations inside the QCD transition using the linear evolution of fluctuations from Eq.~\ref{eq:2_perturbations} to set up the initial conditions.
\subsection{Initial curvature profile}
\begin{figure}[t]
\centering
\includegraphics[width=2.85 in]{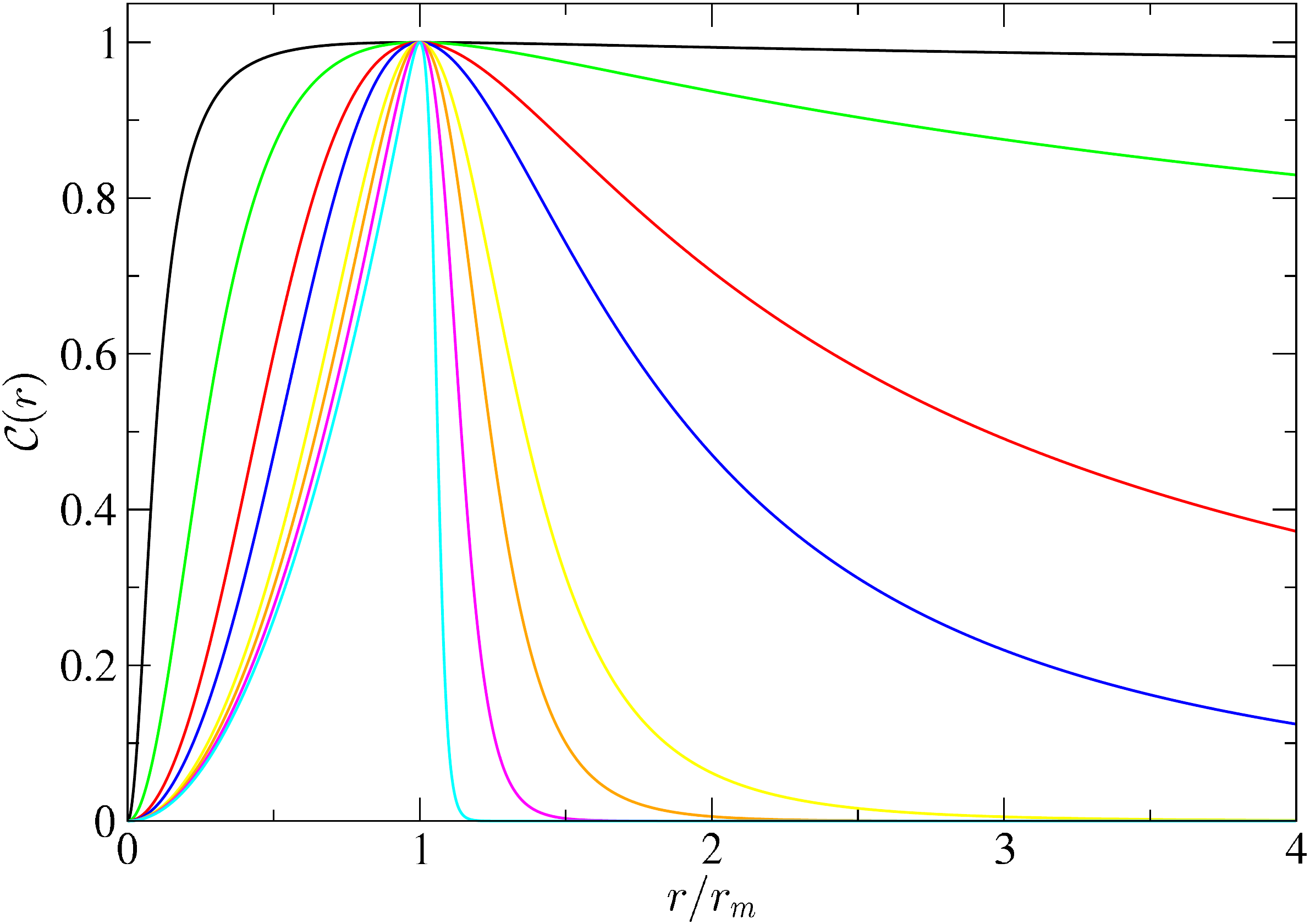}
\includegraphics[width=3.0 in]{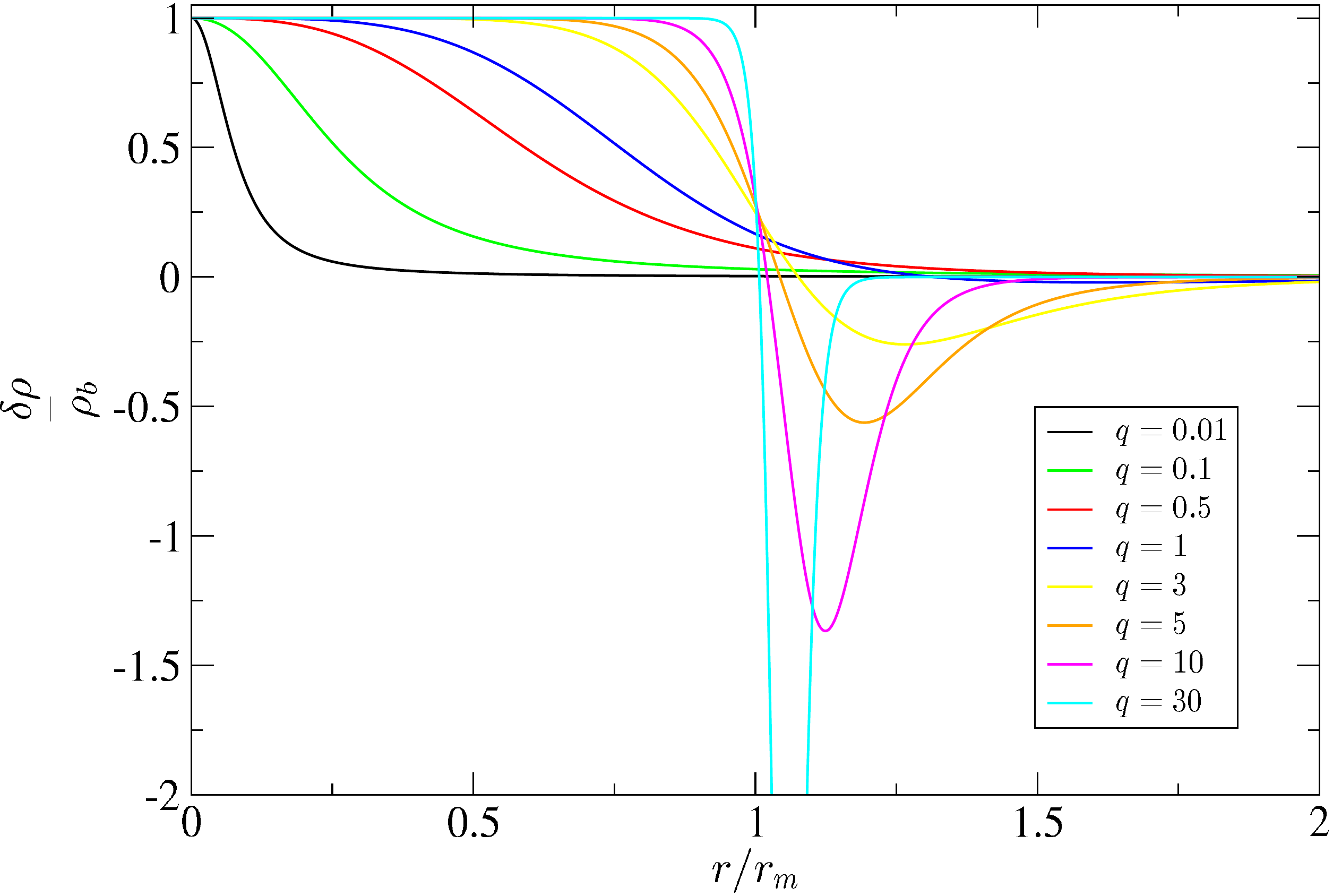}
\caption{Left panel: Radial profiles of the compaction function $\com(r)$ obtained from Eq.~\eqref{eq:basis_pol}, for different values of $q$ and corresponding $\delta_{\rm m}$ values normalised to $1$. Right panel:  Radial profiles of the density contrast $\delta \rho / \rho_{\rm b}$ for the same curvature profiles as in the left panel, normalised such that $\delta \rho (r=0) /\rho_{\rm b} = 1$. }
\label{fig:C_profiles}
\end{figure}

 We consider two types of initial profiles for the curvature fluctuations: polynomial~\cite{Escriva:2020tak} and exponential~\cite{musco2018,refrencia-extra-jaume}, respectively described by:
\begin{align}
 K_{\rm pol}(r) &= \frac{3}{2}\frac{\delta_{\rm m}}{r_{\rm m}^2}\frac{1 + 1/q}{1+\frac{1}{q}\left(\frac{r}{r_{\rm m}}\right)^{2(q+1)}}, \label{eq:basis_pol} \\
 K_{\rm exp}(r) &= \frac{3}{2}\frac{\delta_{\rm m}}{ r_{\rm m}^2}\,
\left(\frac{r}{r_{\rm m}}\right)^{2\lambda}\,
e^{\frac{(1+\lambda)^{2}}{q}\left(1 - \left(\frac{r}{r_{\rm m}}\right)^{\frac{2q}{1+\lambda}}\right)}. \ \label{eq:profile_exp}
\end{align}
We recover the so-called Gaussian profile by taking $\lambda=0$ and $q=1$ in Eq.~\ref{eq:profile_exp} and a polynomial Gaussian profile with $q=1$ in Eq.~\ref{eq:basis_pol}. The parameter $\lambda$ allows to shift the peak in the curvature $K$ to larger values of $r$. The dimensionless parameter $q$~\cite{universal1}, defined as
\begin{equation}
\label{eq:q_factor}
q \equiv -\frac{\mathcal{C''}(r_{\rm m})r^{2}_{\rm m}}{4 \mathcal{C}(r_{\rm m})},
\end{equation}
allows to span all possible threshold values using Eq.~\ref{eq:basis_pol} (we refer such profile as a basis-profile) in the range $q \in [0, \infty[$. Specifically, $q \rightarrow 0$ leads to a broad profile in $\mathcal{C}$ and to a minimum threshold value, which is $0.4$ for a radiation-dominated Universe. For $q \rightarrow \infty$, one gets a sharp profile in $\mathcal{C}$, leading to a maximum threshold value $2/3$ in radiation. One can notice that this analysis is not possible 
with Eq.~\ref{eq:profile_exp} because for $q<0.5$ (with $\lambda=0$), the curvature profile does not fulfill regularity conditions. For this reason, we will mainly use Eq.~\ref{eq:basis_pol} to characterize the thresholds, while Eq.~\ref{eq:profile_exp} is used to compare with previous results for the case $w=1/3$~\cite{escriva_solo}. Examples of curvature profiles obtained with Eq.~\ref{eq:basis_pol} for different values of $q$, with the corresponding density contrast, are shown in Fig.~\ref{fig:C_profiles}.
At the initial time with perturbations at super-horizon scales, we define $\beta_{\rm m}$ following Eq.~\ref{eq:epsilon} as $\beta_{\rm m} = 1/\epsilon(t_0) = a(t_0)r_{\rm m}/R_{\rm H}(t_0)$. We always consider super-horizon fluctuations with $\beta_{\rm m} =10$ at the beginning of the simulations. We also define the time of horizon crossing $t_{\rm H}$ as the time when $\epsilon=1$, which means that $a(t_{\rm H})r_{\rm m}=R_{\rm H}(t_{\rm H})$. Indeed, the mass of the horizon at the horizon crossing time $M_{\rm H}(t_{\rm H})$ can be related to the Hubble scale as $M_{\rm H}(t_{\rm H}) = 1/2H_{\rm H}$, where $H_{\rm H}$ is the Hubble scale at the time of horizon crossing.
In the case of constant $w$, it will be simply given by $M_{\rm H}(t_{\rm H}) = [3(1+w)t_0/4](a_0 \beta_{\rm m})^{3(1+w)/(1+3w)}$~\cite{Escriva:2021aeh}. However, this is not true for the QCD crossover, where Eq.~\ref{eq:background_solution} needs to be solved numerically.

\subsection{Simulations of PBH formation with pseudo-spectral methods}\label{sec:numerics}

We have used the publicly available code of~\cite{escriva_solo,web} based on pseudo-spectral methods (see \cite{novak,spectralmatlab,spectrallloyd} for more details about the technique) to perform the numerical simulations of PBH formation at the QCD epoch with a time-varying equation of state. Here we give some main insights, but we refer the reader to check for more details about the implementation for simulations of PBH formation in~\cite{escriva_solo}.

In spectral methods, one uses a fixed grid where the node points $x_{k}$ (which are conveniently mapped from the spectral to the physical domain \cite{escriva_solo}) satisfy $T'_{k}(x_k)=0$, with $T_{k}(x)$ being the Chebyshev polynomials of order $k$. These points $x_k$ are called the Chebyshev collocation points.   In order to compute the derivative of a field $u_{k}$ with spectral accuracy (which means that the error decays exponentially increasing the number of points) one has to multiply the node values of the field by the Chebyshev differentiation matrix, $u'_{k}= D \cdot u_{k}$, where $D $ is defined as
\begin{align}
D^{(1)}_{i,j} &= \frac{\bar{c}_{i}}{\bar{c}_{j}}\frac{(-1)^{i+j}}{(x_{i}-x_{j})} , \hspace{6 mm} i,j = 1,\ldots,N_{\rm cheb}-1   (i \neq j),\\
D^{(1)}_{i,i} &= -\frac{x_{i}}{2(1-x_{i}^2)} , \hspace{6 mm} i=1,\ldots,N_{\rm cheb}-1,\\ 
D^{(1)}_{0,0} &=-D^{(1)}_{N_{\rm cheb},N_{\rm cheb}} = \frac{2N_{\rm cheb}^2+1}{6}\,.
\end{align}
One can improve the rounding errors~\cite{spectralmatlab} by using the following identity for the diagonal elements,
\begin{equation}
D^{(1)}_{i,i} = -\sum_{j=0,j \neq i}^{N_{\rm cheb}} D_{i,j}^{(1)}.
\end{equation}
The method, initially designed for a constant $w$, has been modified to handle situations with a time-varying equation of state 
$w(\rho)$, using the tabulated values of~\cite{qcd_table} and a cubic spline interpolation. In terms of stability and performance, we find that the new code is equally robust and accurate.
The main difficulty arises when solving Eq.~\ref{eq:lapse} for the lapse function $A(r,t)$. In the case of constant $w$, it can be solved analytically, but this is not the case for $w(\rho)$. However, it can be solved pseudo-analytically with spectral accuracy, using the Chebyshev differentiation matrix.  At each time step, the lapse function is then given by
\begin{equation}
\label{lapse}
\ln\left(\frac{A}{A_{0}}\right) = D^{-1}_{\rm cheb} \cdot \left( - \frac{\rho'}{\rho} \frac{c^{2}_{\rm s}}{1+w}  \right),  \end{equation}
and we consider the boundary condition $A_{0}(r_{\rm f})=1$ at the final point of the grid $r_{\rm f}$.  This is consistent with our assumption to recover the FLRW solution at $r \rightarrow \infty$.

From a technical point of view, we use a multi-grid spectral domain with two layers of approximately $300$ Chebyshev points.
The time evolution is performed independently in each subdomain, i.e. the spatial derivatives are computed using the Chebyshev differentiation matrix $\tilde{D_{l}}$ associated to each subdomain, using a standard RK4 integration method. We have chosen the coordinates to be in units of $r_{\rm m}$ and have used the conformal time as the integration variable, instead of the cosmic time.

\section{PBH formation dynamics}

\begin{figure}[t]
\centering
\includegraphics[width=4 in]{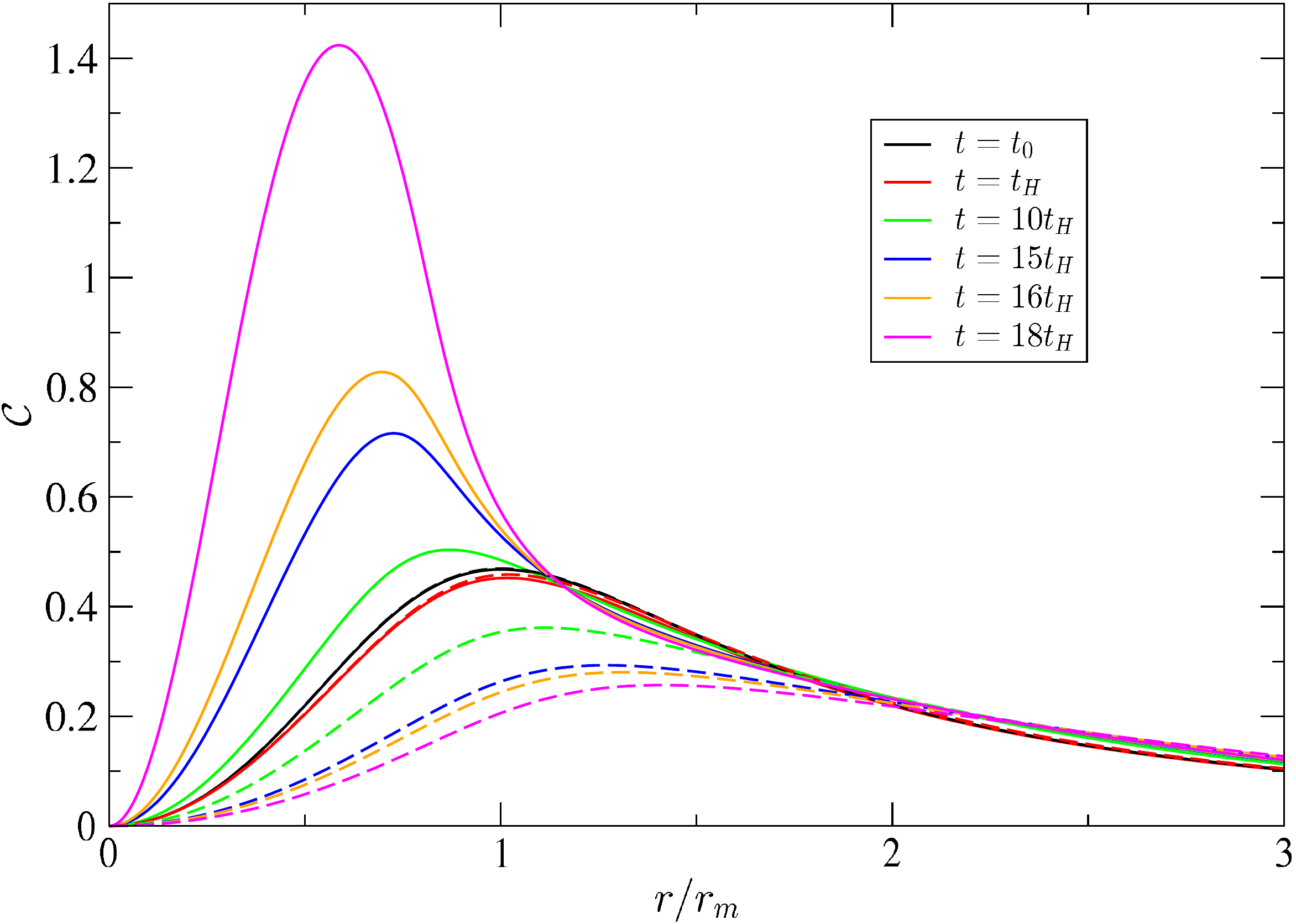}
\caption{Snapshots of the compaction function profile at different times, in a simulation of PBH formation with an initial polynomial Gaussian profile. The solid lines represent the case where the equation of state evolves during the QCD crossover transition, for a fluctuation with $M_{\rm H}(t_{\rm H}) \approx 2.5 M_{\odot}$. The dashed lines show the evolution with a fixed equation of state $w=1/3$. In both cases, we take $\delta_{\rm m} = 0.47$.}
\label{fig:C_evolution}
\end{figure}

In this Section, we study in detail how the QCD crossover affects the dynamics of PBH formation with a varying equation of state, until the formation of the first apparent horizon.  In particular, we analyse the evolution of a fluctuation in different cases, the time needed for a large enough fluctuation to form an apparent horizon and how the threshold changes for different curvature profiles when the QCD crossover transition is fully taken into account.  Finally, we estimate how the PBH mass evolves after the horizon formation.

\subsection{Dynamics of the gravitational collapse}\label{sec:dynamics}

The dynamics of PBH formation obtained with the same numerical method were analyzed exhaustively in~\cite{escriva_solo} for the case $w=1/3$, but the fact that they have never been explored with the time-varying $w(\rho)$ at the QCD epoch incites us to study the modifications that they imply.  We consider a polynomial Gaussian curvature profile and distinguish three regimes, as in~\cite{escriva_solo}: the super-critical regime ($\delta_{\rm m} > \delta_{\rm c}$), the sub-critical regime ($\delta_{\rm m} < \delta_{\rm c}$) and the critical regime ($\delta_{\rm m} \approx \delta_{\rm c}$).

We obtain that the dynamical evolution of $w(\rho)$ and $c^{2}_{\rm s}(\rho)$ both in time and space leads to non-trivial features.  The evolution of each fluctuation depends on the specific profile, on the amplitude $\delta_{\rm m}$ and on the horizon mass at the time of Hubble-horizon re-entry $t_{\rm H}$, i.e, $M_{\rm H}(t_{\rm H})$.
An example of how the QCD crossover affects the evolution of the compaction function and the PBH formation process is shown in Fig.~\ref{fig:C_evolution}, for an amplitude $\delta_{\rm m} = 0.47$.  This example illustrates that a fluctuation can collapse and form a PBH at the QCD epoch whereas it does not lead to PBH formation in a radiation-dominated Universe.  Indeed, if $w=1/3$ one has $\delta_{\rm m} < \delta_{\rm c} \approx 0.5$, but when considering realistically the transient reduction of $w$ and $c_{\rm s}^2$ (we chose $M_{\rm H} \approx 2.5 M_{\odot}$,  lying at the minimum value of $w$ according to Fig.~\ref{fig:tabulated_data}), the threshold of PBH formation is reduced to $\delta_{\rm c} \approx 0.45 < \delta_{\rm m}$, leading to the collapse and formation of an apparent horizon.

\begin{figure}[t]
\centering
\includegraphics[width=3. in]{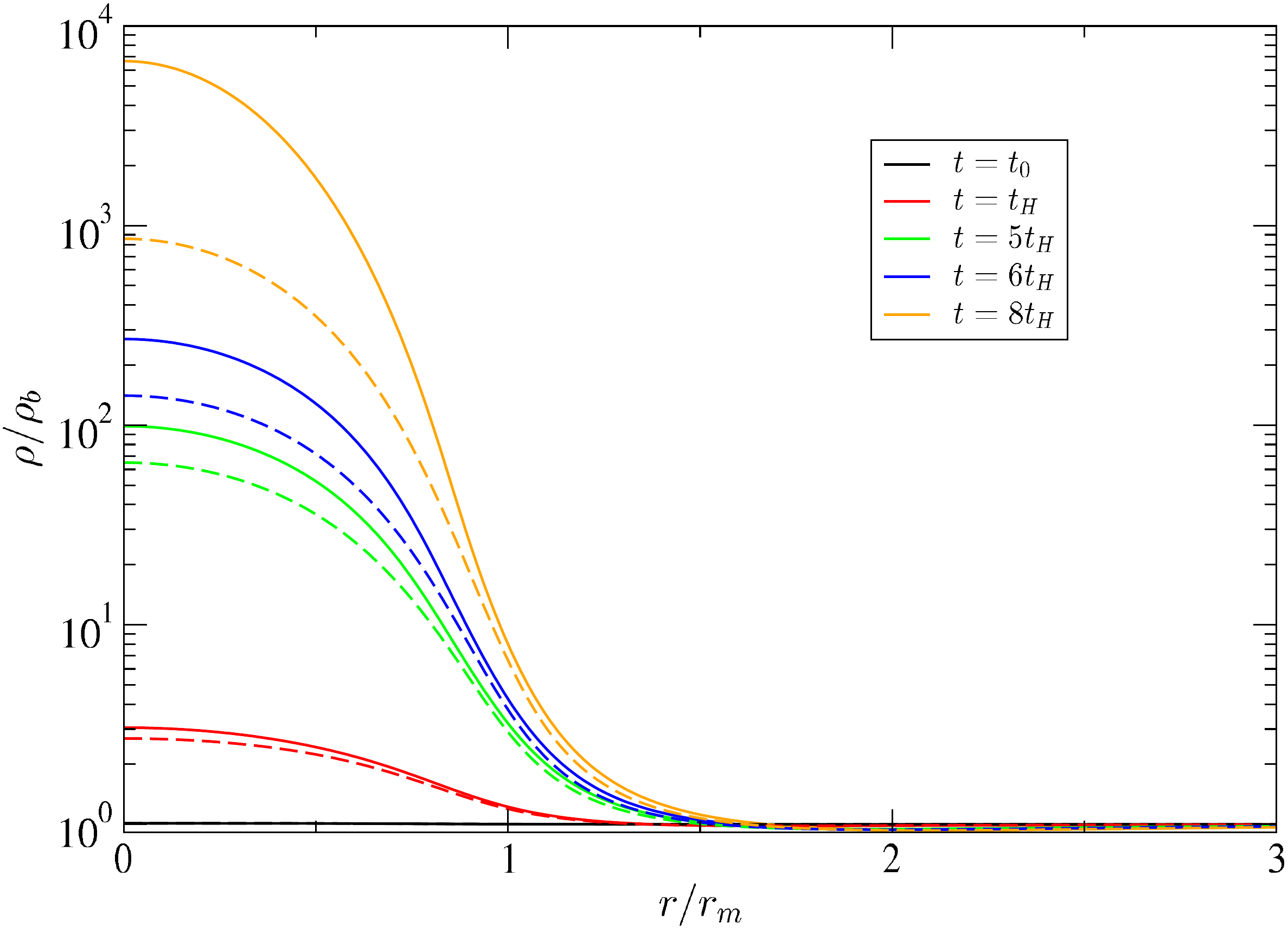}
\includegraphics[width=2.95 in]{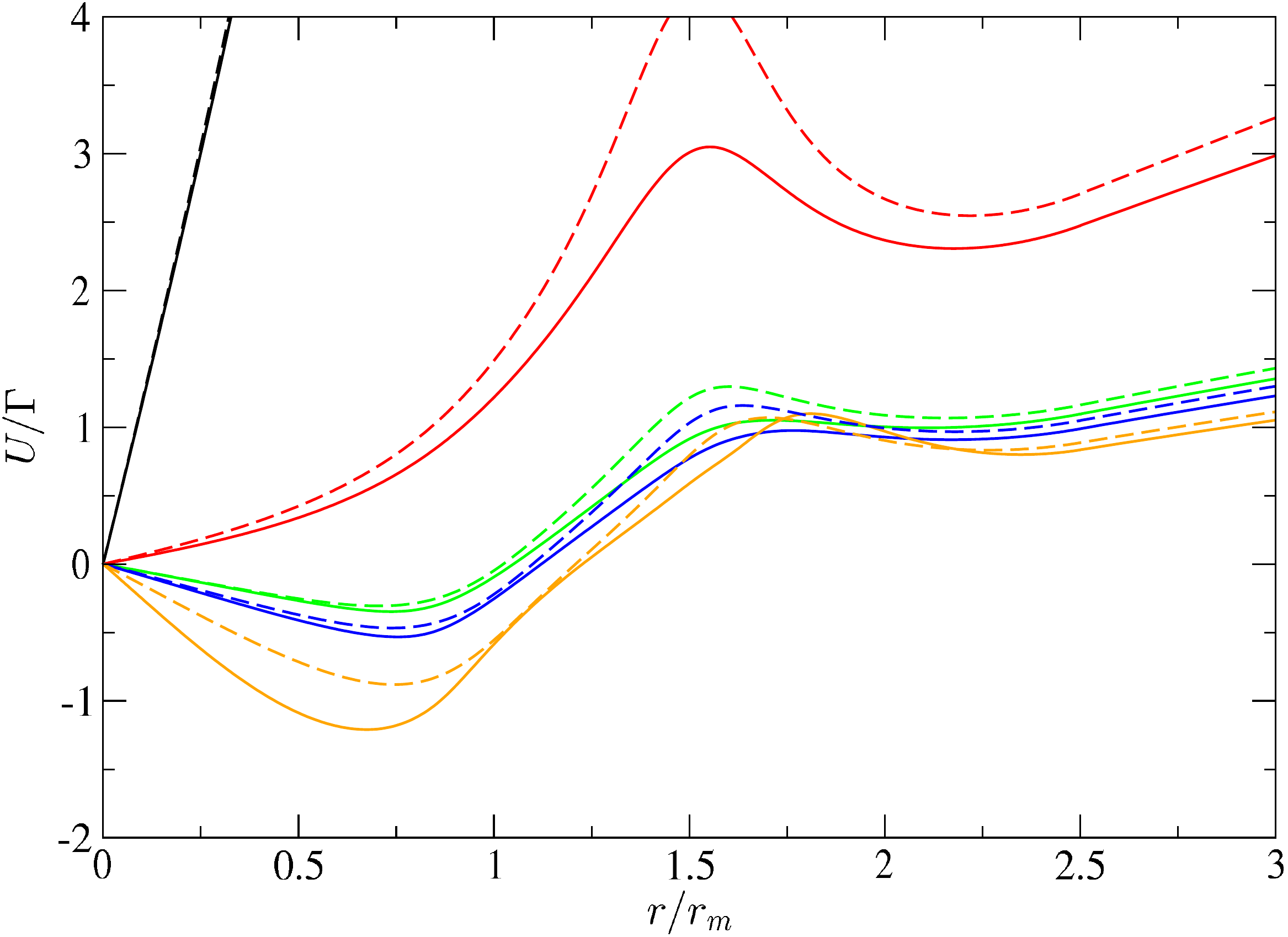}
\includegraphics[width=2.9 in]{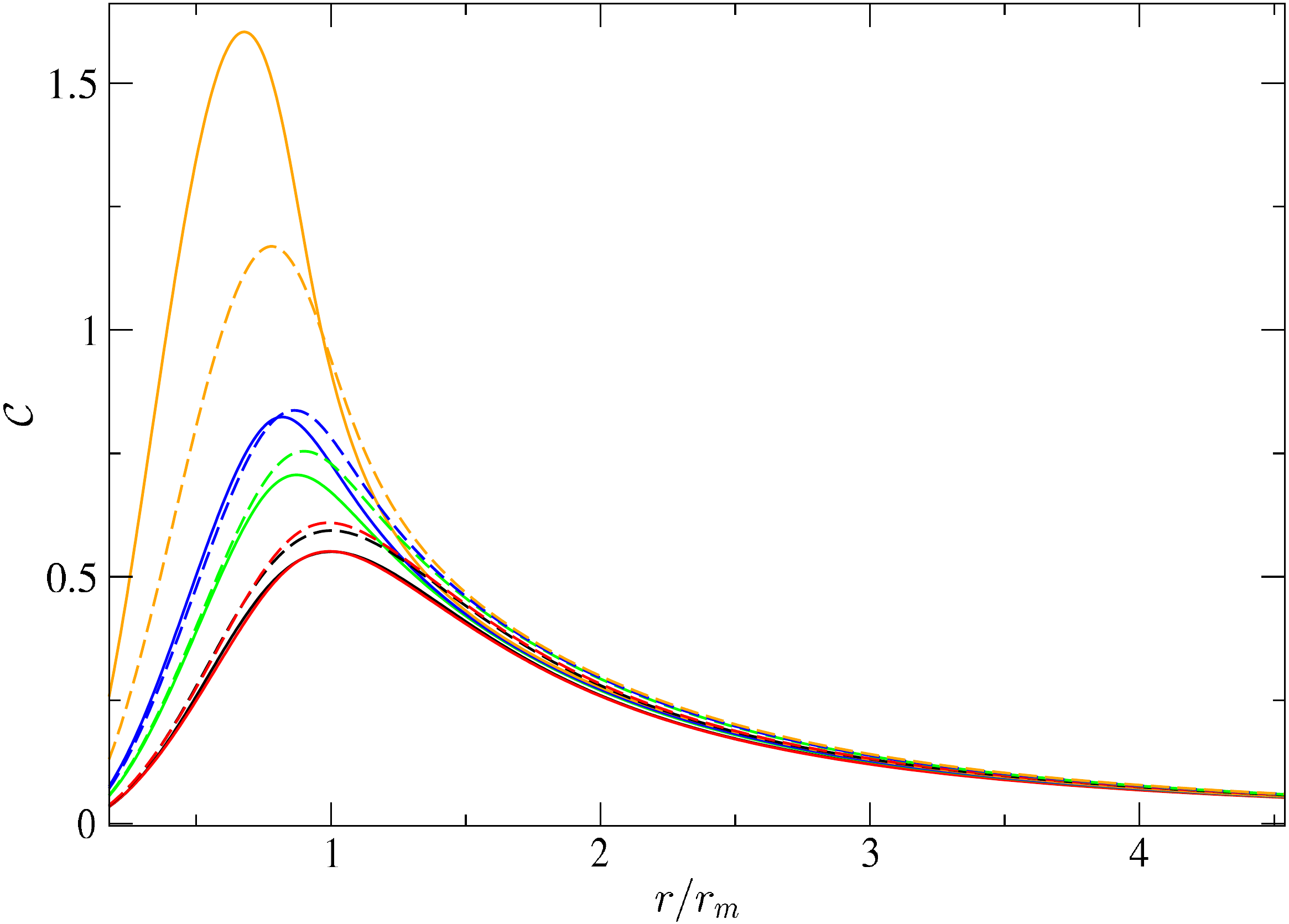}
\includegraphics[width=3. in]{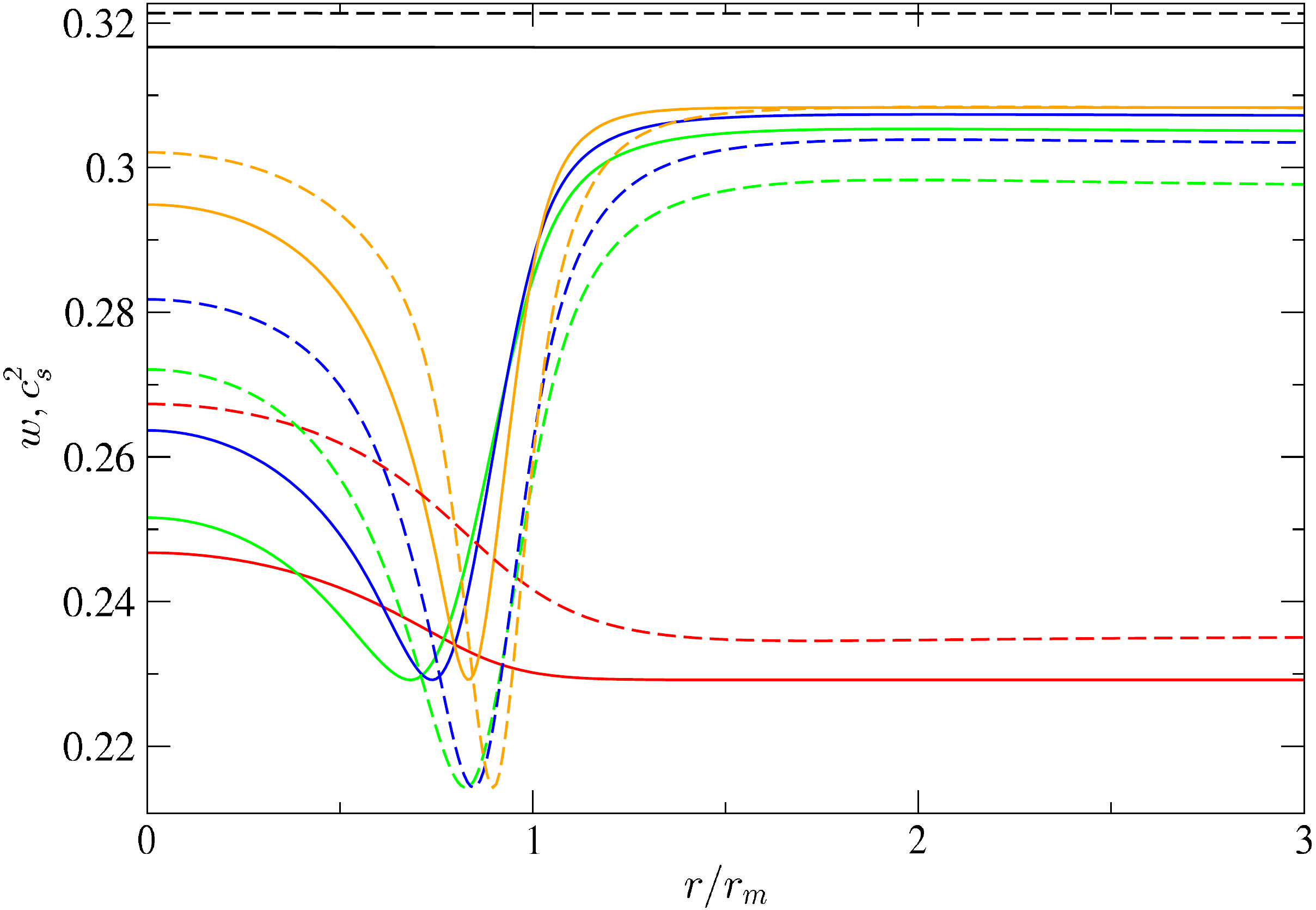}
\vspace{-2mm}\caption{Snapshots of different quantities for the polynomial Gaussian profile at specific times during the gravitational collapse. Top left panel: $\rho/\rho_{\rm b}$, top right panel: $U/\Gamma$, bottom left panel: $\mathcal{C}$. For these panels, the solid line represents the evolution for the QCD crossover with $M_{\rm H}(t_{\rm H}) \approx 2.5 M_{\odot}$, the dashed-line represents the evolution for the radiation-dominated Universe. The bottom right panel shows the case of $w$ (solid line) and $c^2_{\rm s}$ (dashed-line) for the QCD crossover only (for radiation-dominated Universe $w=c^2_{\rm s}=1/3$). In all cases, $\delta_{\rm m}-\delta_{\rm c} = 10^{-1}$ (super-critical).}
\label{fig:w_supercritical}
\end{figure}

\begin{figure}[t]
\centering
\includegraphics[width=3. in]{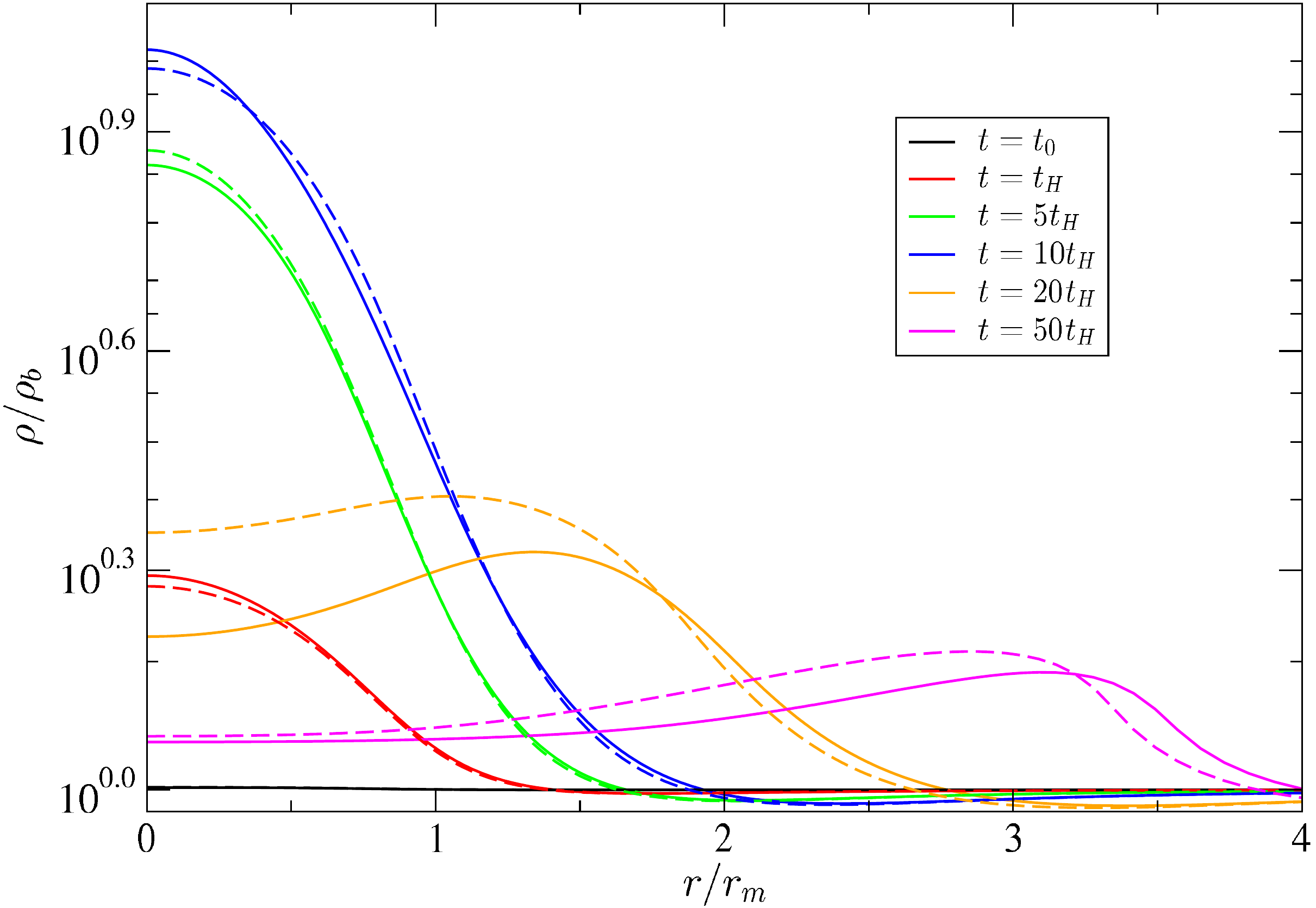}
\includegraphics[width=2.95 in]{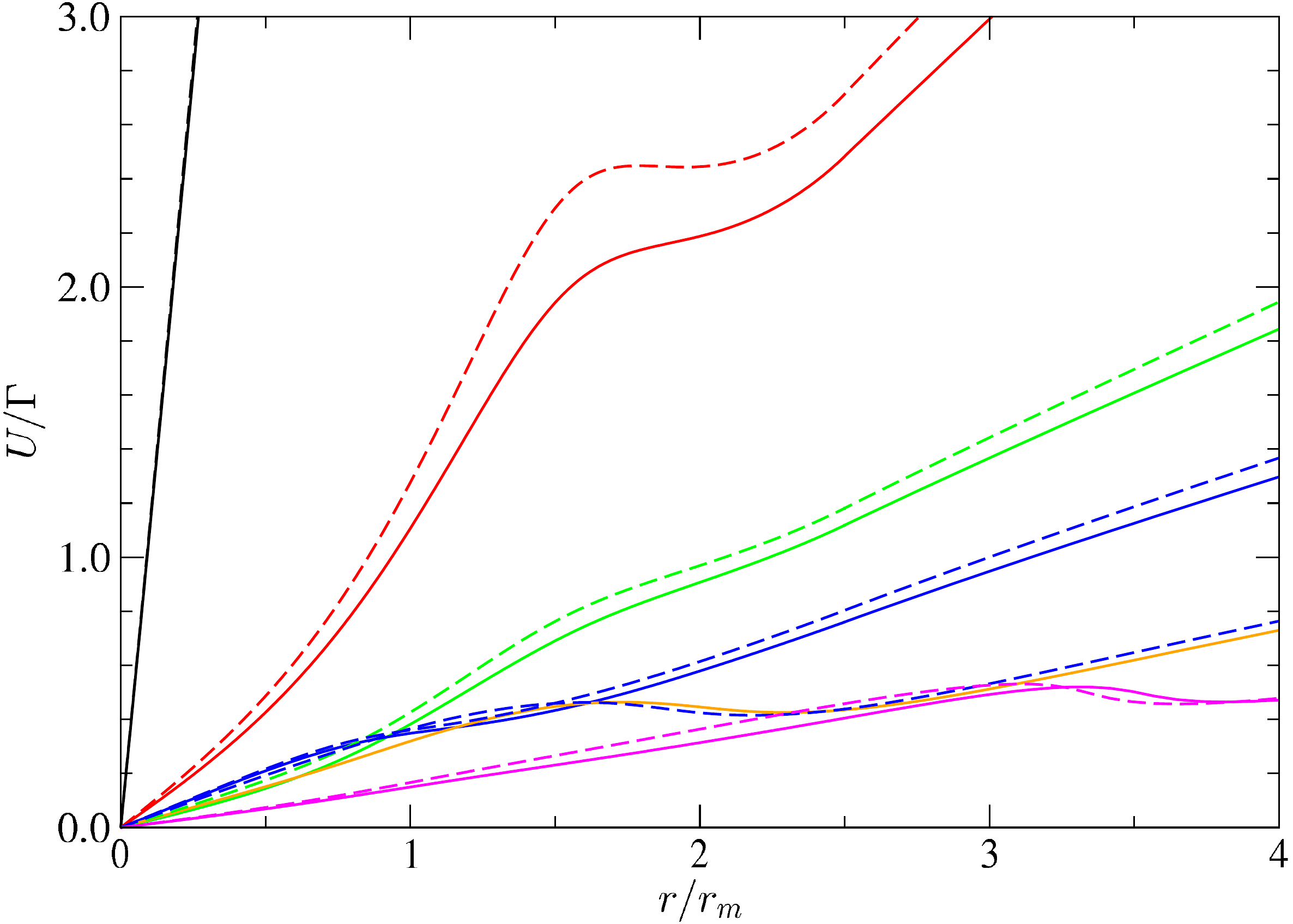}
\includegraphics[width=3. in]{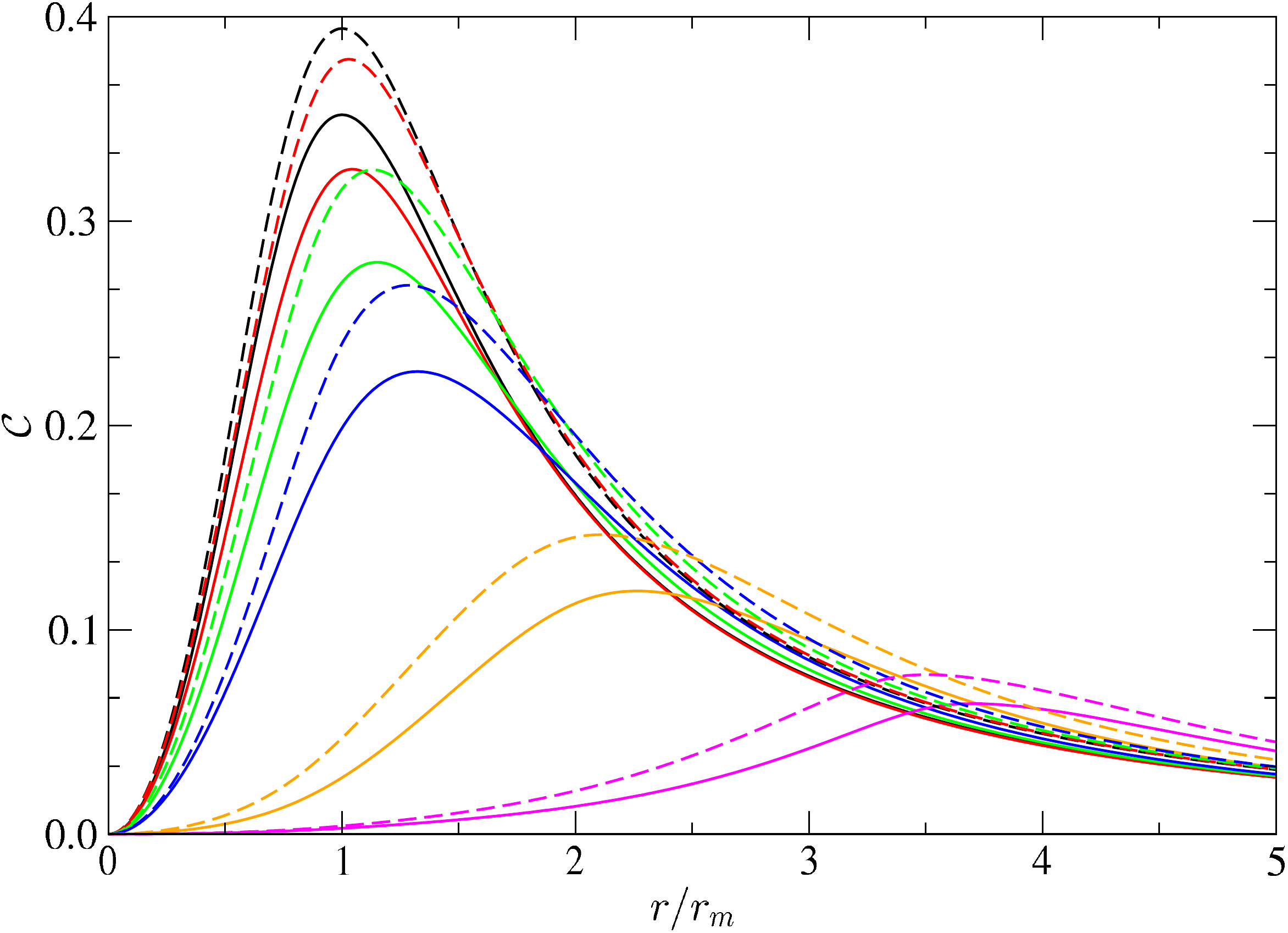}
\includegraphics[width=3. in]{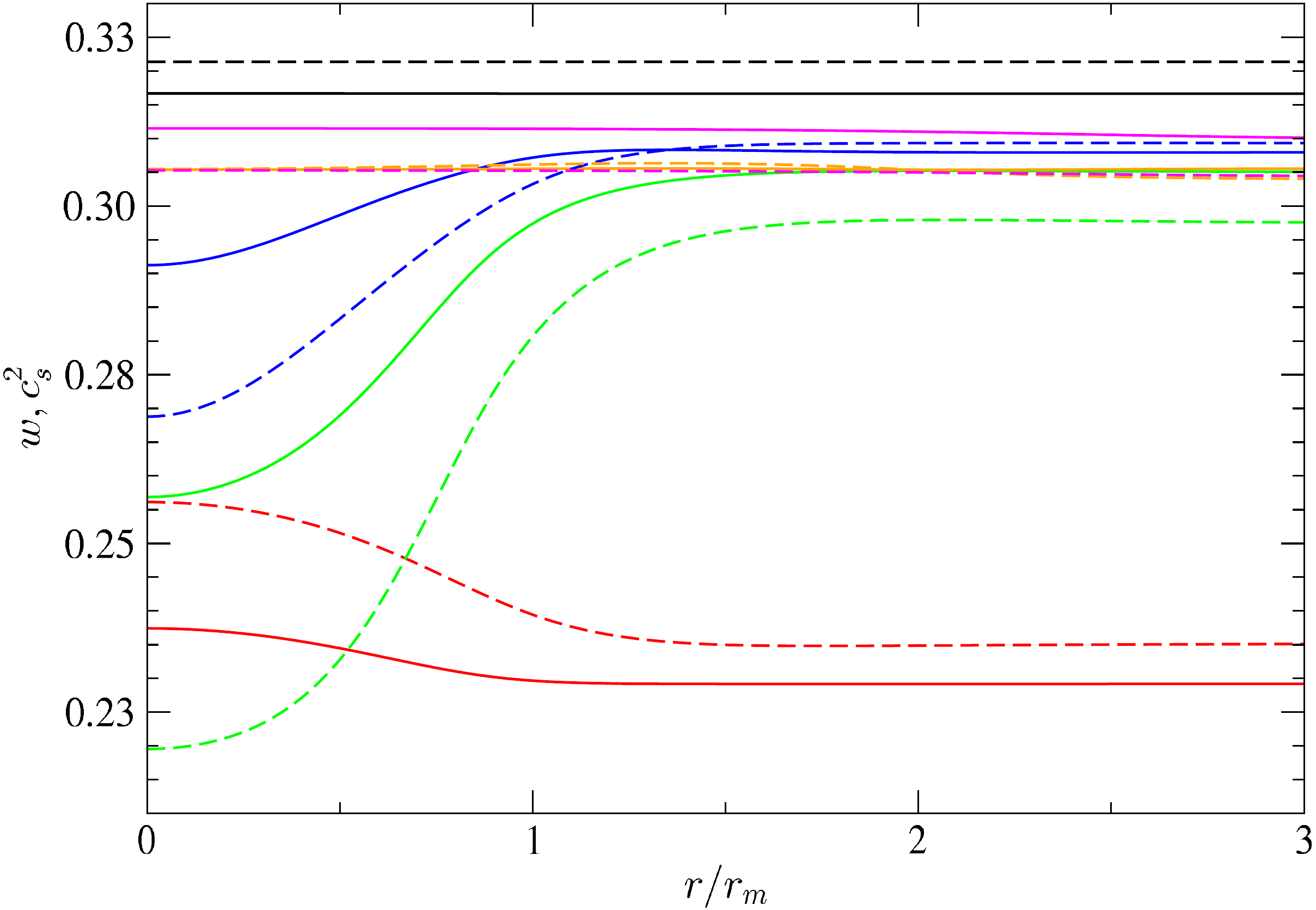}
\vspace{-1mm}\caption{Same caption information as in Fig.~\ref{fig:w_supercritical} with $\delta_{\rm m}-\delta_{\rm c} = -10^{-1}$ (sub-critical).}
\label{fig:w_subcritical}
\end{figure}

\begin{figure}[t]
\centering
\includegraphics[width=3. in]{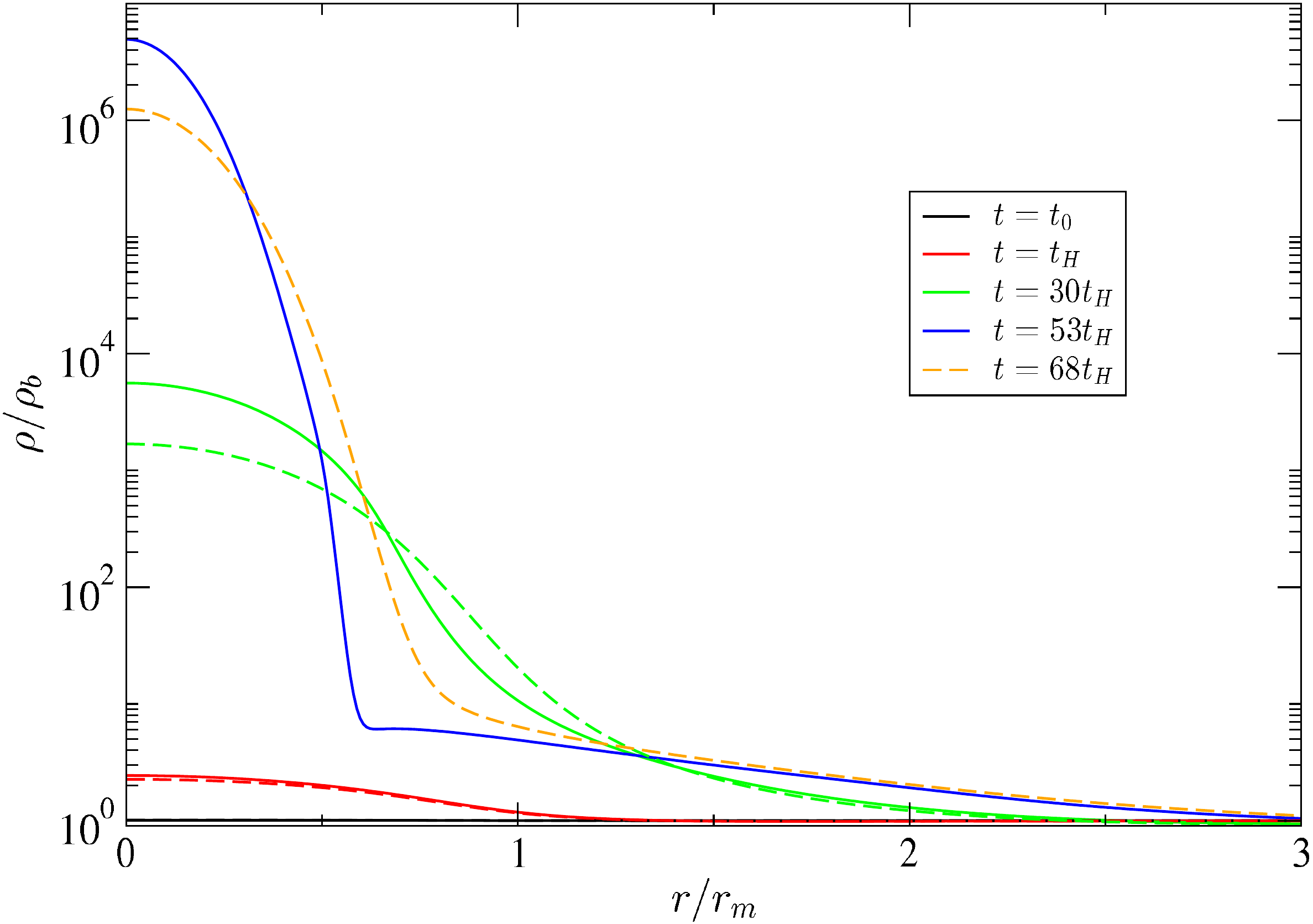}
\includegraphics[width=3. in]{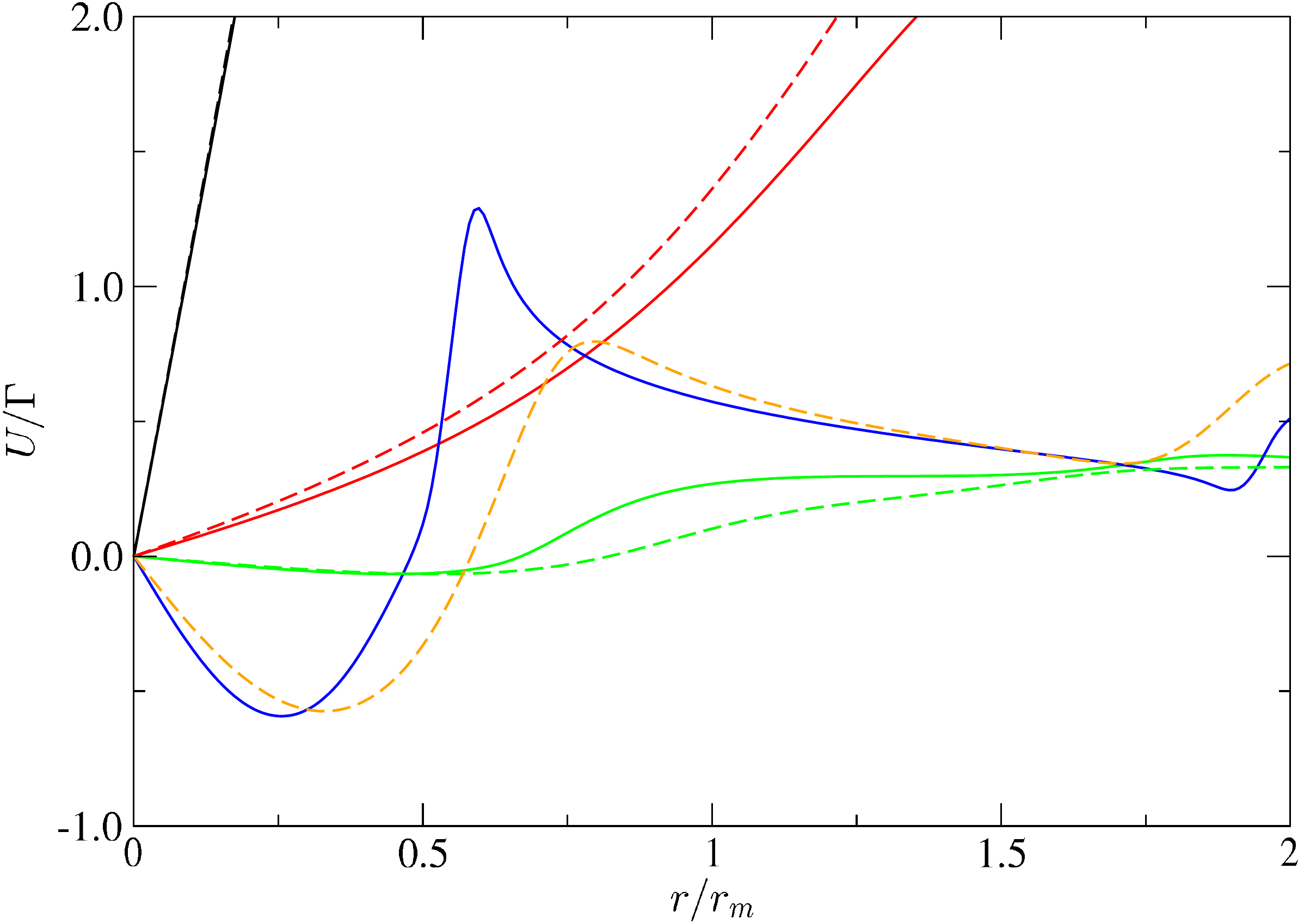}
\includegraphics[width=2.95 in]{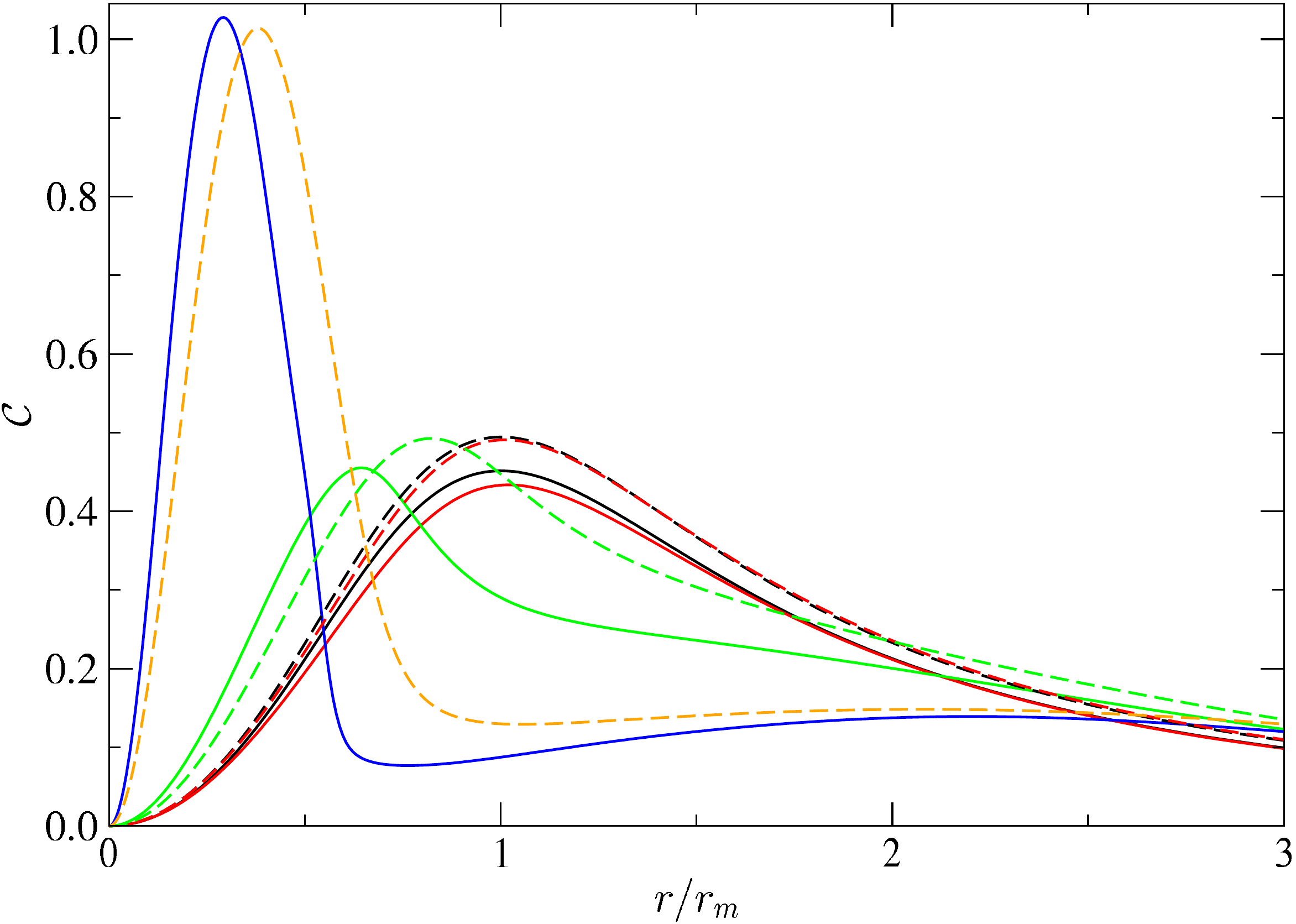}
\includegraphics[width=3. in]{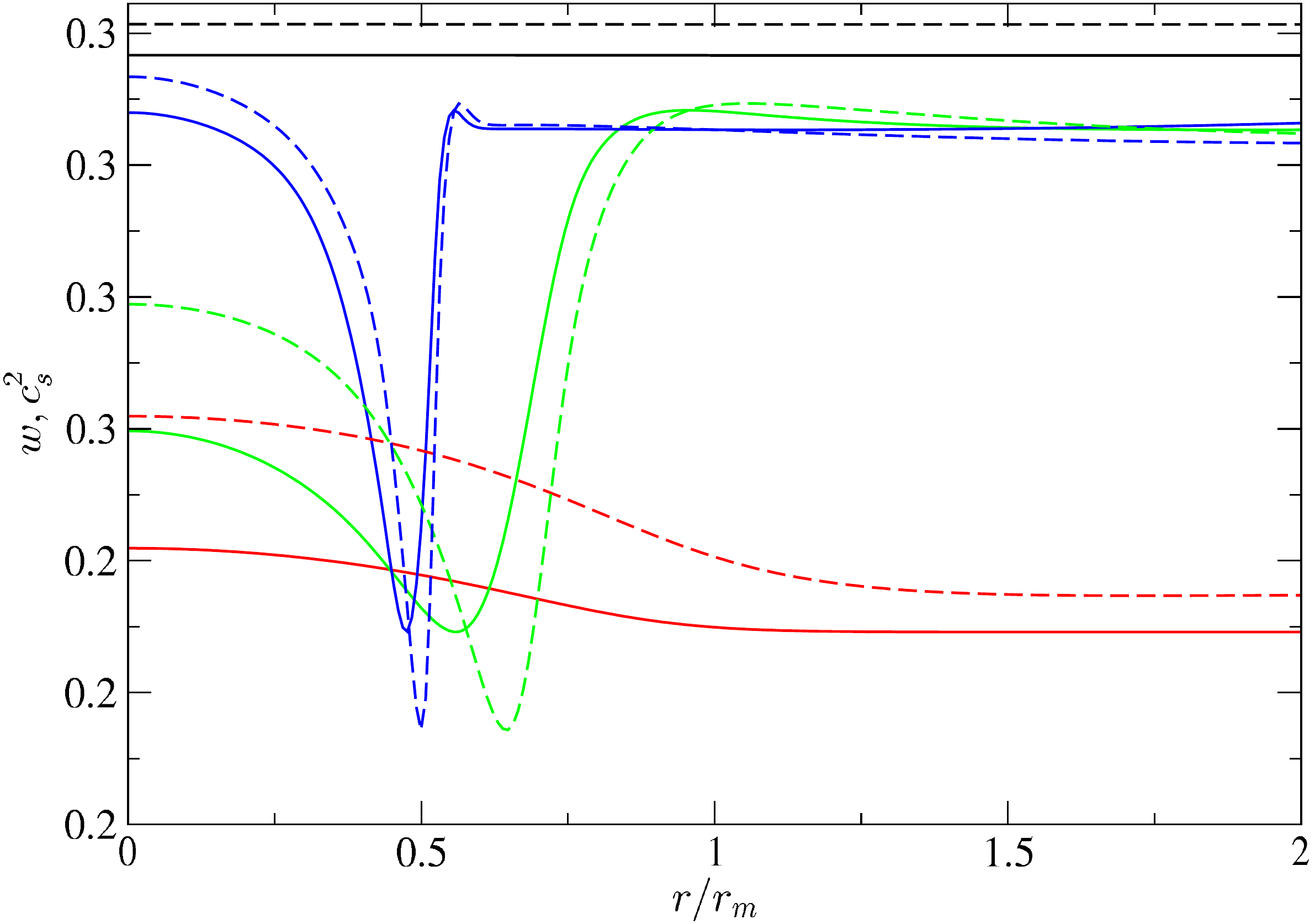}
\caption{Same caption information as in Fig.~\ref{fig:w_supercritical} with $\delta_{\rm m}-\delta_{\rm c} \approx 5 \cdot 10^{-5}$ (critical).}
\label{fig:w_critical}
\end{figure}

The specific evolution of $\rho/\rho_{\rm b}$, $U/\Gamma$, $\mathcal{C}$, $w$ and $c^{2}_{\rm s}$ shown in Figs.~\ref{fig:w_supercritical},~\ref{fig:w_subcritical} and~\ref{fig:w_critical} for the three above-mentioned regimes, leads to the following observations:

\begin{itemize}
    \item In Fig.~\ref{fig:w_supercritical} (super-critical case), the gravitational collapse leads to a continuous growth of the energy density in the central region and
    an apparent horizon forms. The same behaviour is obtained in Fig.~\ref{fig:w_critical} (critical case). But the situation is different in Fig.~\ref{fig:w_subcritical} (sub-critical case), where the peak of the energy density increases up to a specific time and starts to decrease later, while moving outwards from the central region, which means that the perturbation is dispersed in the FLRW background.  The difference with the case $w=1/3$ is noticeable, even the evolution remains qualitatively similar.
    
    \item In Fig.~\ref{fig:w_supercritical}, one  sees that $U/\Gamma$ decreases quickly in time. Instead, in Fig.~\ref{fig:w_subcritical} (sub-critical case), only a small negative value $U/\Gamma$ is reached at early times, and later no negative value is found, which means that the whole perturbation is dispersing instead of collapsing. The most remarkable behaviour is found in the critical case (see Fig.~\ref{fig:w_critical}). The fluid splits into two parts, one going inwards (negative $U$) and one outwards (positive $U$), generating an under-dense region between them. The under-dense region then re-attracts the fluid, leading to an alternation between rarefaction and compression, which is an accelerating process.
  
    \item The typical behaviour of $\rho/\rho_{\rm b}$, $U/\Gamma$ and $\mathcal{C}$ is qualitatively similar to the case of a constant $w=1/3$~\cite{escriva_solo} for the three considered regimes and the chosen values of $\delta_{\rm m}-\delta_{\rm c}$.

    \item As expected, in comparison with the case $w=1/3$, one gets time and radial evolutions for $w$ and $c^{2}_{\rm s}$, as shown in the bottom right panels of Figs.~\ref{fig:w_supercritical},~\ref{fig:w_subcritical} and~\ref{fig:w_critical}. The determinant behaviour of the gravitational collapse will be precisely given by the values of $w, c^2_{\rm s}$ at $t_{\rm H}$, since it is around this time-scale that the pressure gradients and gravitational effects will be more dominant in determining if the cosmological fluctuation will collapse and form a PBH. Moreover, the fact that there are regions where $w,c^2_{\rm s}$ are smaller or larger in comparison with others (especially near the central region) will lead to non-trivial dynamics affecting the PBH mass, as we will see later. 
\end{itemize}

\begin{figure}[t]
\centering
\includegraphics[width=3.0 in]{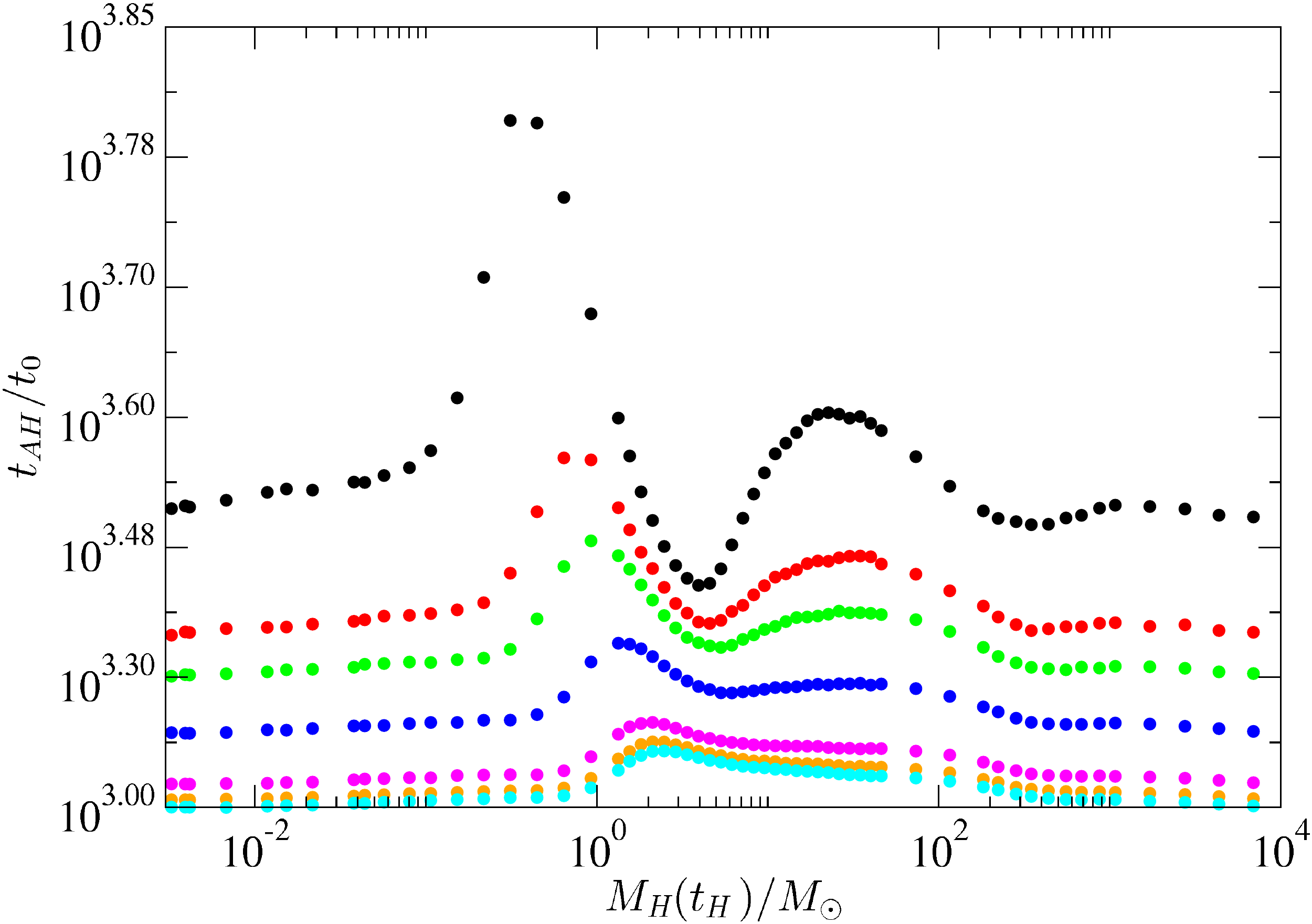}
\includegraphics[width=2.9 in]{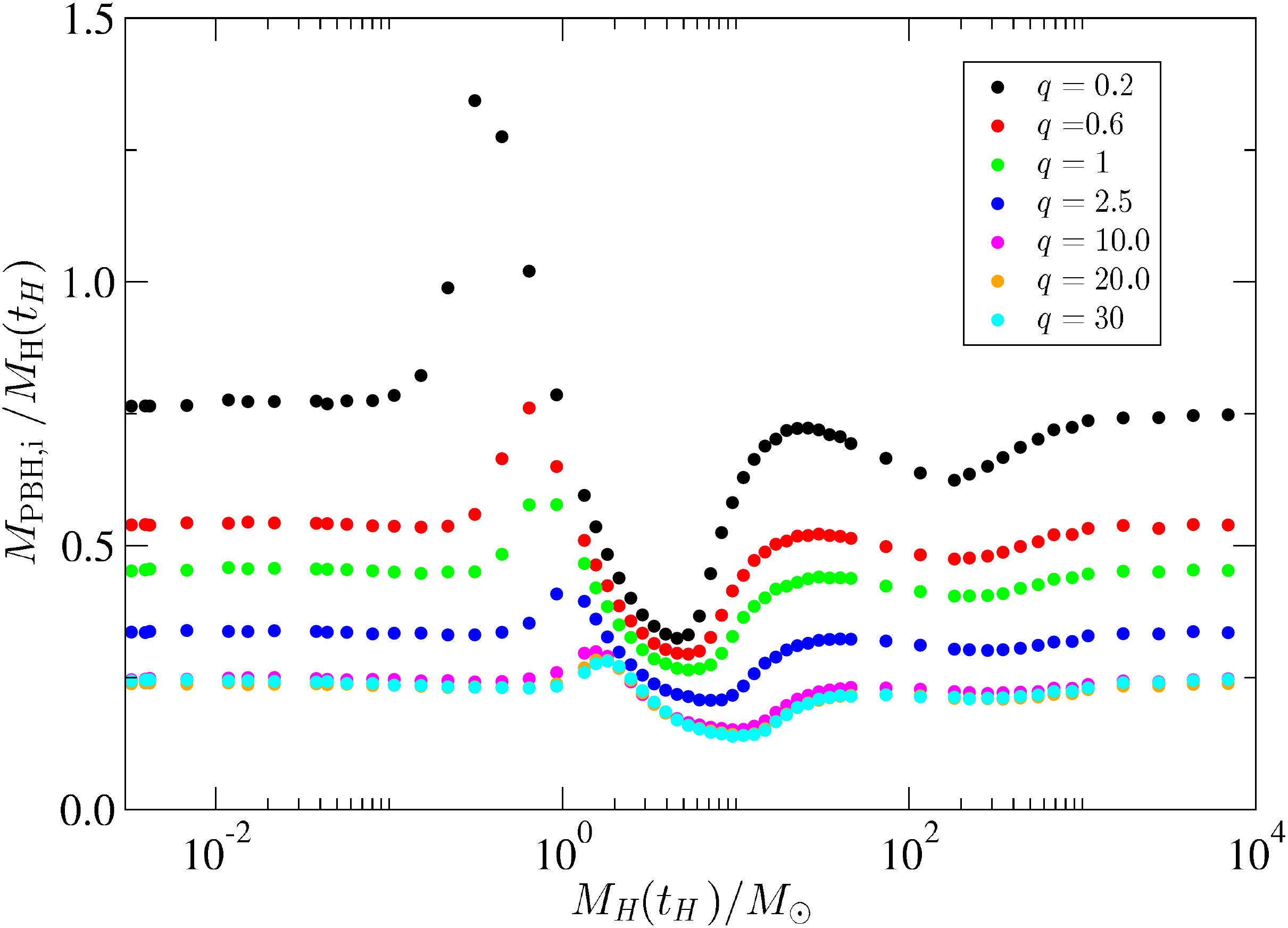}
\caption{Left panel: The time $t_{\rm AH}/t_0$ it takes for the cosmological fluctuations characterized by the profiles of Eq.~\ref{eq:basis_pol} to form the apparent horizon. Right panel: Mass of the apparent horizon when is formed $M_{\rm PBH,i}$ in terms of the Hubble mass $M_{\rm H}(t_{\rm H})$. In all cases, $\delta_{\rm m}-\delta_{\rm c,QCD}=10^{-2}$.}
\label{fig:tAH_formation}
\end{figure}

Fluctuations that satisfy $\delta_{\rm m}>\delta_{\rm c}$ will form an apparent horizon at much later times than $t_{\rm H}$. The time taken for a cosmological fluctuation at super-horizon scales to collapse and form the first apparent horizon $t_{\rm AH}$ will depend on the strength of the fluctuation, the shape, and the equation of state. For a substantial study in the case of a radiation-dominated Universe, see~\cite{Escriva:2021pmf}. As was pointed out in~\cite{Escriva:2020tak}, it is important to notice that pressure gradients are also a form of gravitational energy. So, while they initially work against the collapse, once it is triggered, they mostly favour it. Therefore, in the case of the QCD crossover, we expect that it will have a clear impact. This is precisely what is observed in the left panel of Fig.~\ref{fig:tAH_formation} for different profiles. First of all, there is the effect of the different $q$ profiles: the sharper the profile (larger $q$),  the larger the pressure gradients, and therefore the smaller $t_{\rm AH}/t_{0}$. But we also observe a non-trivial behaviour in the range of $M_{\rm H}(t_{\rm H})$ between 1 and 10 $M_\odot$ for small $q$, where $t_{\rm AH}/t_{0}$ suffers from a transition from larger to smaller values compared with the case $w=1/3$. Although one could expect a larger $t_{\rm AH}/t_{0}$ when $w$ is decreased during the QCD crossover, it is clear that the time-evolution leads to non-trivial behaviours that are different from what one could a priori expect. This is also observed for the case of the mass ratio $M_{\rm PBH,i}/M_{\rm H}(t_{\rm H})$ (we define $M_{\rm PBH,i}$ as $M_{\rm PBH,i}= M_{\rm PBH}(t_{\rm AH})$), where a shorter formation time $t_{\rm AH}$ implies a smaller mass at the moment of formation for the apparent horizon, as shown in the right panel of Fig.~\ref{fig:tAH_formation}.

\subsection{Threshold values}\label{sec:thresholds}

Once the dynamics of PBH formation have been solved numerically, one can compute the threshold values $\delta_{\rm c}$ using a bisection method as a function of $M_{\rm H}(t_{\rm H})$ and study how they are modified by the QCD crossover transition.  We chose a set of representative curvature profiles according to Eq.~\ref{eq:basis_pol} with $q=0.2,0.6,1,2.5,5,10,20,30$.   

As already observed in previous works~\cite{escriva_solo,musco2009,refrencia-extra-jaume}, we see that $\delta_{\rm c}$ is mainly affected by the pressure gradients during the gravitational collapse, which depends on the equation of state and the curvature profile considered.  Apart from an effect of the profile, one expects to have two determining effects on the pressure gradients: the dynamical evolution of $w , c^{2}_{\rm s}$ in time and their spatial variation, as shown in Sec.~\ref{sec:dynamics}.  

In comparison with the case $w=1/3$, the QCD crossover (and in general, any time-dependent equation of state) typically introduces an intrinsic physical scale on the problem related to the length-scale of the perturbation, i.e. at which moment of the evolution of the Universe the fluctuations will re-enter the cosmological horizon.

\begin{figure}[t]
\centering
\includegraphics[width=2.9 in]{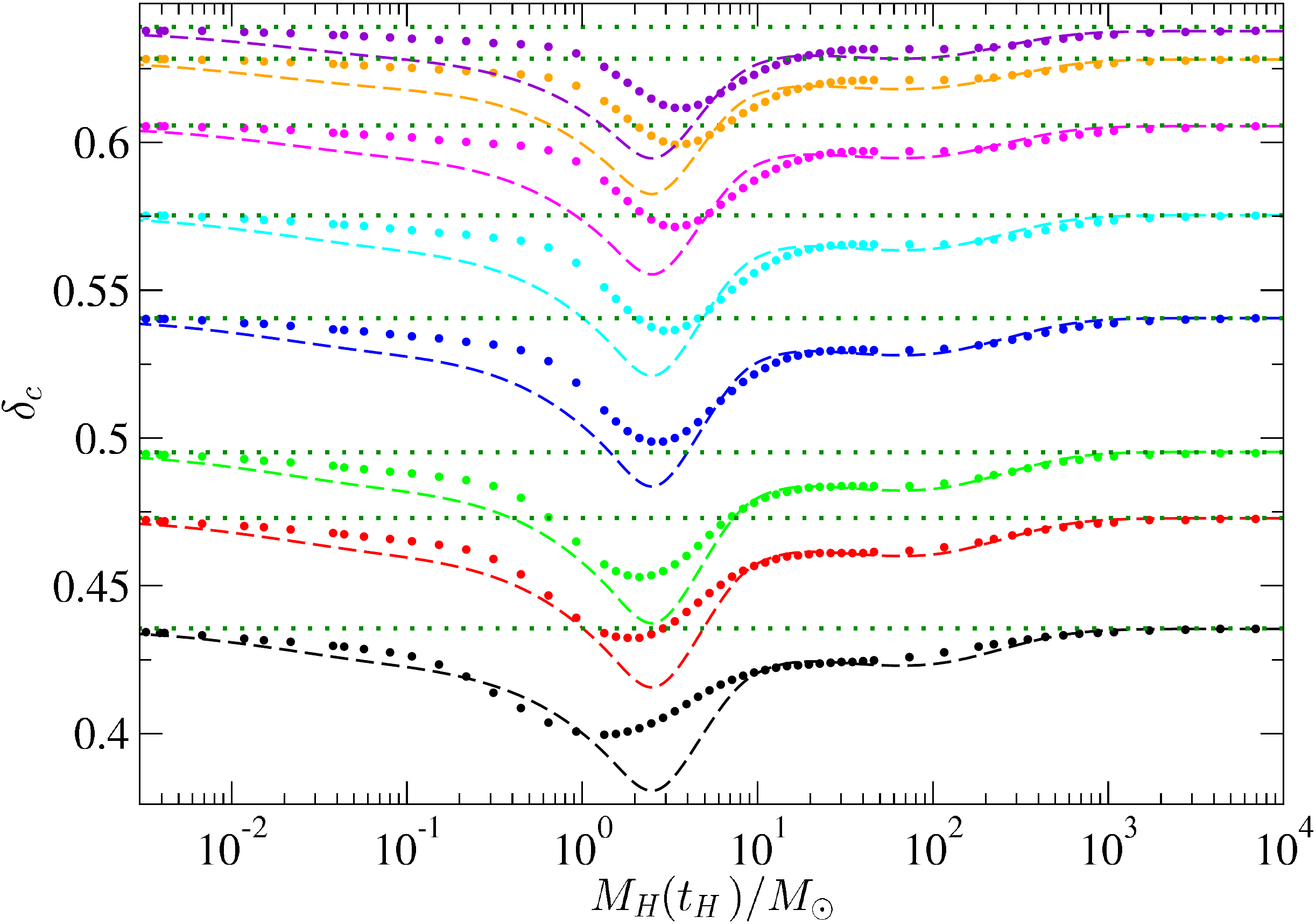}
\includegraphics[width=3.0 in]{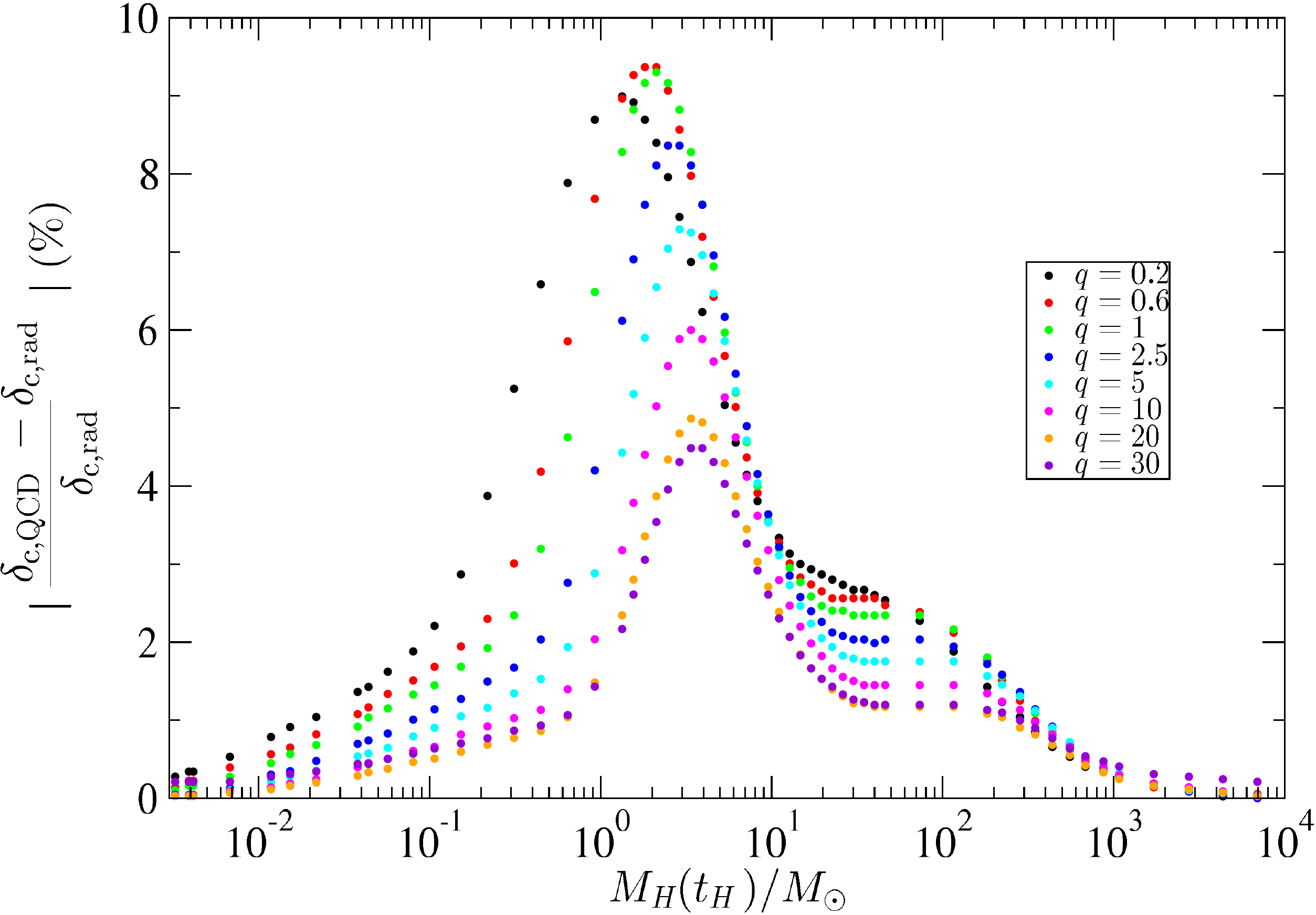}
\caption{Left panel: The dotted lines correspond to the numerical thresholds $\delta_{{\rm c, QCD}}$ of PBH formation with the QCD phase crossover in terms of the horizon mass $M_{\rm H}(t_{\rm H})/M_{\odot}$ at $t_{\rm H}$. The dashed lines correspond to thresholds $ \delta_{{\rm c}, w = \text{cst}}$ considering a constant value $w$ for a given horizon mass value from the tabulated data. The dark green dotted lines represent $\delta_{\rm c,rad}$ for each profile $q$. Right panel: Relative deviation between $\delta_{\rm c, QCD}$ and $ \delta_{{\rm c}, w = \text{cst}}$. In both cases, the colours specify different profiles following Eq.~\ref{eq:profile_exp}.}
\label{fig:thresholds_qcd}
\end{figure}

The values of $\delta_{\rm c}$ obtained from our numerical simulations are shown in Fig.~\ref{fig:thresholds_qcd}, for the three following cases: i) a radiation fluid with $w=1/3$ ($\delta_{\rm c,rad}$); ii) a QCD crossover with $w(t_{\rm H})$ constant ($ \delta_{{\rm c}, w = \text{cst}}$); and iii) a realistic QCD crossover with $w$ varying in each simulation ($\delta_{\rm c,QCD }$).  The relative difference with respect to $\delta_{\rm c,rad}$ is shown on the right panel.  It is about $O(9-10\%)$ for small $q$, but becomes smaller for larger $q$ values, when pressure gradients are stronger. 
For the fluctuations with a length-scale leading to horizon crossing at the QCD epoch, i.e. $10^{-3} \lesssim M_{\rm H}(t_{\rm H})/ M_{\odot} \lesssim 10^3$, the non-linear gravitational collapse and the resulting threshold value are affected by the transient reduction of the equation of state, otherwise the collapse proceeds as in the case of a radiation fluid.  

We observe that the minimum value of $\delta_{\rm c}$ is obtained when $M_{\rm H}(t_{\rm H})$ equals roughly $\mathcal{O}(1)M_\odot$, consistent with the scale at which the minimum of $w, c^2_{\rm s}$ are located.  But the exact location of this minimum depends on the profile and is shifted to larger values of $M_{\rm H}(t_{\rm H})$ when $q$ increases, whereas $ \delta_{{\rm c}, w = \text{cst}}$ remains independent of the curvature profile, with a minimum at $M_{\rm H}(t_{\rm H}) \approx 2.5 M_{\odot}$.  As discussed later, this induces a net effect on the expected PBH mass and merger rate distributions that are importantly modified compared to previous estimations of~\cite{Byrnes:2018clq,Carr:2019kxo,Clesse:2020ghq}. Specifically, we find that for $q \gg 1$, the minimum is located around $M_{\rm H}(t_{\rm H}) \approx 3.5 M_{\odot}$, which seems to coincides with the minimum value of the sound speed $c^{2}_{\rm s}$.  

On the other hand, with the assumption of constant $w$, one finds that the threshold is typically underestimated.
In most cases, $\delta_{{\rm c,QCD} } > \delta_{{\rm c}, w = \text{cst}}$, which means that the gravitational collapse is less efficient than if $c_{\rm s}^{2}=w$ remains constant, because $c_{\rm s}^{2} > w$ during the gravitational collapse.  This will in turn induce a less important QCD peak in the PBH mass function.  Nevertheless, in some cases where $c_{\rm s}^{2}<w$, corresponding to masses between $5 M_\odot$ and $30 M_\odot$ that are of particular interest for gravitational-wave observations, one has $\delta_{{\rm c,QCD} } < \delta_{{{\rm c},w =\text{cst}}}$, leading to slightly boosted abundances in this range. 

Recently, an interesting semi-analytical procedure based on the three-zone model of~\cite{harada} has been proposed in~\cite{Papanikolaou:2022cvo} (see specifically the Fig.~2 of this paper) to estimate the threshold of PBH formation during the QCD crossover \footnote{Notice that the three zone model was shown numerically to lead to non-accurate results for threshold estimations~\cite{Escriva:2020tak}.}. When doing a qualitative comparison with our numerical results, one sees some discrepancies between the left panel of our Fig.~\ref{fig:thresholds_qcd} and the Fig.~2 of \cite{Papanikolaou:2022cvo}. This illustrates the importance of  numerical simulations to take into account the effect of pressure gradients, especially when a time-dependent equation of state is considered: i) the spikes are not present, which is consistent with the fact that a priori there is no physical reason to have $\delta_{\rm c,QCD}>\delta_{\rm c,rad}$ in the particular case of the QCD crossover, since in Eq.~\ref{eq:u_simply} the term proportional to pressure gradients $\rho'$ namely $c^{2}_{\rm s}(\rho)/(1+w(\rho))$ is always $c^{2}_{\rm s}(\rho)/(1+w(\rho)) \leq w_{\rm rad}/(1+w_{\rm rad})$ with $w_{\rm rad}=1/3$ for any $\rho$. This could be different for other equations of state even when $w,c^{2}_{\rm s} \leq 1/3$; ii) the minimum of $\delta_{\rm c,QCD}$ in \cite{Papanikolaou:2022cvo} roughly corresponds to $4.7 M_{\rm H}(t_{\rm H})/M_{\odot}$ and the maximum to $1.2 M_{\rm H}(t_{\rm H})/M_{\odot}$, but for $ q \rightarrow 0 $ (profiles for which the estimation of \cite{Papanikolaou:2022cvo} is assumed to be valid) we find the opposite behaviour: for $q=0.2$ we have the minimum around $1 M_{\rm H}(t_{\rm H})/M_{\odot}$ and we expect that for $q \rightarrow 0$ will be slightly smaller; iii) the asymptotic limit to a radiation fluid in~\cite{Papanikolaou:2022cvo} is also discrepant with simulations, with the three model leading to $\delta_{\rm c} \approx 0.4135$ whereas numerical simulations~\cite{Escriva:2020tak} lead to $\delta_{\rm c}=0.4$ for $q \rightarrow 0$.

\subsection{Comparison of different curvature profiles}\label{sec:other_profiles}

It was shown in~\cite{universal1,Escriva:2020tak} for simulations based on a constant $w \geq 1/3$ that the threshold values obtained for different profiles characterized by the same parameter $q$ are almost identical, with variations that do not exceed $O(6\%)$.  This universal behavior of the averaged compaction function has been used to propose analytical estimations of the threshold~\cite{Escriva:2020tak}.  Notice however that for some specific profiles (those with large negative deviations from Gaussianity~\cite{Escriva:2022pnz}), this analytical approach can be inaccurate, but for our profiles of Eqs.~\ref{eq:basis_pol} it gives correct estimates.

\begin{figure}[t]
\centering
\includegraphics[width=3.0 in]{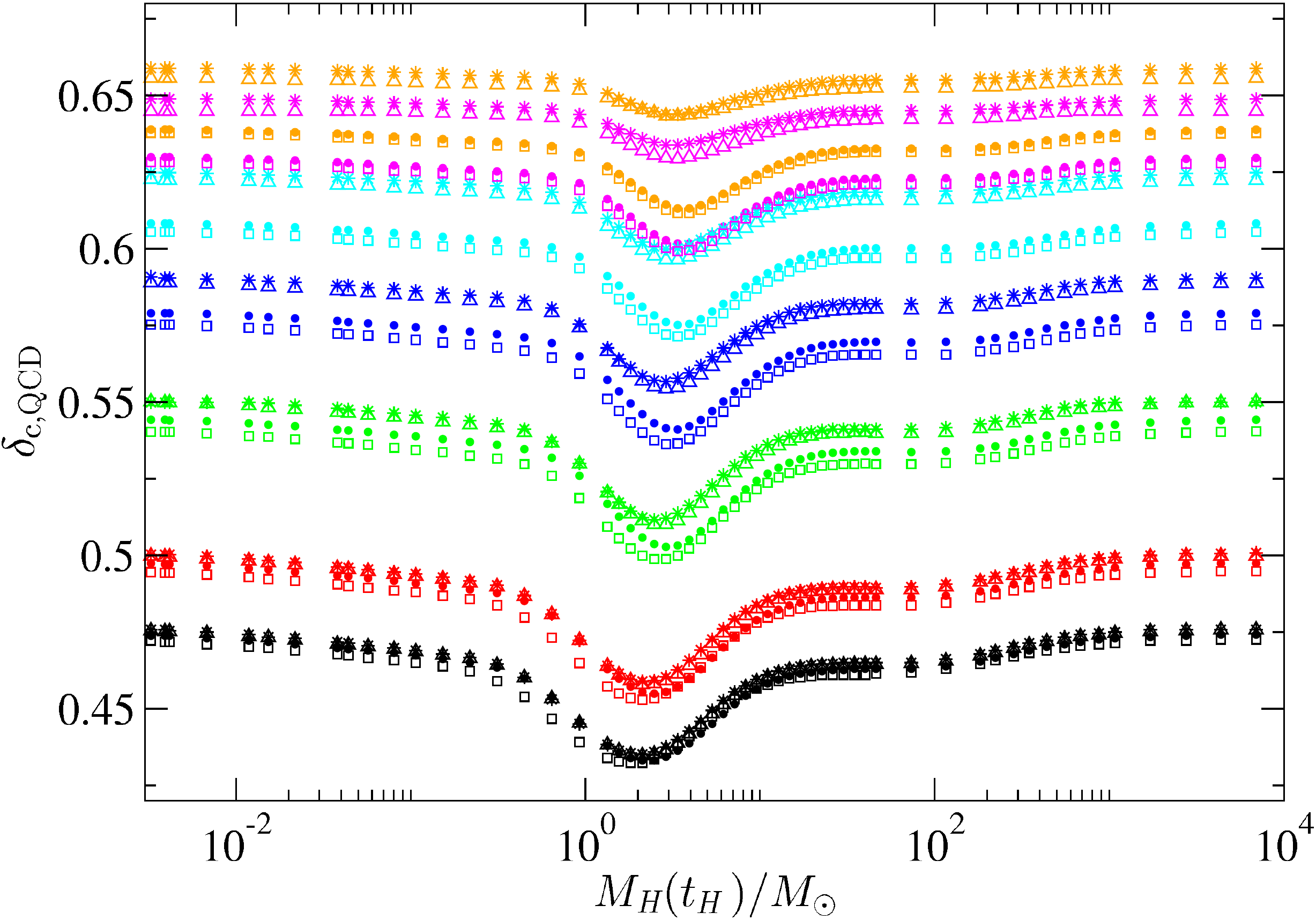}
\includegraphics[width=2.9 in]{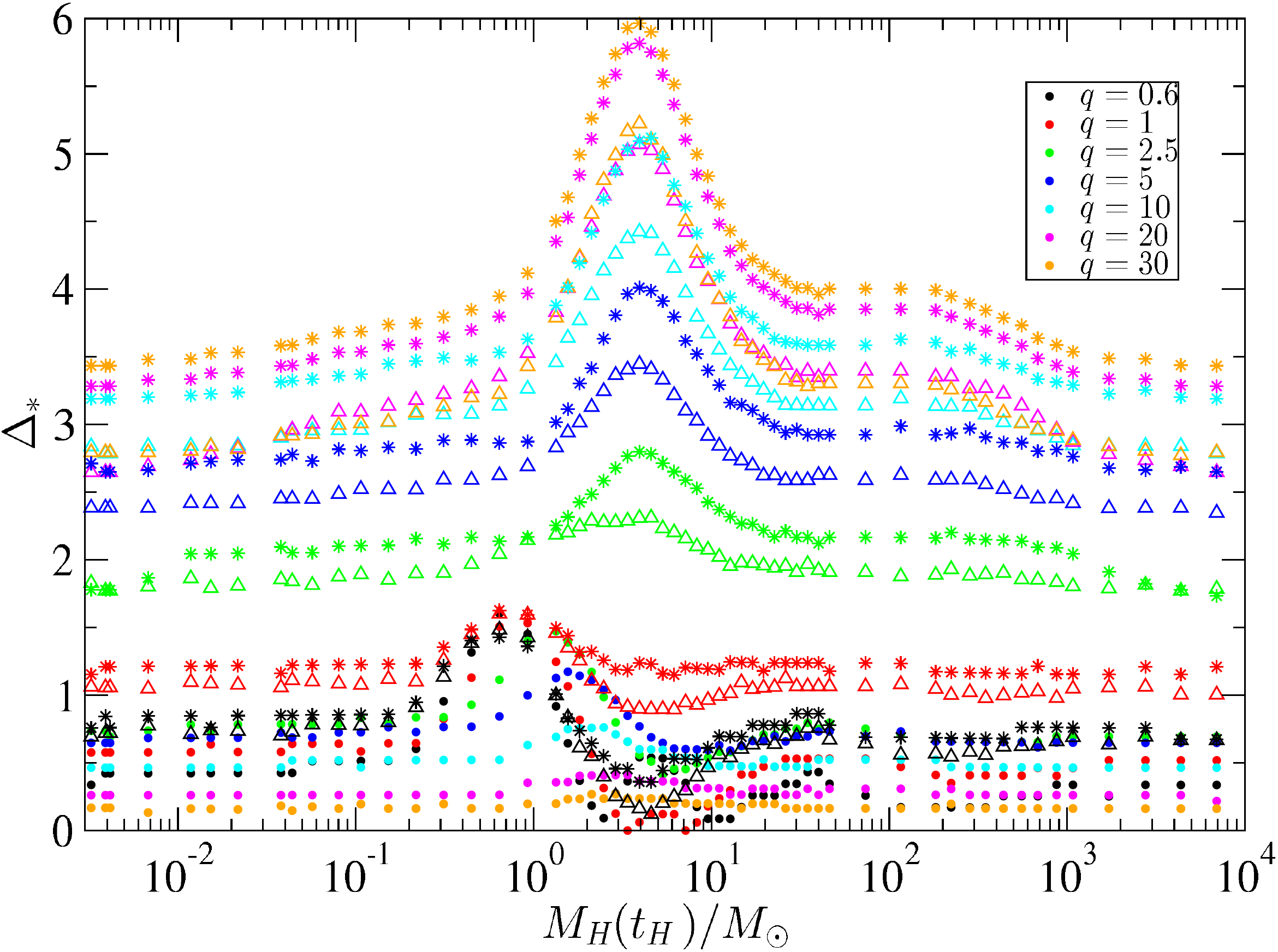}
\caption{Left panel: Numerical thresholds for the profiles of Eqs.~\ref{eq:basis_pol} and~\ref{eq:profile_exp}. The squares represent the values for Eq.~\ref{eq:basis_pol}, the dots the values for the exponential profile Eq.~\ref{eq:profile_exp} with $\lambda=0$, the triangles with $\lambda=2$ and the stars the with $\lambda=10$. Right panel: Relative deviation between the threshold values: exponential - polynomial (dotted), $\lambda= 2$ - polynomial (stars) and $\lambda=10$ - polynomial (triangles).}
\label{fig:thresholds_qcd_comparison}
\end{figure}

Taking profit of the numerical results obtained for profiles described by Eq.~\ref{eq:basis_pol} with different $q$ values, we now perform a comparison between the threshold values for the other types of profiles described by Eq.~\ref{eq:profile_exp}, with $\lambda=0,2,10$. We estimate the relative difference between the thresholds for the same $q$,
\begin{equation}
 \Delta_{*}(\%) =100\left\lvert \frac{\delta_{\rm c,QCD (pol)}-\delta_{\rm c,QCD (exp)}}{\delta_{\rm c,QCD (pol)}}  \right\rvert ,
\end{equation}
where $\delta_{\rm c,QCD (pol)}$ and $\delta_{\rm c,QCD(exp)}$ denote the threshold obtained for the profiles of Eqs.~\ref{eq:basis_pol} and~\ref{eq:profile_exp}, respectively. Our results are shown in Fig.~\ref{fig:thresholds_qcd_comparison}. The relative differences appear to be less than $\mathcal{O}(6\%)$ for the cases considered.  When comparing  polynomial and exponential profiles, we do not observe any substantial deviation and threshold values agree well at $\mathcal{O}(1\%)$ level. 
The main difference comes for the profiles of Eq.~\ref{eq:profile_exp} with $\lambda=2$ and $\lambda=10$, especially for large $q$, in the region where the reduction of the equation of state is the most important. For instance, for the profile with $q=30$, we get roughly two times more deviations at $M_{\rm H}(t_{\rm H})/M_{\odot} \approx 5$ than in the case of a radiation fluid. 

Given that the deviations for different profiles are not substantial, it could be interesting to follow the approach of~\cite{Escriva:2020tak} and use the averaged compaction function to produce an analytical formula for the threshold estimation at the QCD epoch. 
This could be useful for statistical estimations considering all possible profile realizations~\cite{nonlinear}.  We leave this issue for possible future research.

\subsection{Post-horizon dynamics and PBH mass}\label{sec:pbhmass}
Once the apparent horizon has formed, the initial PBH mass, i.e. the mass of the PBH $M_{\rm PBH,i}$ at the moment of formation of the first apparent horizon $t_{\rm AH}$, starts to grow until it reaches an almost stationary value $M_{\rm PBH,f}$ that we denote $M_{\rm PBH}$ for simplicity.  Here we are interested in obtaining the PBH mass in the regime where PBHs are more statistically significant, i.e. $M_{\rm PBH} \approx M_{\rm H}(t_{\rm H})$ \cite{Germani:2018jgr}, and we study the post-horizon dynamics for the particular case of a Gaussian profile.

In order to estimate the final PBH mass, one should in principle follow the evolution of the apparent horizon by using an excision technique (see \cite{escriva_solo} based on the same method as in this work) until the horizon remains static at very late times.  However, in the context of our work, this would be computationally too expensive and would need a substantial increase of the grid resolution.  Therefore, we have followed an approach already used in~\cite{Deng:2016vzb,escriva_solo,Yoo:2021fxs} (based on the Zeldovich-Novikov formula from~\cite{acreation1}) that considers the Bondi-Hoyle accretion~\cite{acreation1,acreation2,acreation3} and assumes that the energy density decreases like in a FLRW background right outside the apparent horizon. 
It is crucial to point out that this assumption is not valid at the formation time $t_{\rm AH}$, but one can apply it to sufficiently late times ($t \gg t_{\rm AH}$), considering an effective accretion efficiency rate $F$~\cite{acreation2,acreation3}. The procedure therefore consists in applying the excision until sufficiently late times in order to subsequently fit the numerical evolution to the Zeldovich-Novikov formula.  In particular, for late enough times $t \gg t_{\rm AH}$, the mass accretion rate should follow
\begin{equation}
\label{eq:2_ZN_formula}
\frac{{\rm d}M}{{\rm d}t} = 4 \pi F R^2_{\rm PBH} \rho_{\rm b}(t),
\end{equation}
where $F$ is a numerical factor found to be of order $\mathcal{O}(1)$, and $R_{\rm PBH}=2M_{\rm PBH}$ is the radius of the PBH.
Contrary to the case of a constant equation of state $w$, Eq.~\ref{eq:2_ZN_formula} does not have in general an analytical solution.  Therefore, our approach to obtain the final PBH mass consists in the following procedure: 
\begin{enumerate}
    \item for a given numerical evolution $M_{\rm PBH}(t)$ with the corresponding $t_{0}$ and $\rho_{\rm b}(t_0)$, we solve the background evolution using Eq.~\ref{eq:background_solution};
    \item using $\rho_{\rm b}(t)$ we solve numerically Eq.~\ref{eq:2_ZN_formula} for a set of values in the range $F \in(1,5)$;
    \item we compute the $\chi^{2}$ value, $\chi^{2}=\sum_{i}\frac{M_{\rm PBH}^{\rm N}(t_{i})-M_{\rm PBH}^{\rm A}(t_{i})}{M_{\rm PBH}^{\rm N}(t_{i})}$,
    by comparing the analytical evolution of the PBH mass with the numerical one, in the range of $t_i$ where the condition  $\dot{M}^{N}_{\rm PBH}(t_i)/[H(t_i) M^{N}_{\rm PBH}(t_i)] <0.1$ is fulfilled~\cite{escriva_solo} ; 
    \item  we find the value of $F$ such that it minimizes $\chi^{2}$, and we solve again Eq.~\ref{eq:2_ZN_formula} to find the final PBH mass at $t \rightarrow \infty$. 
\end{enumerate}

\begin{figure}[t]
\centering
\includegraphics[width=3.0 in]{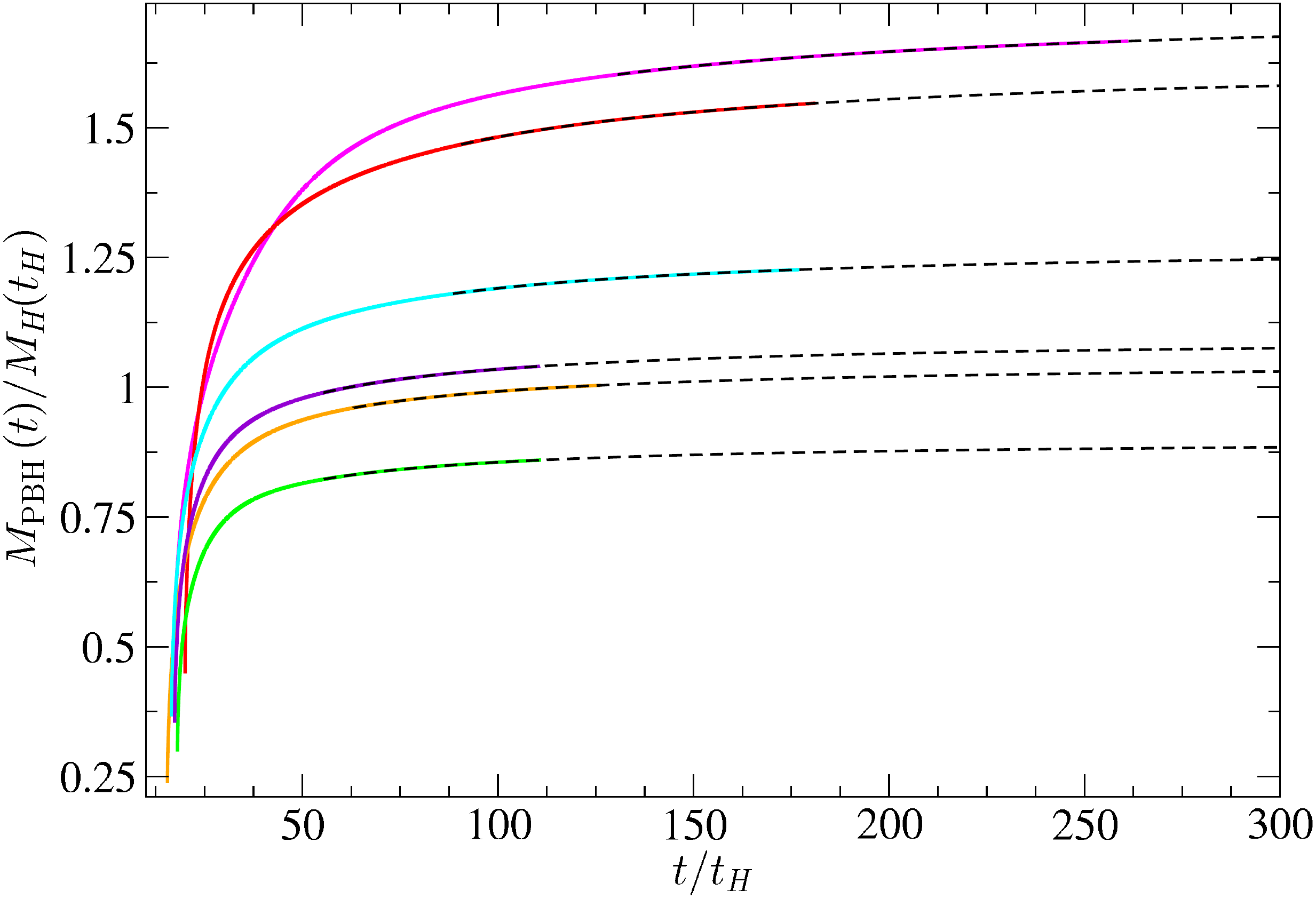}
\includegraphics[width=3.0 in]{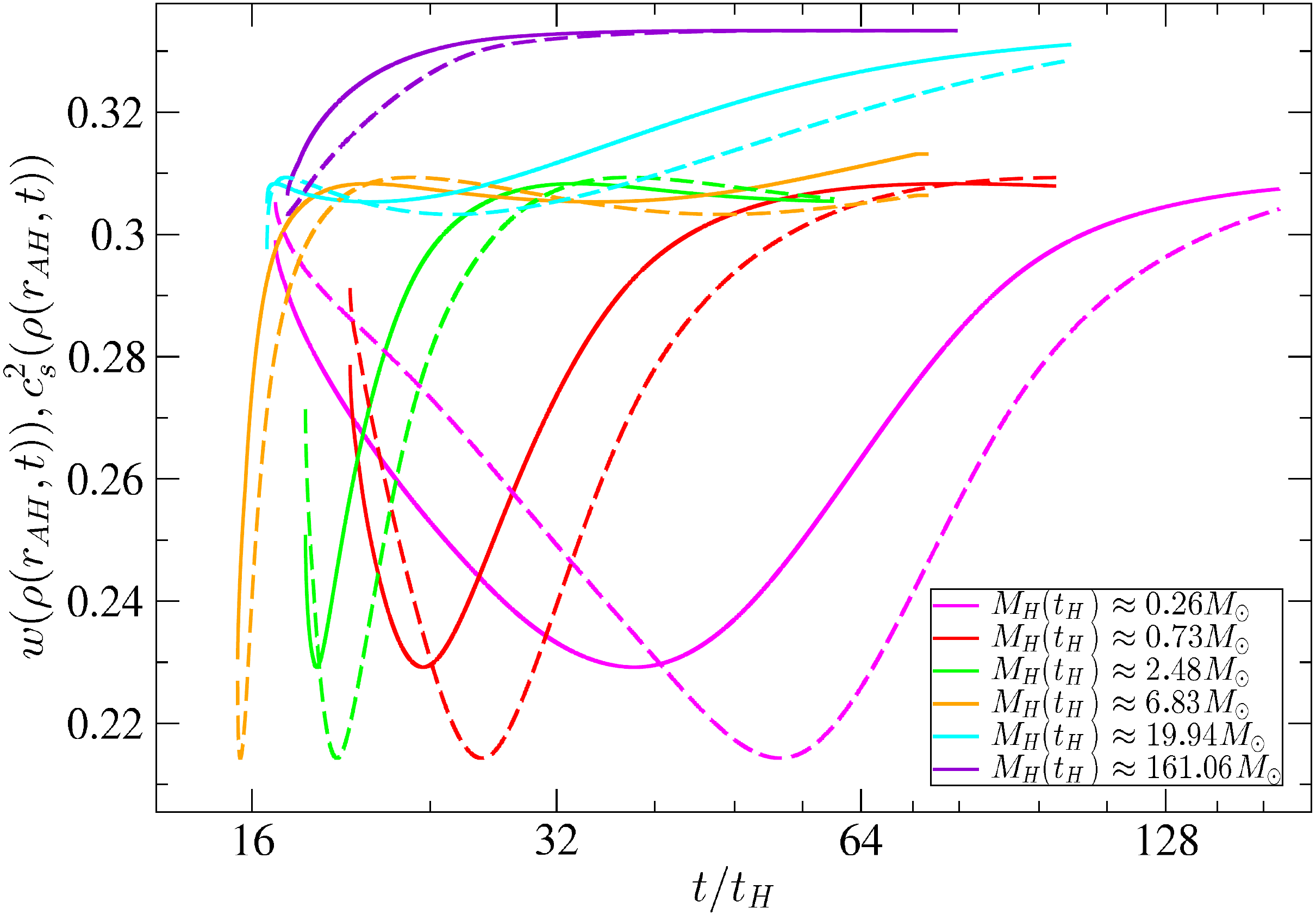}
\caption{Left panel: Evolution of the PBH mass with time for different values of $M_{\rm H}$, the dashed black lines correspond to the numerical fitting from Eq.~\ref{eq:2_ZN_formula}. Right panel: Evolution of $w$ (solid line) and the sound speed $c_{\rm s}^{2}$ (dashed line) evaluated at the apparent horizon location $r_{\rm AH}$, for different values of $M_{\rm H}$. In both cases, $\delta_{\rm m} -\delta_{\rm c,QCD} = 10^{-2}$.}
\label{fig:mass_accretion}
\end{figure}

Typically, we find that $F \sim 4$ minimizes the $\chi^{2}$ value, which is consistent with what was obtained for different constant values of $w$ in~\cite{Escriva:2021aeh}.  If $w$ decreases, the accretion rate $F$ increases, and for $w=1/3$ it was found that $F \approx 3.5$.  One should notice that our approach assumes a constant rate $F$, but a time-dependent rate could ideally be considered since $w$ is not constant. However, for the Gaussian profiles and the range of $\delta_{\rm m}-\delta_{\rm c, QCD}$ considered in this work, we observe that $w$ is practically almost constant for very late times and near to $w=1/3$ (see the right panel of Fig.~\ref{fig:mass_accretion}) and so {we assume that} Eq.~\ref{eq:2_ZN_formula} can be approximately applied to our case when a substantial mass has already been accreted (see left panel of Fig.~\ref{fig:mass_accretion}).

\begin{figure}[t]
\centering
\includegraphics[width=3.5 in]{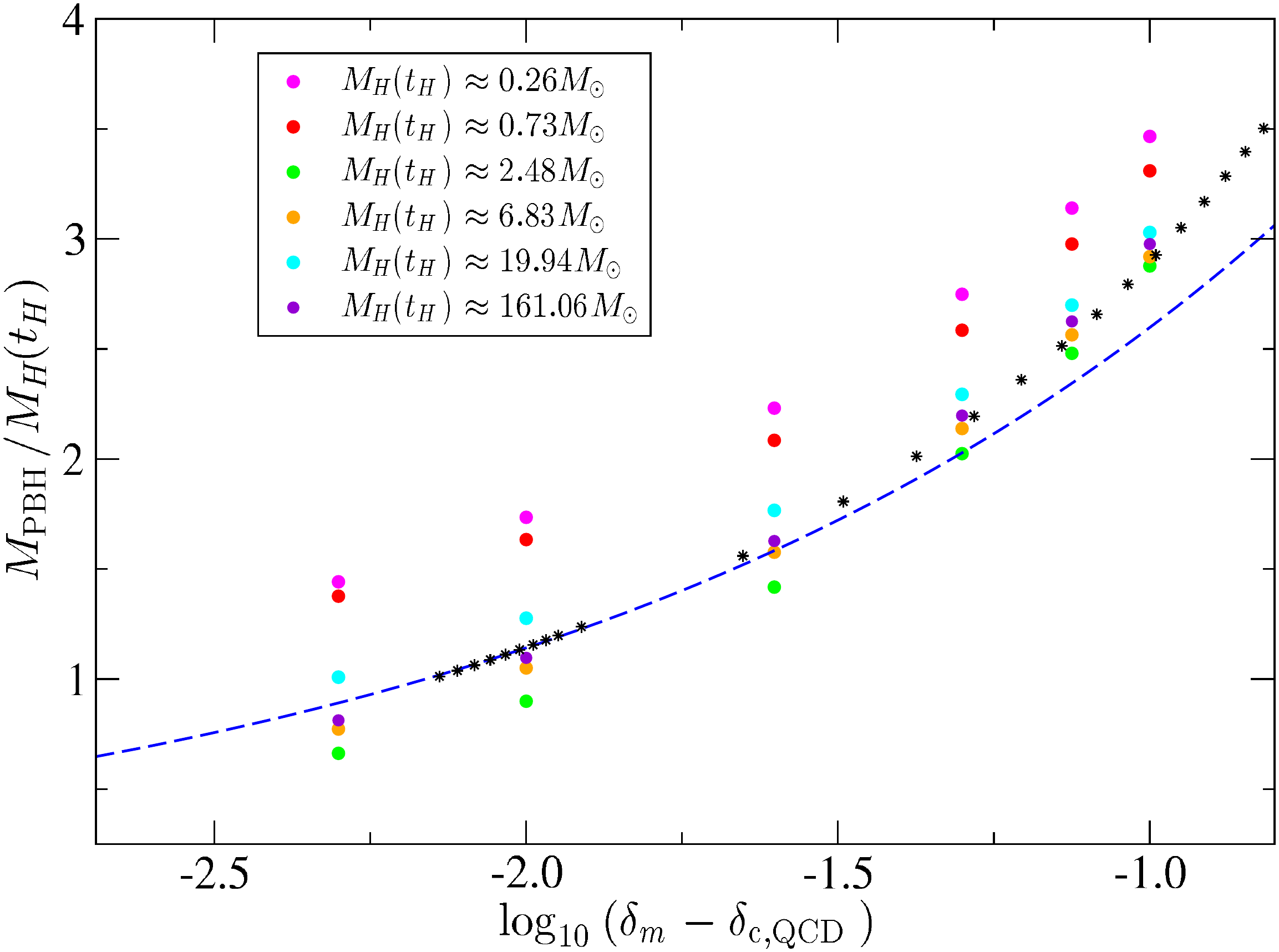}
\caption{PBH mass as a function of $\delta_{\rm m}$ for different values of $M_{\rm H}(t_{\rm H})$ (colored dots). The black dots correspond to the case found in~\cite{escriva_solo} using a Gaussian profile and $w=1/3$. The dashed line is the scaling law mass.}
\label{fig:PBHmass}
\end{figure}
In Fig.~\ref{fig:PBHmass} we show the final PBH mass as a function of the horizon mass $M_{\rm H}(t_{\rm H})$ in different cases for $\delta_{\rm m}-\delta_{\rm c,QCD} \gtrsim 10^{-2.5}$. In the case of a constant $w$, the final black hole mass follows a scaling when $\delta_{\rm m}$ is very close to the critical value $\delta_{\rm c}$ (critical regime)~\cite{musco2009,Niemeyer2,hawke2002},
\begin{equation}
M_{\rm PBH} = {\cal K} M_{\rm H}(t_{\rm H}) (\delta_{\rm m}-\delta_{{\rm c},w={\rm cst}})^{\gamma},
\label{eq:2_scaling}
\end{equation} 
where $\gamma \approx 0.356$ in radiation domination and $\mathcal{K}$ depends on the curvature profile with values of $\mathcal{O}(1)$. The values of $\gamma$ in terms of the equation of state $w$ were found semi-analytically in~\cite{Maison:1995cc}, and confirmed numerically in~\cite{musco2013}. In the case of the QCD crossover, we see that the final mass of the PBH depends on $M_{\rm H}(t_{\rm H})$ and differs from the case $w=1/3$, but this dependence seems to be relatively weak. 
Interestingly, as $\delta_{\rm m}$ increases, the differences in $M_{\rm PBH}$ for different $M_{\rm H}(t_{\rm H})$ are slightly reduced. We also observe in accordance with Fig.~\ref{fig:tAH_formation}, that $M_{\rm PBH}$ is higher or smaller than the case of $w =1/3$ depending on the $M_{\rm H}(t_{\rm H})$.

Although this goes beyond the scope of this paper, in order to explore consistently the behaviour of the critical regime for the QCD crossover, one would need to run simulations with a much higher accuracy and smaller $\delta_{\rm m}-\delta_{\rm c,QCD}$, which is computationally expensive. But assuming that a scaling law still holds in the regime $\delta_{\rm m}-\delta_{\rm c,QCD} \approx 10^{-2}$, we make a numerical fit using the values near the horizon mass to find that $\gamma \approx [0.3, 0.38]$ (with $\gamma \sim 0.34$ as mean value) for the different $M_{\rm H}(t_{\rm H})$ cases, with $\mathcal{K}\approx [4,6]$. It should be noticed that due to the resolution on the threshold estimation (in this case $\mathcal{O}(0.01\%$)) and probably also due to our approach based on Eq.~\ref{eq:2_ZN_formula} (where we consider that $F$ is constant when performing the numerical fit), we have an uncertainty on the value of $\gamma$ of $\sim \mathcal{O}(0.02)$ and the corresponding $\mathcal{K}$.  Therefore, we take the crude estimation $\gamma=0.34$ in Sec.~\ref{sec:broad_pbh}. 

The critical behaviour for much smaller values than $\delta_{\rm m}-\delta_{\rm c,QCD} \approx 10^{-2}$ could be much more complicated and non-trivial than what is found in Fig.~\ref{fig:PBHmass} \cite{Musco_talk}\footnote{We thank Ilia Musco for presenting his preliminary work with collaborators regarding the critical behaviour during the QCD crossover at the Mini-workshop on PBHs in Brussels \cite{minipbh}.}. Moreover, it would be interesting to test the critical behaviour for other different curvature profiles, which could be different from the Gaussian profile tested. Future studies of these aspects will be crucial to better determine the critical behaviour of PBH formation during the QCD transition~\cite{Musco_talk}.

\section{PBH mass function with QCD features} \label{sec:pbh_distribution}

We compute the inevitable effects of the QCD crossover transition on the PBH mass function, following ~\cite{Byrnes:2018clq,Carr:2019kxo}. The aim is to compare previous results with mass distributions relying on i) our new results for the threshold $\delta_{\rm c,QCD}$; and ii) different possible shapes for the curvature profile, from Eq.~\ref{eq:basis_pol}.  Then, we study the effects of the scaling law regime from the information obtained in Sec.~\ref{sec:pbhmass}. 

There are alternative or more accurate approaches to estimate the PBH mass function~\cite{Suyama:2019npc,DeLuca:2020ioi,Tokeshi:2020tjq,Kim:1996hr,Green:2004wb,Young:2014ana,Yoo:2018kvb,germani-vicente}, such as using peak theory~\cite{peak_theory,Germani:2018jgr} or non-linear statistics~\cite{nonlinear}. Most often, this impacts the total abundance of PBHs but does not affect the normalized PBH mass distribution.  But the PBH abundance can be simply rescaled by an overall shift of amplitude of the primordial power spectrum of curvature fluctuations.  Our approach here is to \textit{assume} a PBH abundance $f_{\rm PBH} \equiv \rho_{\rm PBH}/\rho_{\rm DM}$ today (where $\rho_{\rm PBH}$ and $\rho_{\rm DM}$ are the PBH and dark matter densities, respectively), such that one can use the simple standard method of~\cite{Byrnes:2018clq,Carr:2019kxo} to estimate the \textit{normalized} PBH mass distribution $f(M_{\rm PBH})$, defined such that $\int {\rm d} \ln M_{\rm PBH} f(M_{\rm PBH}) = 1$.

\subsection{Standard approach}\label{mass_function3}

We suppose that primordial density fluctuations $\delta_{\rm m}$ are Gaussian, with a probability density function 
\begin{equation}
P(\delta_{\rm m}) = \frac{1}{\sqrt{2 \pi \delta^2_{\rm rms}} }e^{-\frac{\delta^{2}_{\rm m}}{2\delta^2_{\rm rms} }},
\end{equation}
where $\delta_{\rm rms}(M_{\rm PBH})$ is the root-mean-square amplitude, related to the primordial power spectrum.  We assume a nearly scale-invariant power spectrum on PBH length scales, with a spectral index within the range $n_{\rm s}=0.965-0.975$, such that~\cite{Bagui:2021dqi,Carr:2019kxo} 
\begin{equation}
    \delta_{\rm rms}(M_{\rm PBH}) = \bar{A} \left(\frac{M_{\rm PBH}}{\Msun} \right)^{\frac{1-n_{\rm s}}{4}}.
    \label{eq:sigma_fpbj}
\end{equation}
This range of $n_{\rm s}$ is well motivated for two reasons.  First, smaller or larger values would lead to an overproduction of PBHs with large or tiny masses compared to the solar-mass scale associated with the QCD transition.  Scenarios with $n_{\rm s} = 1$, for instance, require a cut-off scale in the power spectrum~\cite{DeLuca:2020agl} to avoid this overproduction.  Second, this is a generic outcome of slow-roll inflation models.  Note that we are assuming a transition in the power spectrum between large cosmological scales, probed in CMB observations, and small scales associated to stellar-mass PBH formation, which would therefore correspond to two different inflationary phases, based on two distinct scalar fields.

Different shapes for the primordial power spectrum are possible (log-normal, Gaussian, broken power laws,...) as well as non-Gaussian statistics~\cite{DeLuca:2019qsy,Young:2019yug}, depending on the underlying theoretical model, but different choices typically impact only the general shape of the PBH mass distribution and not the QCD-induced features in the $[1-100] M_\odot$ range. Slow-roll single field inflation models, for instance, can hardly make such a transition in the power spectrum~\cite{Byrnes:2018txb}, except if there is a phase where quantum diffusion becomes important~\cite{Ezquiaga:2018gbw,Garcia-Bellido:2017mdw} and induces non-Gaussian tails, like in critical Higgs inflation~\cite{Ezquiaga:2017fvi}, thereby boosting the PBH formation.

{For a scale invariant power spectrum of Gaussian primordial fluctuations, one would expect curvature profiles characterized by $q\approx 3$~\cite{Musco:2020jjb}.  Nevertheless, for the sake of generality, we have explored the effects of different values of $q$, which could arise from non-Gaussian tails, different power spectra, or still unknown physical phenomena.   }

The abundance of PBHs at formation (per unit of logarithmic mass) is then obtained by applying the Press-Schechter formalism and integrating $P(\delta_{\rm m})$ above the threshold $\delta_{\rm c} (M_{\rm PBH})$, which gives
\begin{equation}
\beta(M_{\rm PBH}) \equiv \frac{{\rm d} \rho_{\rm PBH}}{{\rm d} \ln M_{\rm PBH}} = {\rm erfc} \left[\frac{\delta_{\rm c}(M_{\rm H})}{\sqrt{2}\delta_{\rm rms}(M_{\rm H})}\right].
\label{eq:beta}
\end{equation}
Finally, the present mass distribution $f(M_{\rm PBH})$ is given by:
 \begin{equation}
f(M_{\rm PBH}) = \frac{1}{f_{\rm PBH} \, \rho_{\rm DM}}\frac{{\rm d}  \rho_{\rm PBH}}{{\rm d} \ln(M_{\rm PBH})} \approx \frac{2.4}{f_{\rm PBH}} \beta(M_{\rm PBH}) \sqrt{\frac{M_{\rm eq}}{M_{\rm H}}},
\label{eq:mass_function}
\end{equation}
where the factor $2.4$ comes from $2(1+\Omega_{\rm b}/\Omega_{\rm CDM})$, with $\Omega_{\rm CDM} = 0.245$ and $\Omega_{\rm b}=0.0456$ being the CDM and baryon density parameters~\cite{Planck:2018vyg}, respectively, and $M_{\rm eq}$ is the Hubble mass at radiation-matter equality, $M_{\rm eq}= 2.8 \times 10^{17} \Msun$.  Finally, we adopt an effective approach for the ratio between the Hubble-horizon mass and the PBH mass, described by an effective parameter $\alpha \equiv M_{\rm PBH} / M_{\rm H}$ within the range $[0.7-1]$, taking into account the uncertainties due to the absence of a complete study of the critical collapse for the QCD crossover transition.  
In Sec.~\ref{sec:broad_pbh}, we motivate this range and this approach using a refined method based on the the above-mentioned scaling law.

\begin{figure}[t]
\centering
\includegraphics[width=3.0 in]{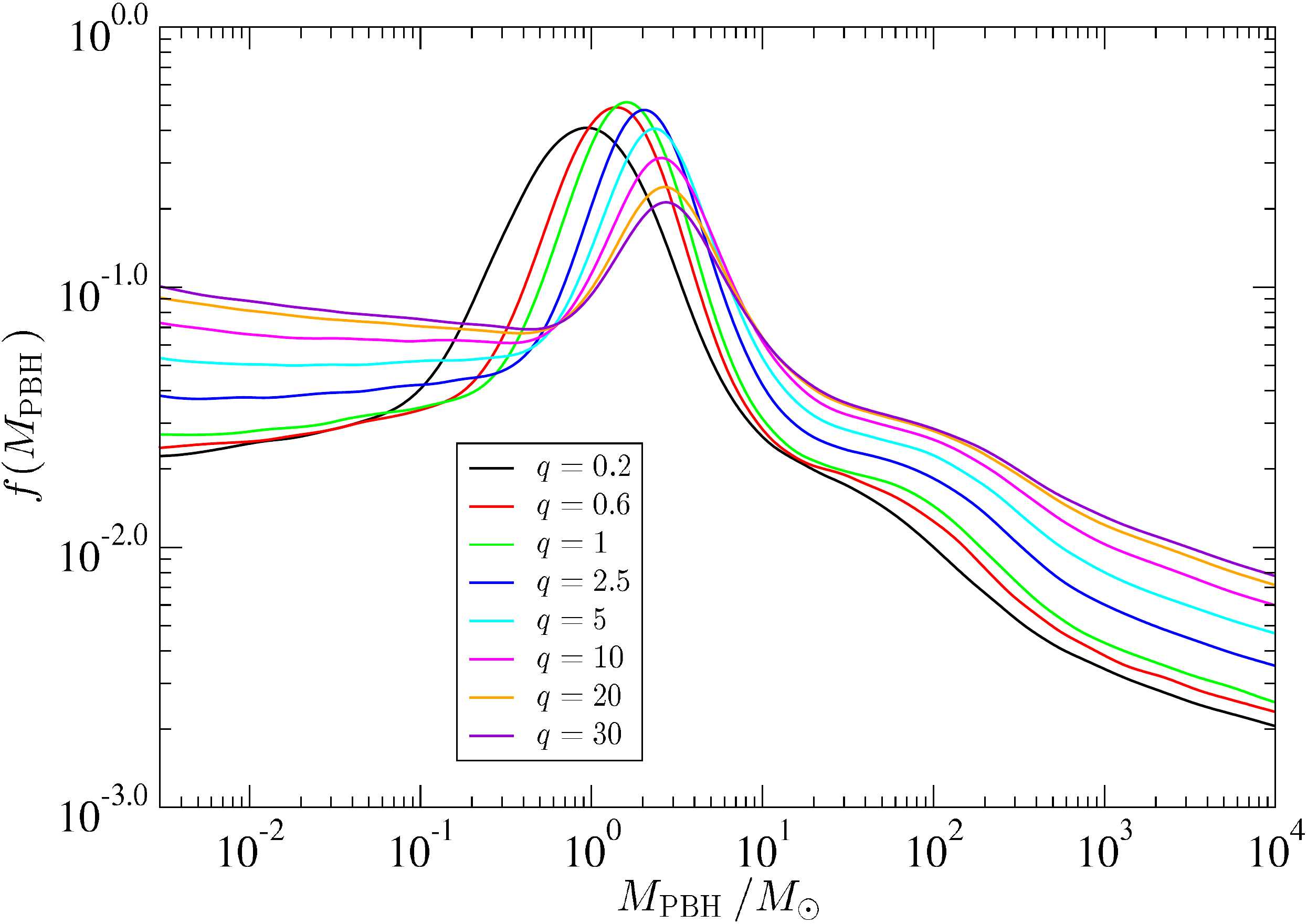}
\includegraphics[width=3.0 in]{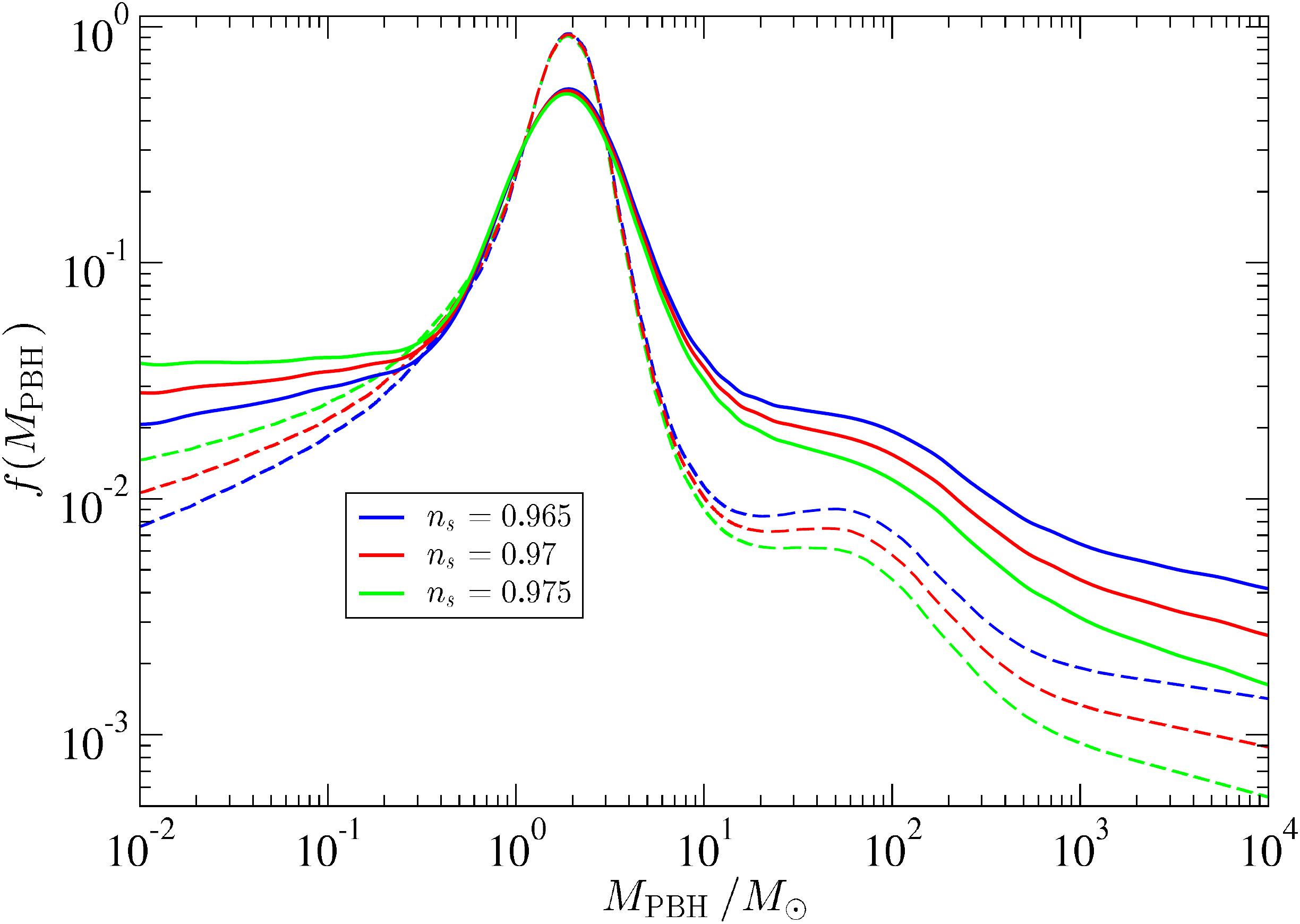}
\caption{Left panel: Mass functions using Eq.~\ref{eq:mass_function} and the numerical results of $\delta_{\rm c,QCD}$ for different profiles of Eq.~\ref{eq:basis_pol} and $n_{\rm s}=0.97$. Right panel: Comparison between the mass function used in~\cite{Bagui:2021dqi,Carr:2019kxo} (dashed lines) with the one obtained in this work (solid lines), using the numerical results of $\delta_{\rm c,QCD}$ for three different $n_{\rm s}$ values. The profile used is a Gaussian profile. In both panels we fix $f_{\rm PBH}=1$.}
\label{fig:mass_function}
\end{figure}

We have computed the PBH mass distribution $f(M_{\rm PBH})$ following Eq.~\ref{eq:mass_function} for the profiles of Eq.~\ref{eq:profile_exp} using $\delta_{\rm c,QCD}$. Our results are shown in the left panel of Fig.~\ref{fig:mass_function}. We see that the dominant peak of the mass function is around $M_{\rm PBH} \approx [1-3]\Msun$, depending on the value of $q$.  When $q$ increases, we observe that the peak of the PBH mass function is moved to higher values of $M_{\rm PBH}$. As expected, we also observe a lump in the region $M_{\rm PBH} \approx [30-100] \Msun$ whose importance also depends on the considered $q$ value. 
Interestingly, for sharper profiles, the relative amplitude between the main peak and the lump is reduced, which will have an impact on the merger rates (see Secs.~\ref{sec:merger_rates} and~\ref{sec:catalog}).
We have also compared the mass functions obtained with $\delta_{\rm c, \rm QCD}$ (solid line) to the ones obtained in~\cite{Bagui:2021dqi,Carr:2019kxo} (see right panel of Fig.~\ref{fig:mass_function}), in which $\delta_{{\rm c}, w={\rm cte}}$ is obtained from the numerical results of~\cite{musco2013}, where the threshold was computed for a Gaussian profile with a constant $w$.
There is a substantial difference in the mass functions that is directly coming from the differences in between $\delta_{\rm c, \rm QCD}$ and $\delta_{{\rm c},w={\rm cte}}$.  As expected, the small variations in the threshold impact importantly the resulting PBH mass distribution.  First, one observes that the dominant peak is approximately two times smaller and significantly broader when the time-varying equation of state is taken into account in the simulations.  Also, the lump appears to be less pronounced, which has thus a direct impact for predictions in the LIGO/Virgo range.  
 
 Our results therefore show the importance of using threshold values computed from simulations of PBH formation that fully take into account the variations of the equation of state at the QCD epoch as well as the curvature profile.  {Beyond the models considered here, they could be also used to compute more accurately the PBH mass distributions in models including exotic physics, e.g. lepton flavor violation~\cite{Bodeker:2020stj} and solitosynthesis~\cite{Garcia-Bellido:2021zgu}.}
 
\subsection{PBH mass distribution with a scaling law}\label{sec:broad_pbh}

\begin{figure}[t]
\centering
\includegraphics[width=4. in]{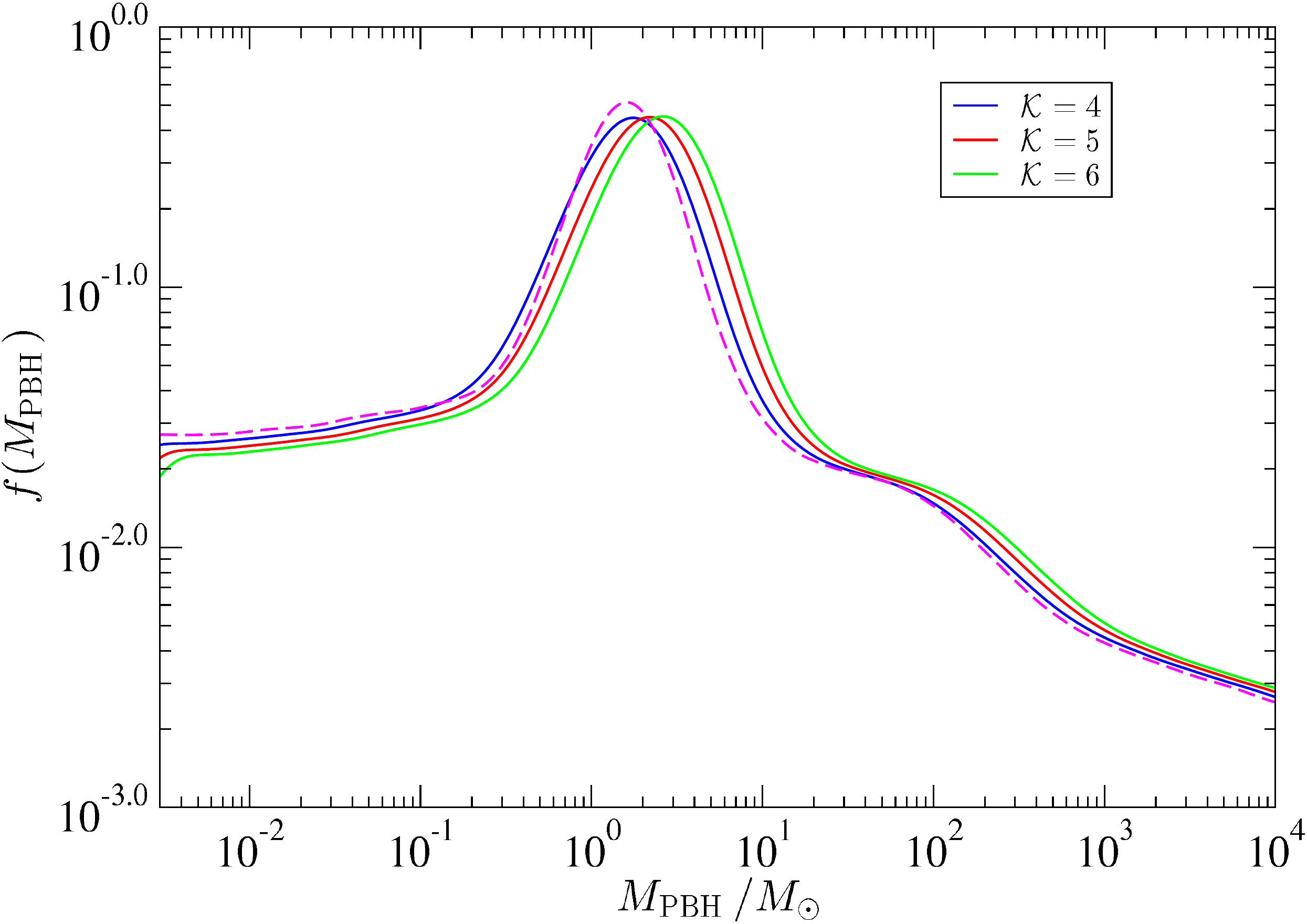}
\caption{Comparison between the PBH mass function obtained with Eq.~\ref{eq:mass_function} (magenta dashed line) with $\alpha = 0.8$ and with Eq.~\ref{eq:integral_fpbh} (solid lines) based on the PBH mass scaling law. The curvature profile used in both cases is a Gaussian profile ($q=1$) with $n_{\rm s}=0.97$.}
\label{fig:mass_function_comparison}
\end{figure}

Before concluding this Section, we make a comparison between the previous effective approach and the PBH mass distribution that can be obtained assuming the scaling law regime of Eq.~\ref{eq:2_scaling}.  For this purpose, we follow the same method as in~\cite{Byrnes:2018clq}.   One can first compute the PBH abundance at formation in this regime,
\begin{equation}
\beta(M_{\rm PBH}) = 2 \int_{\delta_{\rm c}}^{\delta_{\rm max}} \frac{M_{\rm PBH}}{M_{\rm H}}P(\delta_{\rm m}) {\rm d} \delta_{\rm m} = 2 \int_{\delta_{\rm c}}^{\delta_{\rm max}} \mathcal{K} (\delta_{\rm m} - \delta_{\rm c})^{\gamma}P(\delta_{\rm m}){\rm d} \delta_{\rm m} ,
\label{eq:betaextended}
\end{equation}
where $\delta_{\rm max}$ is the maximum allowed value of $\delta_{\rm m}$ (in case of $w=1/3$,  $\delta_{\rm max}=2/3$).  Then  one can make a change of variables from $\delta_{\rm m}$ to $M_{\rm H}$ using Eq.~\ref{eq:2_scaling}.
In particular, we use:
\begin{align}
\label{2_expansion_fpbh}
\tau &= \frac{M_{\rm PBH}}{\mathcal{K} M_{\rm H}},\nonumber\\
\delta_{\rm m}  &=  \tau^{1/\gamma}+\delta_{\rm c}(M_{\rm H}),\nonumber\\ 
\frac{{\rm d} \delta_{\rm m}}{{\rm d}\ln M_{\rm H}} &=-\frac{1}{\gamma}\tau^{1/\gamma}   .
\end{align}
The final result for the PBH mass distribution today reads
\begin{align}
  f(M_{\rm PBH}) &= \frac{1}{\Omega_{\rm CDM} f_{\rm PBH}} \int_{-\infty}^{+\infty}\frac{2}{\sqrt{2 \pi \delta_{\rm rms}^{2}(M_{\rm H})}} 
  e^{-\frac{(\tau^{1/\gamma}+\delta_{\rm c}(M_{\rm H}))^{2}}{2 \delta_{\rm rms}^{2}(M_{\rm H})}}  \frac{M_{\rm PBH}}{\gamma M_{\rm H}} \tau^{\frac 1 \gamma}\sqrt{\frac{M_{\rm eq}}{M_{\rm H}}} {\rm d} \ln M_{\rm H} \label{eq:integral_fpbh},
\end{align}
where we have assumed that $\mathcal{K}$ is a constant, so we ignore the possible weak dependence on the horizon mass discussed in Sec.~\ref{sec:pbhmass} (the same applies for $\gamma$). This integral is solved numerically. 
We have then compared the mass distributions obtained with Eqs.~\ref{eq:integral_fpbh} and~\ref{eq:mass_function} in order to check the validity of approximating $M_{\rm PBH} = \alpha M_{\rm H}$. 
The results are shown in Fig.~\ref{fig:mass_function_comparison} for the specific case $n_{\rm s} = 0.97$, $\gamma = 0.34$, $\alpha=0.8$ and three different values of $\mathcal{K}$.  The two approaches lead to very similar mass functions, with only small differences: i) the main peak in $f(M_{\rm PBH})$ is slightly broader when a broad PBH mass in considered; ii) the peak is moved to larger values of $M_{\rm PBH}$ when $\mathcal{K}$ increases, which can be fitted well with slightly different values of $\alpha $ in the other approach, and basically only moves the PBH mass distribution from left to right. 
For the purposes of the estimation of the merger rates and comparison with the catalog GWTC-3, in our work it is enough to consider the approach of Sec.~\ref{mass_function3}.  We obtain the same conclusions for different profiles and spectral indices. 

\section{PBH merger rates} \label{sec:merger_rates}

Using the PBH mass distributions obtained in Sec.~\ref{mass_function3}, one can estimate the PBH merger rates and compare them to the rate limits and measurements from the LIGO/Virgo GW observations. We consider the two PBH binary formation channels that have been already analysed in the context of broad PBH mass distributions with QCD effects~\cite{Carr:2019kxo, Clesse:2020ghq}.  The first channel is the late-time 2-body binary capture of PBHs in dense clusters from close encounters, and the second channel is binary formation in the early Universe, before matter-radiation equality.  In each scenario, the merger rates have a different mass and time dependence. 
We did not consider the rates for three-body captures in PBH clusters~\cite{Franciolini:2022ewd}, which could either be dominant or subdominant depending on assumptions and parameters, so this remains an interesting binary formation mechanism to explore in a future work.

\subsection{Late PBH binaries in clusters}

For late PBHs binaries in clusters, the merger rate distribution per unit logarithmic mass can be described by~\cite{Clesse:2020ghq,Clesse:2016vqa,Bird:2016dcv}
\bea 
    \frac{\dd\tau_{\rm clust}}{\dd \ln m_1 \dd \ln m_2 } & = & R_{\rm clust} \times f(m_1)  f(m_2) f_{\rm PBH}^2 \times \frac{(m_1 + m_2)^{10/7}}{(m_1 m_2)^{5/7}} \rm{yr^{-1}Gpc^{-3}}.
     \label{eq:ratescatpure2}
\eea
The PBH clustering properties and velocity distribution are encoded in the scaling factor $\Rcl$.  Neglecting the additional clustering due to Poisson fluctuations in the PBH spatial distribution typically leads to $\Rcl \sim 1-10$, as assumed in~\cite{Bird:2016dcv}.  However, for stellar-mass PBHs that significantly contribute to the dark matter, one expects that clusters form from Poisson fluctuations at high redshifts.  This would explain the observed spatial correlations between the source-subtracted infrared and X-ray background radiations~\cite{Kashlinsky:2016sdv}.  Today, after the dynamical heating and dilution of the low-mass sub-clusters inside host clusters, they would have masses around $10^6-10^8 M_\odot$~\cite{Clesse:2020ghq} that could be identified to ultra-faint dwarf galaxies.  For such Poisson clusters, one gets possible values of $R_{\rm clust}$ between $100$ and $1000$ if $f_{\rm PBH} \approx 1$.  In particular, a value of about $R_{\rm clust} \approx 460$ was found to explain well the merger rates associated to the observations of GW190425, GW190814 and GW190521~\cite{Clesse:2020ghq}, using a PBH mass distribution coming from simulations with a constant $w$ during the PBH formation. Here we use this value as a benchmark, but one should keep in mind that it can vary depending on the details of the PBH clustering and mass distribution, see e.g.~\cite{Trashorras:2020mwn,DeLuca:2022uvz}.  Our normalized PBH mass function $f(m)$ depends on the shape of the spectrum of primordial fluctuations through the small-scale scalar spectral index $n_{\rm s}$, on the parameter $q$ related to the initial density profile, but is relatively insensitive to the total fraction of PBHs making up the DM, $f_{\rm PBH}$.   

\subsection{Early PBH binaries}

For early PBH binaries, the merger rate distribution is given by~\cite{Raidal:2018bbj}
\bea
\frac{\dd\tau_{\rm prim}}{\dd \ln m_1 \dd \ln m_2 } &=& \frac{3.2 \times 10^6}{\rm Gpc^3 yr} \times f_{\rm sup}(m_1,m_2,z) f_{\rm PBH}^{\frac{53}{37}} f(m_1)  f(m_2) \left[\frac{t(z)}{t_0}\right]^{-\frac{34}{37}}  \nonumber \\
        & \times & \left(\frac{m_1 + m_2}{M_\odot}\right)^{-\frac{32}{37}}  \left[\frac{m_1 m_2}{(m_1+m_2)^2}\right]^{-\frac{34}{37}}, \label{eq:cosmomerg}
\eea
where $t(z) = \int_{z}^{\infty} 1/(H(x)(1+x)) {\rm d}x$ is the age of the Universe at redshift $z$, $H(x)$ is the Hubble parameter, $t_0 = t(z=0)$ is the age of the Universe today.  Compared to~\cite{Raidal:2018bbj} a factor two is added because we consider the merger rate of \textit{mass ordered} binaries assuming $m_1 \ge m_2$\footnote{The authors thank Gabriele Franciolini for pointing out the necessity of this factor two in the merger rates.}.   $f_{\rm sup}$~\cite{Hutsi:2020sol,Raidal:2018bbj} is the suppression factor that was introduced since N-body simulations of PBH binary evolution in the early Universe showed that close PBHs, matter inhomogeneities as well as early Poisson-induced clustering may affect these binaries~\cite{Raidal:2018bbj,Trashorras:2020mwn}, inducing a rate suppression. Here we consider the assumptions on $f_{\rm sup}$ to be the same as in~\cite{Bagui:2021dqi}, which corresponds to the limit of a peaked mass distribution.  We make this choice in order to avoid an unphysical  dependence on the low-mass PBH distribution.  In this limit, for $0.1<f_{\rm PBH}<1$, a useful approximation can be used, $f_{\rm sup} \approx 2.3 \times 10^{-3} f_{\rm PBH}^{-0.65} $.

We detail this calculation, motivate this choice and comment on the possible ambiguities related to the exact definition of the mass distribution in Appendix A.

\subsection{Merger rate distributions and comparison with LIGO/Virgo observations}

\begin{figure}[t]
\centering
\includegraphics[width=6. in]{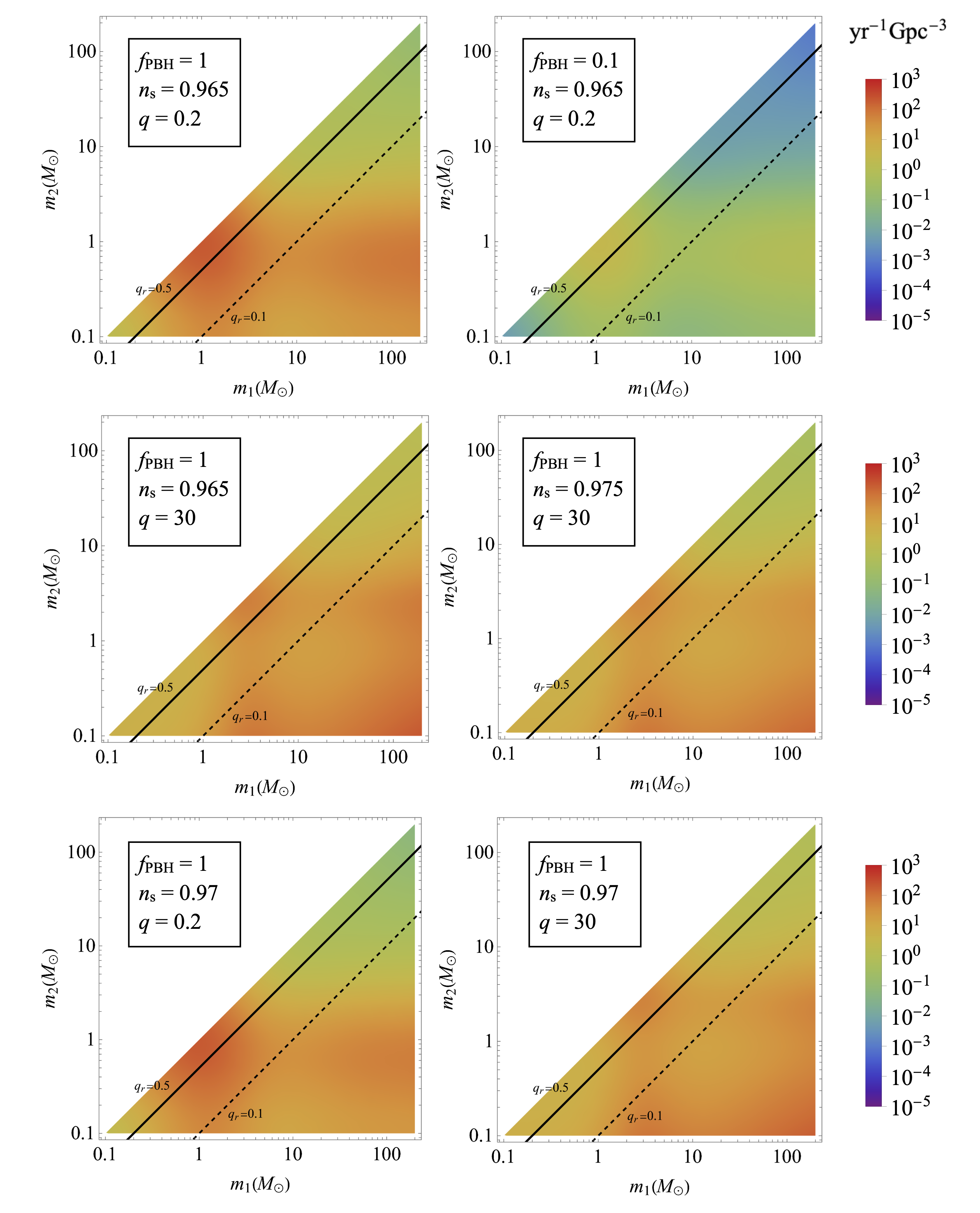}
\caption{Merger rate distribution of late-time PBH binaries as a function of the PBH masses $m_1$ and $m_2$ in solar masses, for two different values of $f_{\rm PBH}= (1, 0.1) $, $n_{\rm s}= (0.965, 0.975) $ and the profile parameter $q = (0.2, 30)$. The solid and dashed black lines correspond to mass ratios of $q_{r} = m_2/m_1 = 0.5$ and $0.1$, respectively. The coloured sidebar gives the rate distribution in units of yr$^{-1}$Gpc$^{-3}$.}
\label{fig:MergratesLB}
\end{figure}

\begin{figure}[t]
\centering
\includegraphics[width=6. in]{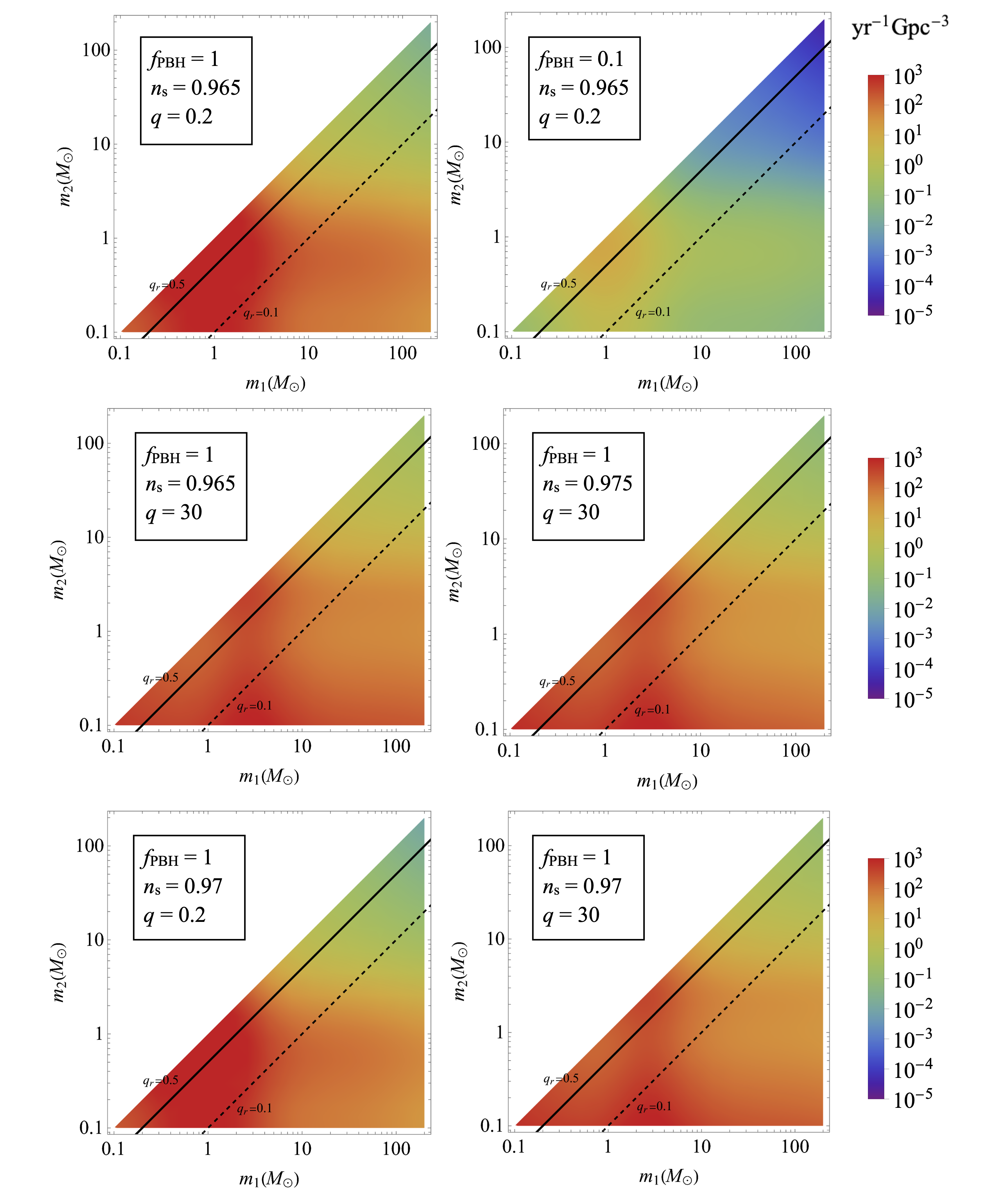} 
\caption{Merger rate distribution of early PBH binaries as a function of the PBH masses $m_1$ and $m_2$ in solar masses, for two different values of $f_{\rm PBH} = (1, 0.1)$ and for $n_{\rm s} = (0.965, 0.975)$, and $q = (0.6, 20)$.  We assume $z=0$ in Eq.~\ref{eq:cosmomerg}.  The solid and dashed black lines correspond to mass ratios of $q_{r} = m_2/m_1 = 0.5$ and $0.1$, respectively. The coloured sidebar gives the rate distribution in units of yr$^{-1}$Gpc$^{-3}$. \vspace{5mm}}
\label{fig:MergratesEB}
\end{figure}

The two-dimensional merger rate distributions as a function of the PBH component masses $m_1$ and $m_2$ are shown in Figs.~\ref{fig:MergratesLB} and~\ref{fig:MergratesEB} for different values of $n_{\rm s}$ and $f_{\rm PBH}$, and for the two extreme values of $q$ we considered, assuming $\alpha = 0.8$. Rate distributions for other values of $q$ typically interpolate between these extremes.  For early binaries, we show the rates at $z=0$.

In the solar-mass range, the rates are dominated by early binaries.  In comparison with the previous analysis of~\cite{Clesse:2020ghq}, these are significantly lower due to the broader and lower peak in the PBH mass distribution.  One can also notice that the region in the $(m_1,m_2)$ plane where the rates are the highest is also more extended.  
Above $10M_\odot$ the PBH merger rates from late binaries become comparable to the ones of early binaries and progressively become higher at larger masses, for mergers that would lie in the pair-instability mass gap, like GW190521. We also observe that, for both early and late binaries, systems composed of a PBH from the peak and a PBH from the lump, i.e. with low mass ratios of order $m_2/m_1 \approx 0.1$, can have comparable merger rates than equal-mass mergers around $30 M_\odot$.  This feature is very different from most predictions of astrophysical scenarios where the rates are lower for lower mass ratios.  The models can therefore explain the observation of binaries with low-mass ratios like GW190814, as already pointed out in~\cite{Clesse:2020ghq}.  
Due to the broader mass distributions, the rates are also typically higher in the sub-solar mass range, but still consistent with the latest rate limits from the searches of subsolar black holes in LIGO/Virgo data~\cite{LIGOScientific:2021job,Nitz:2022ltl}. Nevertheless, some scenarios lead to predictions that are within the range of the next observing runs. 


As $q$ increases from $0.2$ to $30$, one observes that the different features in the rate distributions are typically less pronounced.  The rates of the late PBH binaries (see Fig.~\ref{fig:MergratesLB}) are reduced by at most a factor 10 for PBHs binaries from the QCD peak and with low mass ratios, going from $\sim 100 $ to $30$ yr$^{-1}$Gpc$^{-3}$. This is due to the lower value of $f(m_2) $ in the main peak of the PBH mass function, despite the fact that the lump is higher at larger masses, giving higher values of $f(m_1)$. 
This also explain why for $q = 30$ the merger rates are lower at the QCD peak, compared to $q = 0.2$ (see  Fig.~\ref{fig:mass_function}). 

Furthermore, lowering $f_{\rm PBH}$ by a factor 10 (from 1 to 0.1) reduces the rate of the late PBH binaries, since a smaller PBH abundance implies less frequent PBH mergers.  This reduction is also important for the early PBH binary merger rates, due to the product of $f_{\rm PBH}^{53/37}$ with the suppression factor that has the following specific dependence on $f_{\rm PBH}$, $f_{\rm sup} \propto f_{\rm PBH}^{-0.65}$~\cite{Raidal:2018bbj,Hutsi:2020sol}.  The change in $f_{\rm PBH}$ also induces substantial differences for subsolar-mass (between $10^{-1}$ and $1 M_\odot$) late and early PBHs binaries, where the rate decreases from $\sim 13$ to $6 \times 10^{-2}$ yr$^{-1}$Gpc$^{-3}$ for the late binaries, and from $\sim 100$ to $4 \times 10^{-1}$ yr$^{-1}$Gpc$^{-3}$ for the early binaries.  Such values are compatible with the rate limits obtained from the first half of the third observing run of LIGO/Virgo~\cite{LIGOScientific:2021job}.

We also explore the effect of the spectral index.  When it varies from $n_{\rm s} = 0.965$ to $n_{\rm s} = 0.975$ while $q$ and $f_{\rm PBH}$ remain constant, the merger rates for late binaries increase by a factor 1-2 for subsolar-mass binaries, the rates remain almost unchanged at $\mathcal{O}(1) M_\odot$, and they decrease for higher masses. The same behaviour is observed for early binaries, but in the two plots the changes are very slight. However, these changes are expected by looking at the mass function $f(M)$ for different $n_{\rm s}$ values (see Fig.~\ref{fig:mass_function}). 

\begin{figure}[t]
\centering
\includegraphics[width=3.0 in]{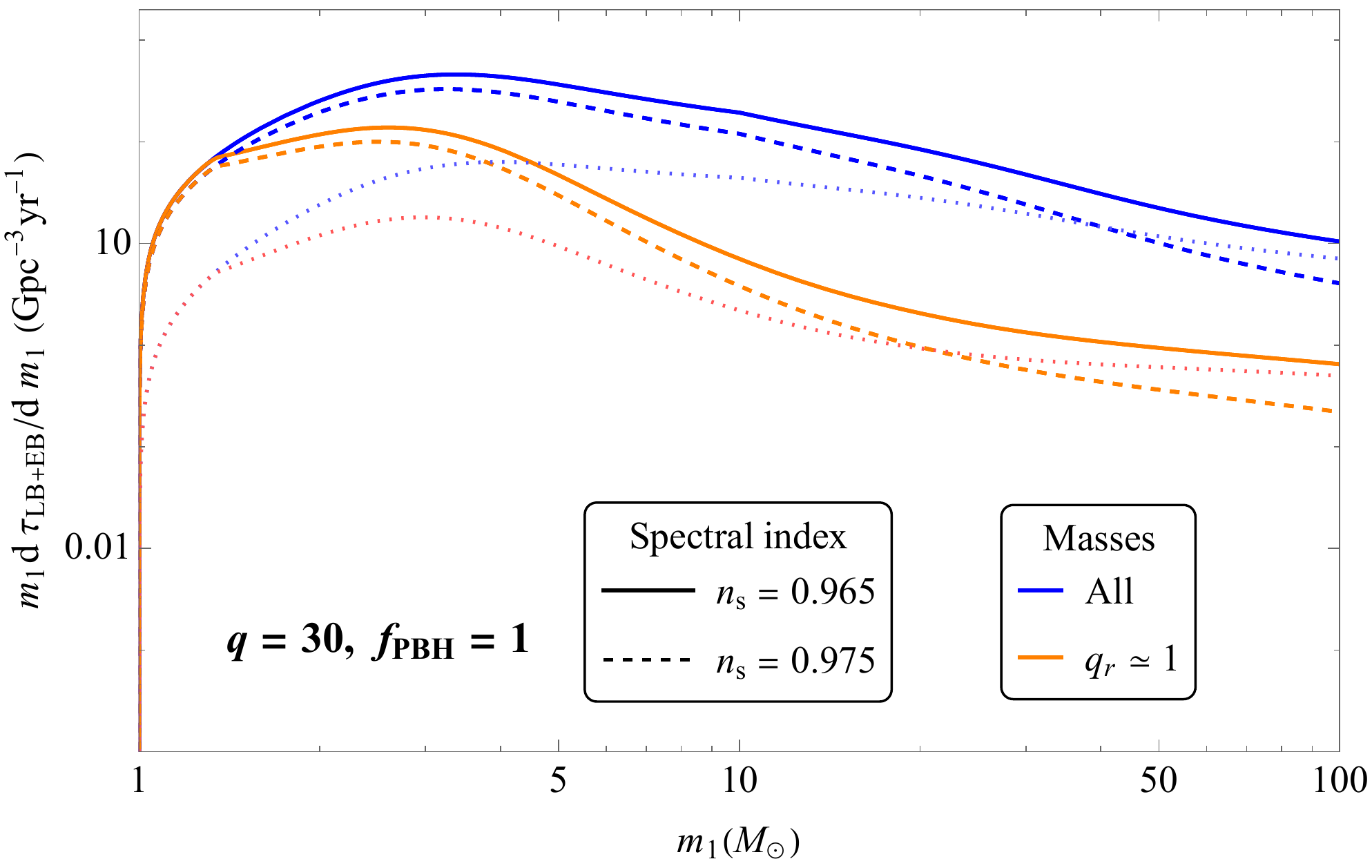}
\includegraphics[width=3.0 in]{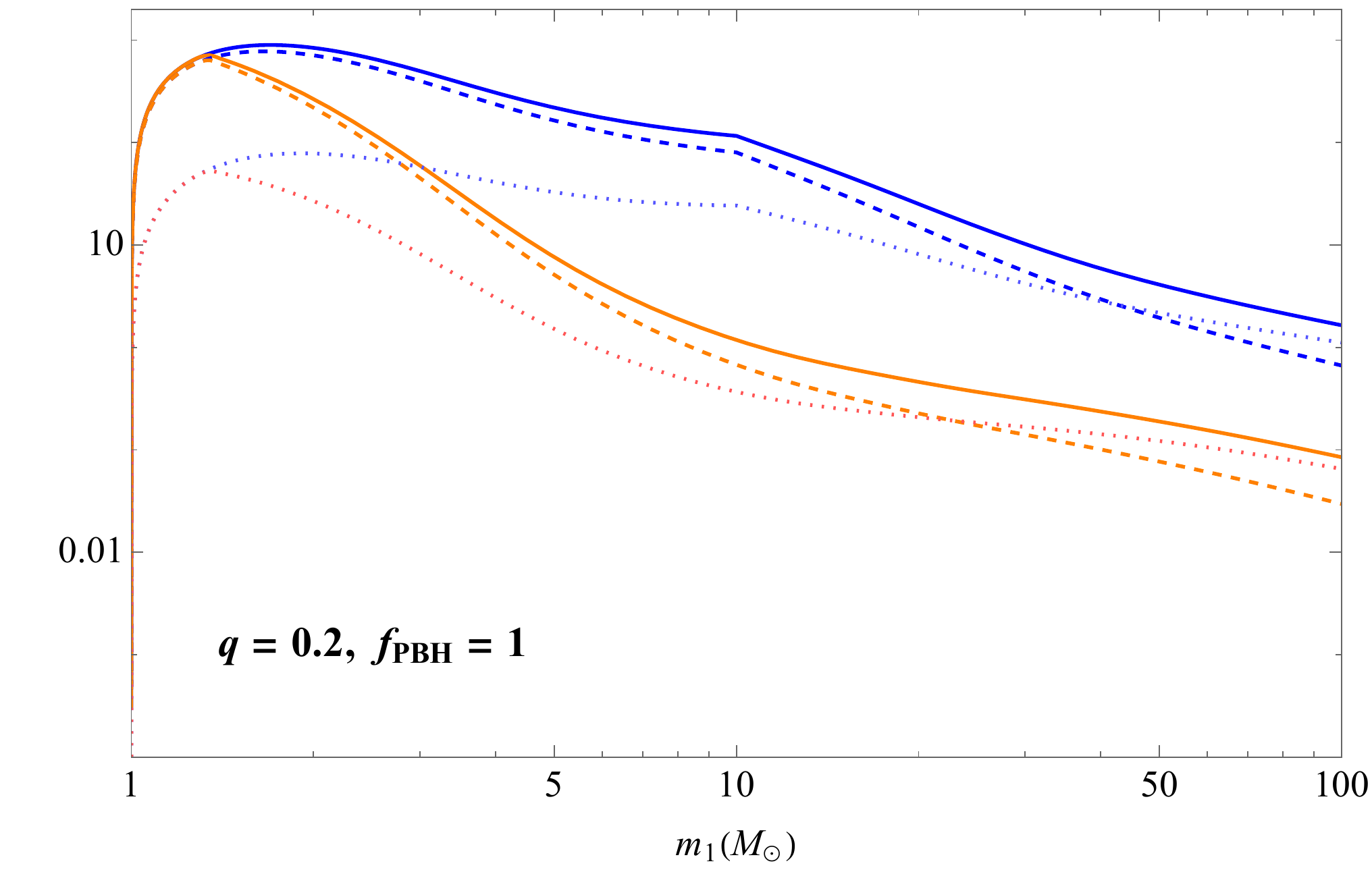}
\includegraphics[width=3.0 in]{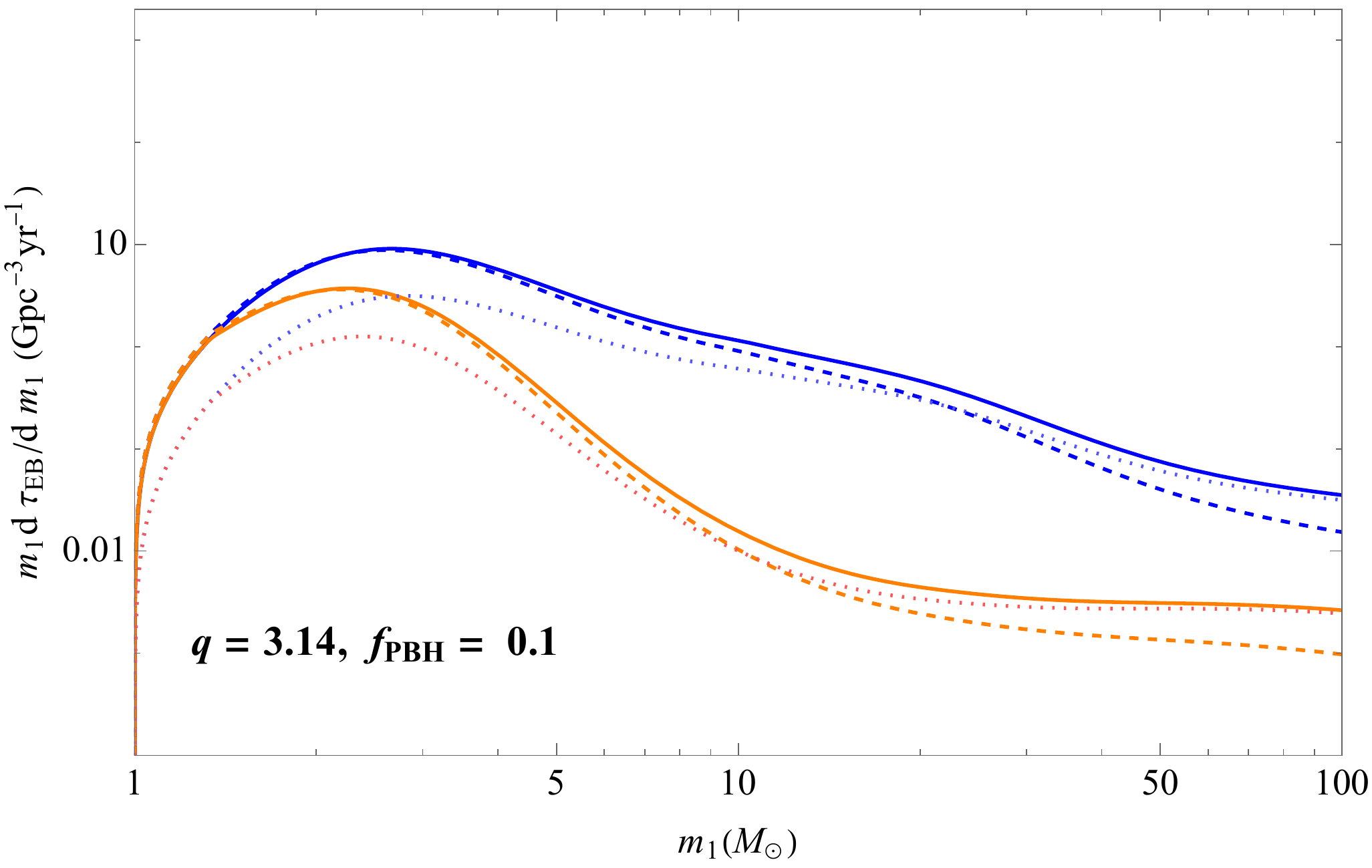}
\includegraphics[width=3.0 in]{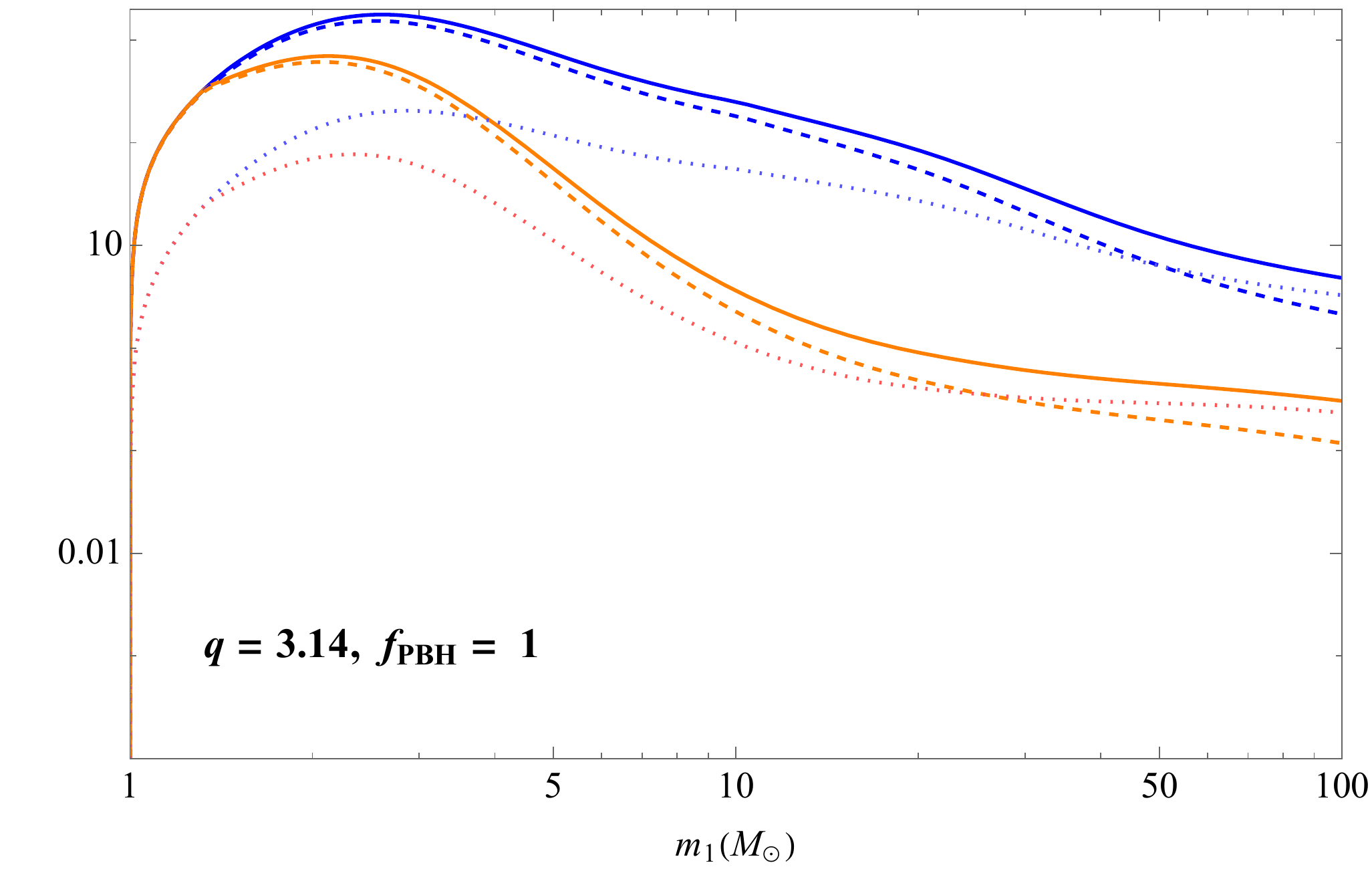}
\caption{Merger rate distribution of the early and late PBH binaries combined as a function of the primary mass $m_1$, for all binary masses (blue curves) and for binaries with mass ratio $q_r \simeq 1$ (orange curves). The plots show three different values of the $q$ parameter, $q = 30$, $0.2$ and $3.14$, and the bottom panels correspond to two different $f_{\rm PBH}$ values. The solid and dashed lines correspond to the spectral indices $n_{\rm s} = 0.965$ and $n_{\rm s} = 0.975$, respectively. The dotted lines indicate the contribution of the late PBH binaries (with $n_{\rm s} = 0.965$) to the total rate distribution for all binary masses (dotted blue), and for mass ratios $q_r \simeq 1$ (dotted orange).}
\label{fig:Mergdistribution}
\end{figure}

Finally, we integrate the two-dimensional merger rate distributions, given by Eqs.~\ref{eq:ratescatpure2} and~\ref{eq:cosmomerg}, over the secondary mass $m_2$.  Then we sum the two contributions to get the quantity ${\rm d}\tau_{\rm LB+EB}/{\rm d} \ln m_1$ as a function of the primary mass $m_1$, in the range 1 to 100$M_\odot$.  Our results are shown in Fig.~\ref{fig:Mergdistribution}. The blue and orange curves respectively represent the rates where we take into account all binary masses down to $m_2/m_1 = 0.1$, and where we limit ourselves to binaries with mass ratios $q_r \approx 1$.  In other words, we integrate the rates over $m_2$ within the range $m_2 =$ max$(1, m_1/10)$ to $m_1$ for the first case, and $m_2 =$ max$(1, 3m_1/4)$ to $m_1$ for the second case, such that we do not integrate over the subsolar region that is not included in the standard searches of LIGO/Virgo.  
These bounds also imply that at around $1 M_\odot$ the rates drop significantly. The plots in Fig.~\ref{fig:Mergdistribution} were obtained by using the PBH mass distributions of Sec.~\ref{mass_function3} with $q = 0.2$ and $30$ (two upper panels of Fig.~\ref{fig:Mergdistribution}). As an example, we also plotted the merger rates for $q=3.14$, which would be the most realistic value derived in~\cite{Musco:2020jjb} for a broad primordial power spectrum of Gaussian  curvature fluctuations~\footnote{There exist small differences of the threshold of order $\mathcal{O}(2-3\%)$, when one considers the effects of the different profiles with $q  \approx 3.14$ shown in Fig.~\ref{fig:thresholds_qcd_comparison}.}.  This value of $q$ corresponds to $\delta_{\rm c} \approx 0.56$~\cite{universal1,Musco:2020jjb}. 

Getting one and two-dimensional rate distributions helps to compare PBH merger rates to estimations or limits from LIGO/Virgo.   For instance, one can compare the results of Figs.~\ref{fig:MergratesLB},~\ref{fig:MergratesEB} and~\ref{fig:Mergdistribution} to the results of the population analysis of~\cite{LIGOScientific:2021psn} for the non-parametric binned Gaussian processed mass model (BGP), see their Figs. 3 and 4.  One should nevertheless remain cautious when doing such a comparison, because assumptions are similar but not identical.  For instance, the BGP model assumes that merger rates over neighboring mass
bins are correlated via a Gaussian process, which can be inconsistent with the strong variations seen in our models in some regions. Moreover, the rate estimations for mass ratios $q_r \approx 1$ are performed in the diagonal mass bins of the $(m_1,m_2)$ plane, whereas we impose a limit on the mass ratio.   Finally, one should remember that both early and late PBH binaries have uncertainties in their merger rates even if the mass dependence usually remains universal, which means that their respective rates could easily be rescaled by order one factors.  As a result, they could compete differently on different mass scales.

We notice in Figs.~\ref{fig:MergratesLB} and~\ref{fig:MergratesEB} that our merger rates reach their highest value ($ \sim 10^3$ yr$^{-1}$Gpc$^{-3}$) at around $1 - 2 M_\odot$, corresponding to the peak in the mass function at the QCD crossover, and there is also an important reduction of the rates at larger masses. These features are also expected from observations of the GWTC-3 catalog (see Fig. 3 of~\cite{LIGOScientific:2021psn}), with rates up to about $10^3$ yr$^{-1}$Gpc$^{-3}$. Moreover, at masses $\sim 100M_\odot$, the rates drop to $10^{-2}$ yr$^{-1}$Gpc$^{-3}$, of same order than what is observed by LIGO/Virgo. We also remark that in the GWTC-3 catalog the merger rates peak again at around $10M_\odot$, which is not observed in Fig.~\ref{fig:MergratesLB}, where they peak at around $80-100M_\odot$ instead, nor in Fig.~\ref{fig:MergratesEB}, where the merger rates for early PBH binaries do not exhibit any other peak. We discuss the implications of this discrepancy in the next Section.

In Fig.~\ref{fig:Mergdistribution}, one sees that the merger rates in the range $1 - 2M_\odot$ can reach $\mathcal{O}(100-1000)$ yr$^{-1}$Gpc$^{-3}$, consistent with the observations of LIGO/Virgo's third observing run (see Fig. 4 of~\cite{LIGOScientific:2021psn}). In the range $2 - 5 M_\odot$ the most likely merger rate observed by LIGO/Virgo drops to about 1 yr$^{-1}$Gpc$^{-3}$ if one includes all secondary masses, and to $10^{-2}$ yr$^{-1}$Gpc$^{-3}$ for $q_r \approx 1$. These features are not observed in the PBH distributions of Fig.~\ref{fig:Mergdistribution}.  Nevertheless, the rates are decreasing when $m_1$ increases, and given the large uncertainties on rate measurements and the fact that one can play with the parameters $f_{\rm PBH}$ and $\alpha$ to move the rate distribution up and down, or left and right, adequate parameter choices can make the model consistent with observations, especially for low values of $q$.  
Furthermore, at $m_1 \approx 10 M_\odot$, the most likely merger rates inferred from observations increase again and reach a maximum of about $60$ yr$^{-1}$Gpc$^{-3}$.  We also observe a small peak at $10 M_\odot$ for PBHs with $q=0.2$, coming from asymmetric binaries that dominate the rates at this mass.
Going to masses around $30M_\odot$, our rates can be comparable (within error bars) with observations for all the considered values of $q$ and $f_{\rm PBH} \approx 1$. The small observed negative slope can also be accommodated for low values of $q$ ($0.2$ and $3.14$), which seems to be therefore favored by observations.  It is essentially driven by the rate of early binaries (despite the fact that they are subdominant at these masses) and the value of the spectral index.
Finally, as we go to even higher masses close to $100M_\odot$, our results again can coincide with the ones inferred from GWTC-3 for low $q$ values and $f_{\rm PBH} \approx 1$.

Overall, when making a comparison with the rates inferred from the GWTC-3 catalog, one observes that PBHs with a mass distribution imprinted by the QCD epoch lead to rate distributions compatible with observations above $10 M_\odot$ for realistic values of $q$, including in the pair-instability mass gap. PBHs can also explain the rates associated to mergers with low mass ratios like GW190814,  mergers involving compact objects around $2.5 M_\odot$, and it is also possible to accommodate the observed merger rates around $1-3 M_\odot$ even for $f_{\rm PBH} = 1$, both for early and late time binaries. 
It seems that the model that is in best agreement with the observations is $q=0.2$ with $f_{\rm PBH}=1$, even if it is not necessarily the most statistically significant one producing such PBHs, since $q=3.14$ is the most realistic value, according to~\cite{Musco:2020jjb} and taking into account the considerations in Sec.~\ref{mass_function3}. Also, one typically gets too many mergers of early binaries between $3$ and $8 M_\odot$, covering the low mass gap. Possibly, these rates could be lower than expected, for instance if the suppression factor has a specific mass dependence for this type of mass distribution that has not been tested with N-body simulations, or if the rate suppression is more important than expected due to late behaviour of Poisson fluctuations that is not accounted for in the calculations. Nonetheless, the rates would remain almost unchanged at large masses, where they are dominated by late binaries. It is therefore difficult to say if the detection of mergers in this low mass gap favors or disfavors the PBH hypothesis compared to realistic astrophysical models having such a mass gap. Finally, since the early binaries dominate the rates at small masses and the late binaries dominate at large masses, our results could fit the LIGO/Virgo observations by carefully adjusting the parameters $R_{\rm clust}, q$, $\alpha$ and $f_{\rm PBH}$.

In order to better explore the predictions of PBHs, we investigate in the next Section how the rate distributions impact the expected distributions of GW events in the $ (m_1,m_2)$ plane, after taking into account the mass-dependent detector sensitivity.  This way, one can directly compare PBH models to GW observations from the GWTC-3 catalog, and pave the way towards a full Bayesian analysis of PBH models with thermal features. 

\section{Comparison with detections of the GWTC-3 catalog}\label{sec:catalog}

The fact that the main peak in the PBH mass distribution is lower and broader than when one assumes a constant $w$ in the computation of the over-density threshold, and that the lump is less pronounced and slightly shifted towards larger masses, has important implications for the expected distribution of actual detections of black hole mergers.  

We have computed the expected distribution of detections per unit of logarithmic mass in the third observing run of LIGO/Virgo, by using the public set of (mixed) injections~\cite{injections} and the associated code to compute the volume-time sensitivity $\langle VT \rangle $.  For this purpose, we have used $40 \times 40$ uniformly spaced mass bins in $\ln m_1$ and $\ln m_2$ between $1 M_\odot$ and $100 M_\odot$.  Practically, the $VT$ is computed at the center of each bin for a sharp log-normal mass function of width $\sigma = 0.1$.  Then we interpolate these values to get the volume-time sensitivity as a function of $m_1$ and $m_2$.  

\begin{figure}[t]
\centering
\includegraphics[width=5. in]{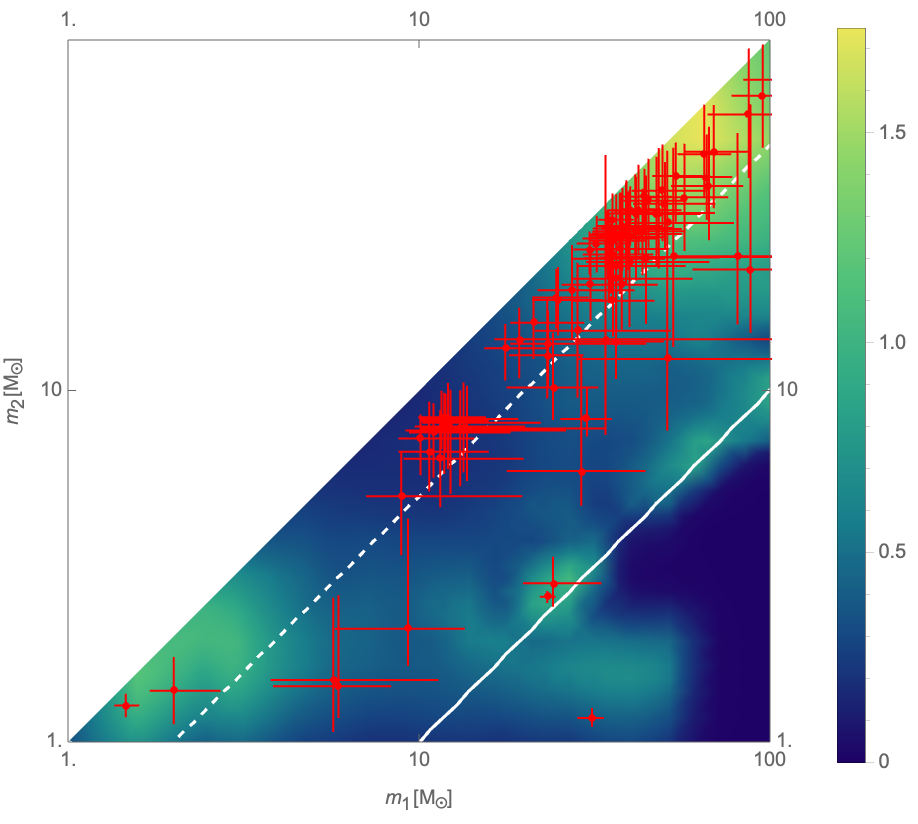}
\caption{Normalized distribution of GW observations expected for the combination of early and late binaries, for the third observing run of LIGO/Virgo,  with $\alpha = 0.7$, $n_{\rm s}=0.975$ and $q=3.14$.  The observed merger events of the GWTC-3 catalog with uncertainties in $m_1$ and $m_2$ are shown in red. }
\label{fig:Ndist}
\end{figure}

In order to get the distribution of detections expected for PBHs, one has to multiply $\langle VT \rangle $ by the theoretical merger rates of both early and late-type binaries from the previous Section,
\begin{equation}
    \left\langle \frac{{\rm d}N}{{\rm d}\ln m_1 {\rm d}\ln m_2} \right\rangle = \langle VT \rangle \times \left(  \frac{{\rm d} \tau_{\rm clust}}{{\rm d} \ln m_1 {\rm d} \ln m_2} +  \frac{{\rm d} \tau_{\rm prim}}{{\rm d} \ln m_1 {\rm d} \ln m_2}\right).
\end{equation}
Interestingly, when we combine the merger rate from early and late binaries using $q=3.14$ and $\alpha = 0.7$, we find that the distribution of the number of events have local maxima in the following regions, assuming a spectral index around $n_{\rm s} = 0.975$.  The normalized distribution obtained in this case is shown in Fig.~\ref{fig:Ndist}  

The region with the highest probability lies between $30 M_\odot$ and $100 M_\odot$ with a distribution that smoothly decreases but is still significant up to $100 M_\odot$.  This could provide an explanation to main population of LIGO/Virgo mergers as well as mergers involving a black hole in the pair-instability mass gap~\cite{Clesse:2020ghq}.  One should note, nevertheless, that this region is slightly shifted to larger masses compared to the observed distribution in the GWTC-3 catalog.  We find that this is a generic feature of all the produced mass distributions, whatever is the curvature profile.  This observation is consistent with the rate distribution as a function of $m_1$ of Fig.~\ref{fig:Mergdistribution} that remains relatively high up to $m_1 \approx 100 M_\odot$.  
But larger values of the spectral index, models including a running of the spectral index typically tend to suppress the merger rates when $m_1$ increases.  This typically moves the region with the highest probability of detection towards smaller masses, increasing the consistency with observations.  These could therefore be favored in a Bayesian model analysis.  

There is another local maximum at the solar-mass scale, where GW190425 was detected, mainly due to mergers of early PBH binaires.   We find that prevalence of this region can change depending on the exact values of $n_{\rm s}$ and $q$, but also of the rate parameters $R_{\rm clust}$ and $f_{\rm sup}$, and of the total PBH abundance $f_{\rm PBH}$.   It is therefore possible to adequately adjust the model parameters  in order optimize the agreement with observations, which should be done with a Makov-Chain-Monte-Carlo analysis in a future work. 

A third local maximum is obtained around $m_1 \approx 20-30 M_\odot$ and $m_2 \approx 2-3 M_\odot$, which corresponds to mergers with a PBH in the QCD peak and a PBH in the lump.  One should note that the probability of detection in this region is not only boosted by the merger rates, but also by a local maximum of the $\langle VT \rangle$.   Two events are observed in this region, including GW190814.  We also notice that this region is smoothly connected to the other two regions, where a few detections have also been made.  

Until now, the situation is rather positive for the PBH hypothesis, able to explain both the dominant population of black hole mergers as well as more exceptional events in the mass gaps or with low mass ratios, when the effects of the QCD cross-over transition are taken into account. However, as also shown in Fig.~\ref{fig:Ndist}, the GWTC-3 catalog contains a series of mergers with $m_1 \approx 10-15 M_\odot$ and $m_2 \approx 5-10 M_\odot$ that seem to be disconnected from the main population around $30-40 M_\odot$ and provide a local maximum in the distribution of detections.  None of the considered PBH models produces a maximum in this region.  We are still investigating if larger values of the spectral index, for instance, could boost the merger rates in this particular region but without overproducing mergers at the solar mass scale.   We have also explored the possibility of larger values of $\alpha$, such that the main peak in the PBH mass distribution is shifted towards this range, but this typically induces larger merger rates above $100 M_\odot$ that are inconsistent with the current limits. Since this region better corresponds to the masses of black holes observed in X-ray binaries, it is also possible that those are astrophysical black holes, whereas the main population at larger mass comes from PBHs.  Nevertheless, these mergers also have low effective spins, which is one important motivation for PBH models~\cite{DeLuca:2019buf,Franciolini:2022iaa,Garcia-Bellido:2020pwq,Koga:2022bij}.

In summary, we confirm the results  obtained in~\cite{Carr:2019kxo}, but using more precise PBH mass distributions taking into account the variation of $w$ during the PBH formation, using different curvature profiles and the LIGO/Virgo detector sensitivity of the third observing run.  If PBHs could explain well most of the GW detections, until now they fail to explain the observation of a significant number of black hole mergers around $5-15 M_\odot$.  But it is important to remain cautious when drawing conclusions because lots of uncertainties remain in the theoretical merger rates, while the statistical significance of the different black hole populations is still relatively weak and could change with future observations. Nonetheless, our results open the way to a better understanding of the possible realistic PBH mass functions that will be used in future Bayesian analyses of PBH models, and in Bayesian comparisons of PBH and astrophysical models.

\section{Discussion and conclusion}\label{sec:conclusions}

PBHs are a plausible explanation for the intriguing properties of compact object coalescences detected by GW observatories.   PBHs with stellar masses would form at the QCD epoch, when the transient reduction of the equation of state of the Universe reduces the critical overdensity threshold, facilitating the formation of a black hole from the collapse of overdensities.  In order to assess a PBH origin of GW observations, it is therefore crucial to compute precisely the impact of the QCD cross-over transition on the PBH mass distribution.
By adapting the numerical code of~\cite{escriva_solo}, we have performed for the first time numerical simulations of PBH formation from the collapse of super-horizon curvature fluctuations at the QCD epoch, taking into account the expected time variations of the equation of state.

First, our results confirm previous works relying on a constant equation of state during PBH formation and show the existence of non-trivial features in the PBH mass distribution, due to the reduction of the threshold by up to  $O(10\%)$, depending on the considered radial curvature profile.
These features take the form of a high peak between one and three solar-masses and a lump at masses between $30 M_\odot$ and $100 M_\odot$.  But, compared to previous calculations, this peak is found to be about two times lower and broader, whereas the lump is less pronounced.  The detailed analysis of the formation process unveils the origin of this phenomenon:  the collapse is not instantaneous and therefore it does not only depend on the equation of state at the time a fluctuation crosses the Hubble radius, but on its complete evolution during the formation process.  This way, the features appear  be \textit{smoothed out}.  For the first time, we also compute PBH mass distributions for various curvature profiles and show that they can be significantly altered, even if the global scheme remains similar.

Our results have important consequences for the expected merger rates of early and late PBH binaires, which can be compared to the  rates inferred from the GWTC-3 catalog.  The rate of early binaries are lower than initially expected and in agreement with observations in the solar mass range for $0.1 < f_{\rm PBH} < 1$.  At larger masses (above $20 M_\odot$), the rates become dominated by late binaries and progressively decrease, as one expects from observations. 

Finally, we used the volume sensitivity of the LIGO/Virgo detectors of the third observing run to compute the expected distrubtion of actual detections as a function of the component masses.   It is qualitatively consistent with the observation of a dominant population around $30-50 M_\odot$ that extents within the pair-instability mass gap, where some exceptional events have been observed.  PBHs also generically lead to a subdominant number of mergers with low-mass ratios like GW190814 and with masses around $2-3 M_\odot$, like GW190425.  However, the rates seem to be higher than observed between $3$ and $8 M_\odot$, but are not able to explain the observation of numerous black hole mergers between $8 M_\odot$ and $15 M_\odot$.

Our analysis therefore ends up on mitigated conclusions. On the one hand, PBHs can explain many of the GW observations in a rather unified way, while astrophysical models encounter difficulties to explain events in the pair instability mass gap and with low mass ratios, so they need to invoke different formation mechanisms.  On the other hand, all the PBH models considered fail to explain the observed population between $8 M_\odot$ and $15 M_\odot$, which could then be of stellar origin.  We remain cautious and do not want to draw definitive conclusions because of the remaining large uncertainties of PBH models, theoretical merger rates and products of population analysis.  

Overall, our work represents the first numerical exploration of PBH formation that fully takes into account the QCD phase crossover. Our approach, in particular the numerical techniques that have been used, could be applied to study the effects of other phase transitions of the very early Universe~\cite{Carr:2019kxo}, of lepton flavor violation~\cite{Bodeker:2020stj}, solitosynthesis~\cite{Garcia-Bellido:2021zgu}, a strongly interacting fermion-scalar fluid~\cite{Chakraborty:2022mwu}, on the formation of PBHs.   More accurate calculations of the gravitational-wave background from PBHs~\cite{Bagui:2021dqi,Braglia:2021wwa,Mukherjee:2021ags} could also be envisaged.
In order to constrain PBH model parameters with the current and future GW observations, realistic and accurate PBH mass distributions also have to be used in combination with Bayesian methods, which is left for a future work.  Bayesian model selection can also be used to assess the preferred models between PBHs and stellar scenarios, or the combination of both. In addition, our results can have important implications for the viability of PBH models with respect to astrophysical and cosmological limits on their abundance, in particular CMB limits~\cite{Ali-Haimoud:2016mbv,Poulin:2017bwe,Serpico:2020ehh} that apply at $ M_{\rm PBH} \gtrsim 10 M_\odot$ and debated microlensing limits~\cite{MACHO:1996qam,MACHO:2000qbb,EROS-2:2006ryy,Hawkins:2015uja,Brandt:2016aco,Green:2016xgy,Green:2017qoa,Garcia-Bellido:2017xvr,Calcino:2018mwh,Hawkins:2020rqu,Hawkins:2020zie,Hawkins:2022vqo,Petac:2022rio,Gorton:2022fyb} for $ M_{\rm PBH} \lesssim 1 M_\odot$.  Because the QCD peak is less important and broader, those seem to exclude our mass distributions with $f_{\rm PBH} = 1$ but could allow smaller dark matter fractions in PBHs.  However, one should also remember that these limits are subject to debate and uncertainties.  Microlensing events can also be seen as a hint for the existence of PBHs in the stellar-mass range~\cite{MACHO:2000qbb,2020A&A...636A..20W}.  It is therefore probably premature to firmly exclude $f_{\rm PBH} = 1$.

\section{Final note} \label{sec:note}

The present paper appeared simultaneously with another study 
\cite{Franciolini:2022tfm} on the same topic.  In this paper, the authors have also used the results of numerical simulations, whose details will be released in an upcoming paper~\cite{Musco_talk} but are not yet available at the time of submitting the second version of the present manuscript, in order to better model the effects of the QCD cross-over transition.  They set constraints on various PBH mass distributions using a Bayesian analysis of the GWTC-3 catalog.  We acknowledge that such an analysis goes beyond the (still qualitative) comparisons in our own paper.  In particular, they take into account the scaling law regime for the PBH mass (which is beyond the scope of our paper) which additionally modifies the exact shape of the QCD-induced peak.  Despite this, our work captures this effect by considering different curvature profiles and different values of $\alpha$, which allows us to obtain a modulation of different mass functions. The main result regarding the fraction of dark matter in the form of PBHs is not substantially affected by only taking into account the effect of the scaling PBH mass, precisely due to this modulation with different parameters.  We believe that our analysis better explores and provides details on the PBH formation process in comparison with~\cite{Franciolini:2022tfm}. We also study the possible effects of different curvature profiles on PBH merger rates. We therefore believe that the two papers are complementary and provide interesting results. 

It must be pointed out that these two papers have been written in total independence, until the very end.  It is worth noticing that we obtain similar results for the computation of the overdensity threshold.  However, we obtain different conclusions on the possible value of $f_{\rm PBH}$ to explain gravitational-wave observations.  In our case, we find that values between $0.1$ and $1$ could partly explain the LIGO/Virgo observations, while a much lower limit is obtained in~\cite{Franciolini:2022tfm}.  Different assumptions could have driven such a difference.  It is probably premature to draw a definitive conclusion on its origin.  Nevertheless, we point out hereafter some differences that could play a role.  

First, even if both studies rely on the same literature for the computation of the merger rate of early binaries, in particular the suppression factor $f_{\rm sup}$, we used a slightly different hypothesis.  In our paper, $f_{\rm sup}$ is calculated in the limit of a sharp mass distribution, motivated by the QCD-induced peak, in order to avoid an unphysical and arbitrary dependence in the low-mass cut in the PBH mass distribution that would suppress merger rates for unequal mass binaries.  
A different prescription was used by Franciolini et al~\cite{Franciolini:2022tfm}, where $f_{\rm sup}$ is not computed in the above-mentioned limit but for each considered mass function, using an additional free parameter to describe the high-mass cut in the PBH distribution.   We have included an Appendix to detail our calculation of this rate suppression factor, to motivate our assumption and to comment on the possible choices made in the literature.  However, we argue that differences in $f_{\rm sup}$ compared to~\cite{Franciolini:2022tfm}, even if significant, are not the main driver of the different conclusions for $f_{\rm PBH}$.

A second difference comes from the range of $n_{\rm s}$ that we consider. The MCMC analysis of~\cite{Franciolini:2022tfm} results in values of $n_{\rm s}$ around $0.8$ favoured by observations, whereas we focus on values around $n_{\rm s} \approx 0.97$ only, in order to avoid an overproduction of light or heavy PBHs.  
For such a low value of the spectral index, the QCD peak is highly suppressed and one has instead a strong increase of $f(M_{\rm PBH}>10 M_\odot)$, resulting in much larger rates for the same value of $f_{\rm PBH}$.  A quick calculation showed that for $n_{\rm s} \approx 0.8$ and considering early binaries only, we would also need $f_{\rm PBH}$ of order $10^{-3}$ to obtain merger rates below the ones inferred by LIGO/Virgo.  In \cite{Franciolini:2022tfm}, values of the spectral index close to unity are disfavored by the MCMC analysis, essentially because they produce too many subsolar-mass black hole mergers.  Even if this has not been detailed in our paper, we have also estimated the expected number of merger detections involving a subsolar-mass black hole, for the O3a $\langle VT \rangle$ sensitivity from~\cite{LIGOScientific:2021job} and found that at most $\mathcal O(1)$ events are expected with our mass distributions for $f_{\rm PBH} \approx 0.1$.  For this calculation, we impose a lower mass cut  $m_2 > 0.2 M_\odot$ and a mass ratio cut $q_r > 0.1$ consistent with LVK searches, whereas \cite{Franciolini:2022tfm} have considered all possible combinations of binaries, leading to a much higher expected number of detections of subsolar-mass black holes.  We argue that binaries with $q_r<0.1$ could indeed be detected in LVK searches but that standard waveforms probably do not optimally describe the merging of such systems.  As a result, we believe that including such low mass ratios is indeed motivated but still relatively speculative.  So each method can be debated. Taking into account all those differences, it is therefore possible that our results are not inconsistent with each other but originate from different hypotheses and methods. 

A third difference comes from the inclusion in our paper of the merger rates of both early and late binaries, whereas \cite{Franciolini:2022tfm} considers early binaries only. This could also indirectly impact the estimation of $f_{\rm PBH}$ through different values of the $n_{\rm s}$ needed to fit the data.

Future investigations are therefore needed to better understand the origin of these differences and to reduce the various sources of uncertainties, which we leave for a future work. Nevertheless, the two papers bring important insights on the crucial effects of the QCD epoch for PBHs and on the theoretical uncertainties and model dependence that inevitably remain in the computation of PBH mass distributions and merger rates.  Both conclude positively on the possibility that a part but not the totality of the black hole mergers observed by LIGO/Virgo are of primordial origin.  The different hypotheses that were used lead to different conclusions about the dark matter fraction that could be made of stellar-mass PBHs.  Future data will certainly also help to test these hypotheses and distinguish the different PBH models.

\appendix

\section{Appendix: Rate suppression of early PBH binaries}

In the merger rates of early PBH binaries, we have introduced a suppression factor $f_{\rm sup}$, following~\cite{Hutsi:2020sol,Raidal:2018bbj}. {It can be written as the product of two functions, $S_1$ and $S_2$, where 
\be
S_1 \approx 1.42 \left[ \frac{(\langle m_{\rm PBH}^2 \rangle/\langle m_{\rm PBH} \rangle^2)}{\bar N + C} + \frac{\sigma_{\rm M}^2}{f_{\rm PBH}^2}\right]^{-21/74} {\rm e}^{-\bar N}
\ee
accounts for binary disruption by matter fluctuations of variance $\sigma^2_{\rm M} \simeq 0.004$ and nearby PBHs.  It strongly depends on the number $\bar N$ of nearby black holes that can disrupt the binary before matter-radiation equality, given by
\be
\bar N \approx \frac{m_1 + m_2}{\langle m \rangle} \frac{f_{\rm PBH}}{f_{\rm PBH} + \sigma_{\rm M}}~.
\ee
 The function $C$ encodes the transition between small and large $\bar N$ limits and an analytical expression can be found in~\cite{Hutsi:2020sol}, 
\bea
C &\simeq& f^2_{\rm PBH} \frac{\langle m^2_{\rm PBH} \rangle / \langle m_{\rm PBH} \rangle^2}{\sigma^2_{\rm M}} \nonumber \\
 & \times & \left\{ \left[\frac{\Gamma(29/37)}{\sqrt{\pi}}U\left(\frac{21}{74},\frac{1}{2}, \frac{5 f^2_{\rm PBH}}{6 \sigma^2_{\rm M}}\right)\right]^{-\frac{74}{21}} - 1\right\}^{-1}~, 
\eea
where $\Gamma$ is the Euler function and $U$ is the confluent hypergeometric function.   One should note that there exists a subtle ambiguity on the definition of the PBH mass function that induces significant differences in the averaging procedure, in the suppression factor, and in the merger rates.  Following~\cite{Hutsi:2020sol,Raidal:2018bbj}, $\langle m \rangle$ are $\langle m^2 \rangle$ are obtained from the PBH number density $n$,
\be
\langle m^p \rangle \equiv (1/n) \int m^p {\rm d} n
\ee
The computation of these quantities depends on the way one defines the PBH mass function and is different if it is defined in terms of the PBH number density $n$ or the PBH density $\rho$, or if the normalisation changes.   In Table~\ref{tab:my_label} we have summarized the possible definitions of the PBH mass function, how these are related, and how to compute $\langle m \rangle$ and $\langle m^2 \rangle$ in each case.  Assuming the same normalization, we point out that there is a conversion factor $m / \langle m \rangle $ that was not considered in~\cite{Franciolini:2022tfm}, leading to an inconsistency with~\cite{Hutsi:2020sol} which uses exactly the same definition and normalization of the PBH mass distribution, but a different merger rate formula that takes into account this conversion. In short, the mass distribution is defined as in the second colummn of Table~\ref{tab:my_label} but the averaged procedure is the one of the fourth column, corresponding to another definition of the mass distribution.   We also spotted a typo in the Eq. A4 of~\cite{Clesse:2020ghq} that has an incorrect mass dependence, but the correct formula was used for the calculations and figures. 

\begin{table}[t]
    \centering
    \begin{tabular}{|c|c|c|c|}
       \hline
        This paper & Raidal et al.~\cite{Raidal:2018bbj} &  &  \\
        Clesse et al.$^{*}$~\cite{Clesse:2020ghq}  & Franciolini et al.$^{*}$~\cite{Franciolini:2022tfm} & Hutsi et al.~\cite{Hutsi:2020sol} & Kocsis et al.~\cite{Kocsis:2017yty} \\
        Bagui et al.~\cite{Bagui:2021dqi} & Hall et al.~\cite{Hall:2020daa} &  & \\
        \hline
        $f\equiv \dfrac{1}{\rho_{\rm PBH}} \dfrac{{\rm d} \rho_{\rm PBH} }{{\rm d}\ln m}   $  & $ \psi_{1} \equiv \dfrac{1}{\rho_{\rm PBH}} \dfrac{{\rm d} \rho_{\rm PBH} }{{\rm d} m}   $  & $ \psi_{2} \equiv \dfrac{1}{n_{\rm PBH}} \dfrac{{\rm d} n_{\rm PBH} }{{\rm d} \ln m}   $ & $ \psi_{3}\equiv \dfrac{1}{n_{\rm PBH}} \dfrac{{\rm d} n_{\rm PBH} }{{\rm d}  m}   $ \\
        \hline
        $f =  m \psi_1  $ & $\psi_1 = f/m $ & $ \psi_2 = f \langle m \rangle /m$   & $ \psi_3= \langle m \rangle f / m^2  $  \\ 
        $ = m \psi_2 / \langle m \rangle $ & $= \psi_2 / \langle m \rangle $  & $ = \langle m \rangle \psi_1$ & $  = \langle m \rangle \psi_1 /m $ \\
        $ = m^2 \psi_3 / \langle m \rangle   $  & $ =  m \psi_3 / \langle m \rangle   $  & $= m \psi_3 $  & $ = \psi_2/m $ \\
       \hline
       $\int f {\rm d} \ln m = 1$ & $\int \psi_1 {\rm d} m = 1$ & $\int \psi_2  {\rm d} \ln m = 1$ & $ \int \psi_3 {\rm d} m = 1$  \\
        \hline
       $\langle m \rangle = \left( \int \dfrac{f}{m} {\rm d} \ln m \right)^{-1} $  & $\left( \int \dfrac{\psi_1}{m} {\rm d} m \right)^{-1}$  & $ \int m \psi_2 {\rm d} \ln m $ & $ \int m \psi_3 {\rm d} m $ \\
       \hline
       $ \langle m^2 \rangle = \langle m \rangle \int  m f {\rm d} \ln m $ & $ \langle m \rangle \int m \psi_1 {\rm d } m $   &  $ \int m^2 \psi_2 {\rm d} \ln m  $ &  $ \int m^2 \psi_3 {\rm d}{m } $ \\
       \hline
    \end{tabular}
    \caption{Different definitions of the normalized PBH mass distribution proposed in various references with their conversion, their normalisation rule and the corresponding $\langle m \rangle$ and $\langle m^2 \rangle$.  The asterisk denotes the references in which an inconsistency has been found (see details in the text).   }
    \label{tab:my_label}
\end{table}

When the mass function is defined in terms of the PBH density, as in the present paper, one obtains that the abundance of low-mass PBHs can strongly change the value of $\langle m \rangle$.  In other terms, for broad mass distributions like ours, there are numerous light black holes that tend to reduce the value of the average black hole mass, compared to a peaked log-normal mass function.  In turn, this increases $\bar N$ and reduces $f_{\rm sup}$ that becomes strongly dependent on the chosen lower mass cut.   However, we argue that such a rate suppression is artificial because it implicitly assumes that tiny black holes nearby a binary are able to disrupt it.  It is therefore suspected that the prescriptions proposed by~\cite{Hutsi:2020sol,Raidal:2018bbj} only apply to peaked mass distributions and that their extension to broader distributions is a complex problem.   Since our mass functions have a high peak induced by the QCD transition, we therefore consider the limit obtained for a monochromatic mass function and assume that $\langle m_{\rm PBH}^2 \rangle/\langle m_{\rm PBH} \rangle^2 \simeq 1$ and $\bar N \approx 2$.  This assumption provides a better estimation of the rates around the QCD-induced peak.  However, we suspect that the merging rate of subsolar-mass binaries is additionally suppressed, while the rate of intermediate-mass binaries is likely less suppressed since the dominant PBHs from the QCD peak would hardly be able to disrupt them.   The latter effect would not dramatically impact our conclusions because the merging rate of intermediate-mass binaries is, in our case, dominated by the capture channel.    If one follows the averaging procedure of~\cite{Franciolini:2022tfm}, one gets instead a strong dependence on the high-mass PBH distribution but which can be cut-off by the transition in the primordial power spectrum.  

The second factor $S_2$ comes from the binary disruption in early-forming clusters and can be approximated today by
\be
S_2 \approx \min \left(1,9.6 \times 10^{-3} f_{\rm PBH}^{-0.65} {\rm e}^{0.03 \ln^2 f_{\rm PBH}} \right),
\ee
which starts to contribute when $f_{\rm PBH} \geq 0.0035$. At higher redshifts, one can simply replace $f_{\rm PBH} \rightarrow (t(z)/t_0)^{0.44}f_{\rm PBH}$ in $S_2$. 
It must also be pointed out that the merger rate of perturbed early binaries (that remain bound after interactions with other PBHs) could be of the same order or even surpass the rate of early binaries.  However, a conservative calculation of these rates has been performed only for a monochromatic or peaked mass function~\cite{Vaskonen:2019jpv} and it is still subject to large uncertainties, so it is unclear how to generalize it to complex broad mass functions like the ones considered in this work.  Consequently, we leave this interesting issue for a future work and assume that the merger rate of perturbed binaries is subdominant.  

Finally, one can get a useful approximation, valid when $0.1<f_{\rm PBH}<1$, by neglecting the exponential factor in $S_2$ and by setting $\langle m \rangle^2 = \langle m^2 \rangle$, $\bar N = 2$ in $S_1$, and neglecting terms in $\sigma_{\rm M}$, 
\be 
f_{\rm sup} \approx 2.3 \times 10^{-3} f_{\rm PBH}^{-0.65} .
\ee
This approximation is useful and when one plug it in the rate formula, one directly sees that $f_{\rm PBH} \approx 0.1$ and  $f(m_{\rm PBH} \approx 2 M_\odot) \approx 1$ give to merger rates comparable to the ones inferred from LIGO/Virgo observations.  

\acknowledgments
  
The authors warmly thank Vicente Atal, Bernard Carr, Gabriele Franciolini, Juan Garc\`ia-Bellido, Jaume Garriga, Cristiano Germani, Javi Gomez-Subils, Florian K\"uhnel, Ilia Musco, Theodoros Papanikolaou, Yuichiro Tada, Hardi Veermae, Shuichiro Yokoyama and  Chulmoon Yoo for usefull discussions and comments.  We also thank Gabriele Franciolini, Ilia Musco, Paolo Pani and Alfredo Urbano for sharing their draft and feeding the discussion on the comparison of our respective results.  We acknowledge support from the Belgian Francqui Foundation through a Francqui Start-up grant and from the Belgian Fund for Research F.R.S.-FNRS through a IISN convention. E.B. is supported by the FNRS-IISN (under grant number 4.4501.19). S.C. and A.E. acknowledges support from the Francqui Foundation.

\bibliographystyle{JHEP}
\bibliography{refs4.bib}

\end{document}